\def\mydouble #1 { }\def\mysingle #1 {#1} 
\def\delete #1 { }  
\delete{
\input{quire}
\shfootline={
}
\shhtotal=13.42in 
\shthickness=0pt 
\shvcorrection=0pt
\shhcorrection=0pt
\shvoffset=-1in 
\shoutline=0pt 
\shstaplewidth=0pt 
\shstaplelength=0pt 
\shcrop=0pt 
\vorigin=0.5in 
\horigin=0.15in 
\htotal=6.655in 
\vtotal=10.285in 
\latexquire
\qtwopages
\mag=810
}
\documentstyle[12pt]{article}
\mydouble {\def\doublespace{\def\baselinestretch{1.8}\Large\normalsize}
\doublespace}
\def\B{\break}
\def\betaE#1{\displaystyle{#1\over k_{\rm B}T}}
\def\half{{\textstyle{1\over2}}}
\def\halfs{{\scriptstyle\frac12}}
\def\quarter{{\textstyle{1\over4}}}
\def\threequarter{{\textstyle{3\over4}}}
\def\fract#1#2{{\textstyle{#1\over#2}}}
\def\fracd#1#2{{\textstyle{\hbox{\footnotesize$#1$}\over
\hbox{\footnotesize\vphantom{\normalsize N}$#2$}}}}
\def\tag#1{\label{#1}}
\def\wb{{\overline W}}
\def\wo{{\overline w}}
\def\mub{{\overline\sigma}}
\def\bA{{\overline\alpha}}
\def\bB{{\overline\beta}}
\def\bCE{{\overline{\cal E}}}
\def\bD{{\overline\Delta}}
\def\bE{{\overline E}}
\def\bF{{\overline f}}
\def\bK{{\overline K}}
\def\bL{{\overline\Lambda}}
\def\bO{{\overline\Omega}}
\def\bP{{\overline\phi}}
\def\bT{{\overline\theta}}

\def\cE{{\cal E}}
\def\cL{{\cal L}}
\def\cM{{\cal M}}
\def\xr{{\rm x}}
\def\yr{{\rm y}}
\def\zr{{\rm z}}
\def\xrb{{\overline{\rm x}}}
\def\yrb{{\overline{\rm y}}}
\def\zrb{{\overline{\rm z}}}
\def\lps{{l_{\psi}}}

\def\bS{{\overline s}}
\def\bU{{\overline u}}
\font\mysxm=msam10 at 10pt\def\lsim{\,\hbox{\mysxm\char'056}\,}
\font\mysym=msbm10 at 12pt\def\bbZ{\hbox{\mysym Z}}
\font\mycal=cmmi12 at 16pt\def\area{\hbox{\mycal a}}
\def\mysig{\hbox{\sl\char'006}}
\def\myref#1{$^{\rm (#1)}$}
\def\cmyref#1{,$\!^{\rm (#1)}$}
\def\pmyref#1{.$^{\rm (#1)}$}
\def\smyref#1{;$^{\rm (#1)}$}
\def\rpages#1#2#3#4{{\bf #1}:#3 (#2)}
\def\rspages#1#2#3{{\bf #1}:#3 (#2)}

\def\ppages#1#2{p.\ #1}
\def\pppages#1#2{pp.\ #1--#2}
\def\nppages#1#2{#1--#2}
\def\book#1{{\it #1}}
\def\jour#1{{\it #1}}
\catcode`\@=11
\def\@sect#1#2#3#4#5#6[#7]#8{\ifnum #2>\c@secnumdepth \def\@svsec{}\else
 \refstepcounter{#1}\edef\@svsec{\csname the#1\endcsname.\ }\fi
 \@tempskipa #5\relax \ifdim \@tempskipa>\z@ \begingroup #6\relax
 \@hangfrom{\hskip #3\relax\@svsec}{\interlinepenalty \@M #8\par} \endgroup
 \csname #1mark\endcsname{#7}\addcontentsline
 {toc}{#1}{\ifnum #2>\c@secnumdepth \else
 \protect\numberline{\csname the#1\endcsname}\fi #7}\else
 \def\@svsechd{#6\hskip #3\relax \@svsec #8\csname #1mark\endcsname
 {#7}\addcontentsline {toc}{#1}{\ifnum #2>\c@secnumdepth \else
 \protect\numberline{\csname the#1\endcsname}\fi #7}}\fi \@xsect{#5}}
\def\section{\@startsection {section}{1}{\z@}{-3.5ex plus -1ex minus
 -.2ex}{2.3ex plus .2ex}{\normalsize\bf}}
\def\subsection{\@startsection{subsection}{2}{\z@}{-3.25ex plus -1ex minus
 -.2ex}{1.5ex plus .2ex}{\normalsize\bf}}
\def\subsubsection{\@startsection{subsubsection}{3}{\z@}{-3.25ex plus
 -1ex minus -.2ex}{1.5ex plus .2ex}{\normalsize\bf}}
\catcode`\@=12

\def\mysp{$\,$}
\font\mybmi=cmmib10 at 12pt
\font\mybsy=cmbsy10 at 12pt
\def\s{\hspace*{.3mm}}
\def\st{\!\!\!}

\def\myfig #1 #2 #3 #4 #5{\begin{figure}[htbp]
 \begin{center}\mbox{\psboxto(#1;#2){#3}}\end{center}%
 \parbox{\hsize}{{\footnotesize{\bf Fig.\ {#4}.}\ {#5}}} \end{figure}}
\def\myskip #1 #2{\item[{\bf Fig.\ {#1}.}]{#2}}
\def\fig #1 #2 #3 #4 #5 #6\par {
\if#1A\def\figcA{\myskip #5 {#6}\par}\def\myfA{\myfig #2 #3 #4 #5 {#6}\par}\fi
\if#1B\def\figcB{\myskip #5 {#6}\par}\def\myfB{\myfig #2 #3 #4 #5 {#6}\par}\fi
\if#1C\def\figcC{\myskip #5 {#6}\par}\def\myfC{\myfig #2 #3 #4 #5 {#6}\par}\fi
\if#1D\def\figcD{\myskip #5 {#6}\par}\def\myfD{\myfig #2 #3 #4 #5 {#6}\par}\fi
\if#1E\def\figcE{\myskip #5 {#6}\par}\def\myfE{\myfig #2 #3 #4 #5 {#6}\par}\fi
\if#1H\def\figcH{\myskip #5 {#6}\par}\def\myfH{\myfig #2 #3 #4 #5 {#6}\par}\fi
\if#1I\def\figcI{\myskip #5 {#6}\par}\def\myfI{\myfig #2 #3 #4 #5 {#6}\par}\fi
\if#1J\def\figcJ{\myskip #5 {#6}\par}\def\myfJ{\myfig #2 #3 #4 #5 {#6}\par}\fi
\if#1K\def\figcK{\myskip #5 {#6}\par}\def\myfK{\myfig #2 #3 #4 #5 {#6}\par}\fi
\if#1L\def\figcL{\myskip #5 {#6}\par}\def\myfL{\myfig #2 #3 #4 #5 {#6}\par}\fi
\if#1M\def\figcM{\myskip #5 {#6}\par}\def\myfM{\myfig #2 #3 #4 #5 {#6}\par}\fi
\if#1N\def\figcN{\myskip #5 {#6}\par}\def\myfN{\myfig #2 #3 #4 #5 {#6}\par}\fi
\if#1O\def\figcO{\myskip #5 {#6}\par}\def\myfO{\myfig #2 #3 #4 #5 {#6}\par}\fi
\if#1P\def\figcP{\myskip #5 {#6}\par}\def\myfP{\myfig #2 #3 #4 #5 {#6}\par}\fi
\if#1Q\def\figcQ{\myskip #5 {#6}\par}\def\myfQ{\myfig #2 #3 #4 #5 {#6}\par}\fi
\if#1R\def\figcR{\myskip #5 {#6}\par}\def\myfR{\myfig #2 #3 #4 #5 {#6}\par}\fi
\if#1S\def\figcS{\myskip #5 {#6}\par}\def\myfS{\myfig #2 #3 #4 #5 {#6}\par}\fi
\if#1T\def\figcT{\myskip #5 {#6}\par}\def\myfT{\myfig #2 #3 #4 #5 {#6}\par}\fi
\if#1U\def\figcU{\myskip #5 {#6}\par}\def\myfU{\myfig #2 #3 #4 #5 {#6}\par}\fi
\if#1V\def\figcV{\myskip #5 {#6}\par}\def\myfV{\myfig #2 #3 #4 #5 {#6}\par}\fi
\if#1W\def\figcW{\myskip #5 {#6}\par}\def\myfW{\myfig #2 #3 #4 #5 {#6}\par}\fi
\if#1X\def\figcX{\myskip #5 {#6}\par}\def\myfX{\myfig #2 #3 #4 #5 {#6}\par}\fi
\if#1Y\def\figcY{\myskip #5 {#6}\par}\def\myfY{\myfig #2 #3 #4 #5 {#6}\par}\fi
\if#1Z\def\figcZ{\myskip #5 {#6}\par}\def\myfZ{\myfig #2 #3 #4 #5 {#6}\par}\fi
\if#1z\def\figcz{\myskip #5 {#6}\par}\def\myfz{\myfig #2 #3 #4 #5 {#6}\par}\fi
}
\def\figtw #1 #2 #3 #4 #5 #6 #7 #8\par {
\def\figcG{\myskip #7 {#8}\par}\def\myfG{\begin{figure}[htbp]
 \begin{center}\mbox{\psboxto(#1;#2){#3}}\end{center}%
 \begin{center}(a)\end{center}%
 \begin{center}\mbox{\psboxto(#4;#5){#6}}\end{center}%
 \begin{center}(b)\end{center}%
 \parbox{\hsize}{{\footnotesize{\bf Fig.\ {#7}.}\ {#8}}} \end{figure}}
\def\myfGa{\begin{figure}[htbp]
 \begin{center}\mbox{\psboxto(#1;#2){#3}}\end{center}%
 \begin{center}(a)\end{center}%
 \parbox{\hsize}{{\footnotesize{\bf Fig.\ {#7}.}\ {#8}}} \end{figure}}
\def\myfGb{\begin{figure}[htbp]
 \begin{center}\mbox{\psboxto(#4;#5){#6}}\end{center}%
 \begin{center}(b)\end{center}%
 \parbox{\hsize}{{\footnotesize{\bf Fig.\ {#7}.}\ {(Continued)}}}
 \end{figure}}}
\def\figth #1 #2 #3 #4 #5 #6 #7 #8 #9\par {
\def\figcF{\myskip #8 {#9}\par}\def\myfF{\begin{figure}[htbp]
 \begin{center}\mbox{\psboxto(#4;#5){#6}}\qquad%
 \mbox{\psboxto(#4;#5){#7}}\end{center}%
 \begin{center}(a){\hspace{6.4cm}}(b)\end{center}
 \vspace{0.5cm}
 \begin{center}\mbox{\psboxto(#1;#2){#3}}\end{center}%
 \begin{center}(c)\end{center}%
 \parbox{\hsize}{{\footnotesize{\bf Fig.\ {#8}.}\ {#9}}} \end{figure}}
\def\myfFa{\begin{figure}[htbp]
 \begin{center}\mbox{\psboxto(#4;#5){#6}}\qquad%
 \mbox{\psboxto(#4;#5){#7}}\end{center}%
 \begin{center}(a){\hspace{6.4cm}}(b)\end{center}
 \parbox{\hsize}{{\footnotesize{\bf Fig.\ {#8}.}\ {#9}}} \end{figure}}
\def\myfFb{\begin{figure}[htbp]
 \begin{center}\mbox{\psboxto(#1;#2){#3}}\end{center}%
 \begin{center}(c)\end{center}%
 \parbox{\hsize}{{\footnotesize{\bf Fig.\ {#8}.}\ {(Continued)}}}
 \end{figure}}}

\fig A {\hsize} {0cm} {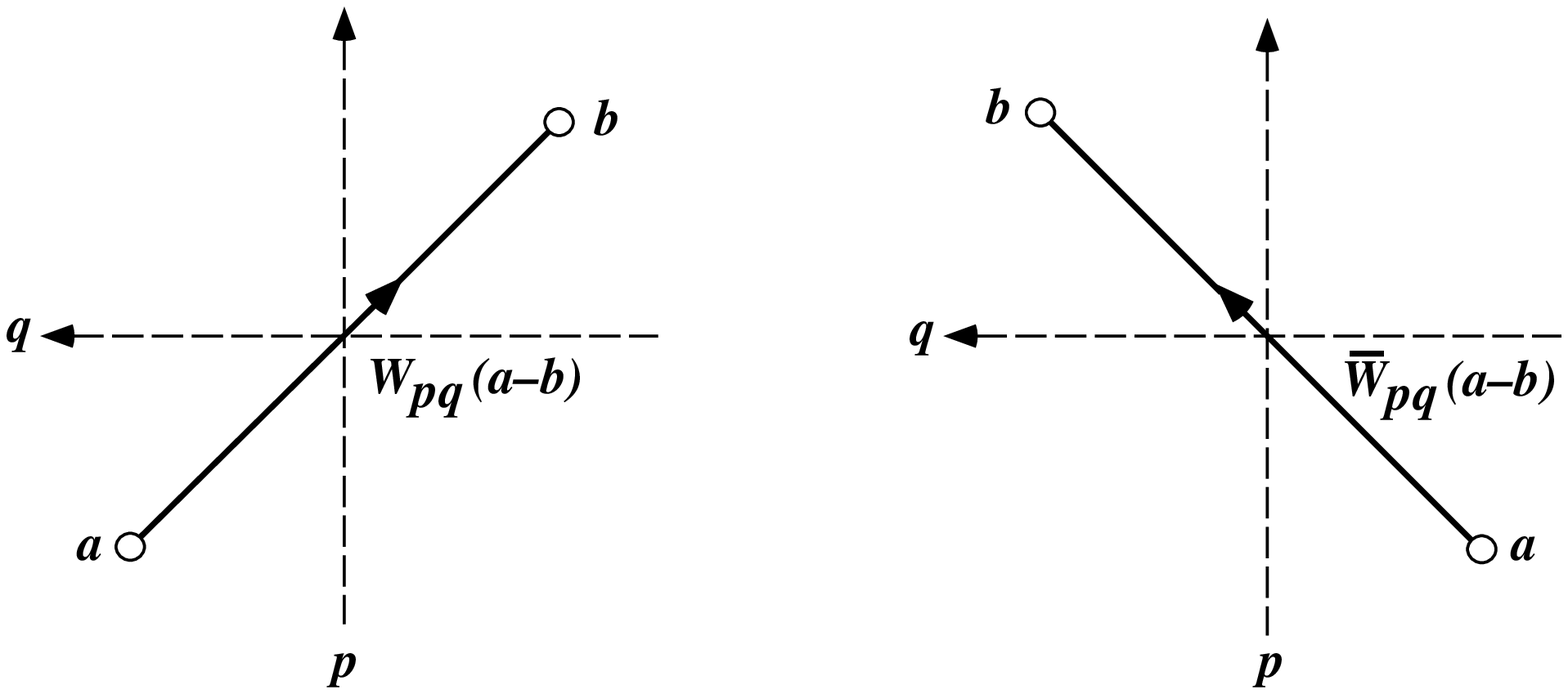} {1} {Boltzmann weights ${W_{pq}(a-b)}$
and ${\wb_{pq}(a-b)}$ for two types of edges connecting spins $a$ and $b$. As
${W(a-b)}\ne {W(b-a)}$, and similarly for $\wb$, we need
to put arrows on the edges to distinguish the two different choices.}\par

\fig B {\hsize} {0cm} {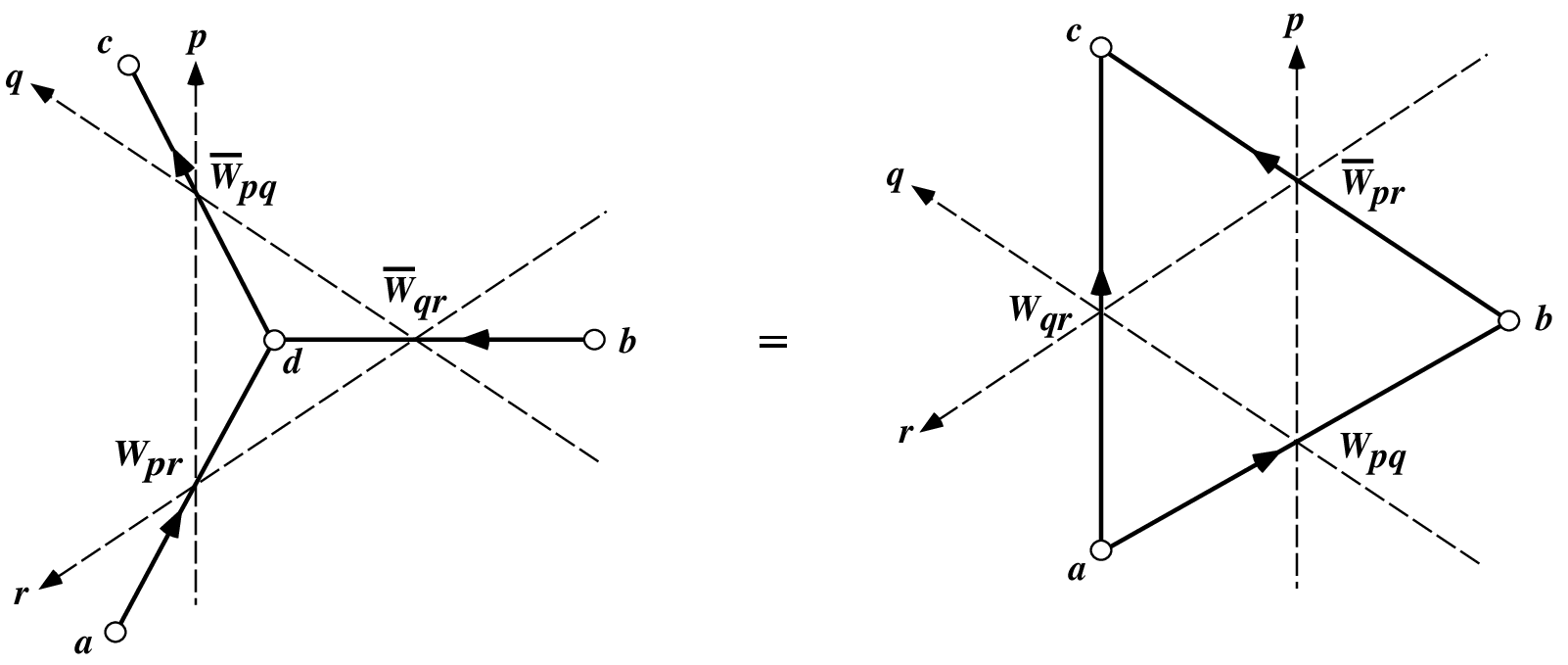} {2} {The star-triangle relations,
allowing one to move the rapidity line $p$ through a vertex, which is the
intersection of the two other rapidity lines $q$ and $r$.}\par

\fig C {6cm} {6cm} {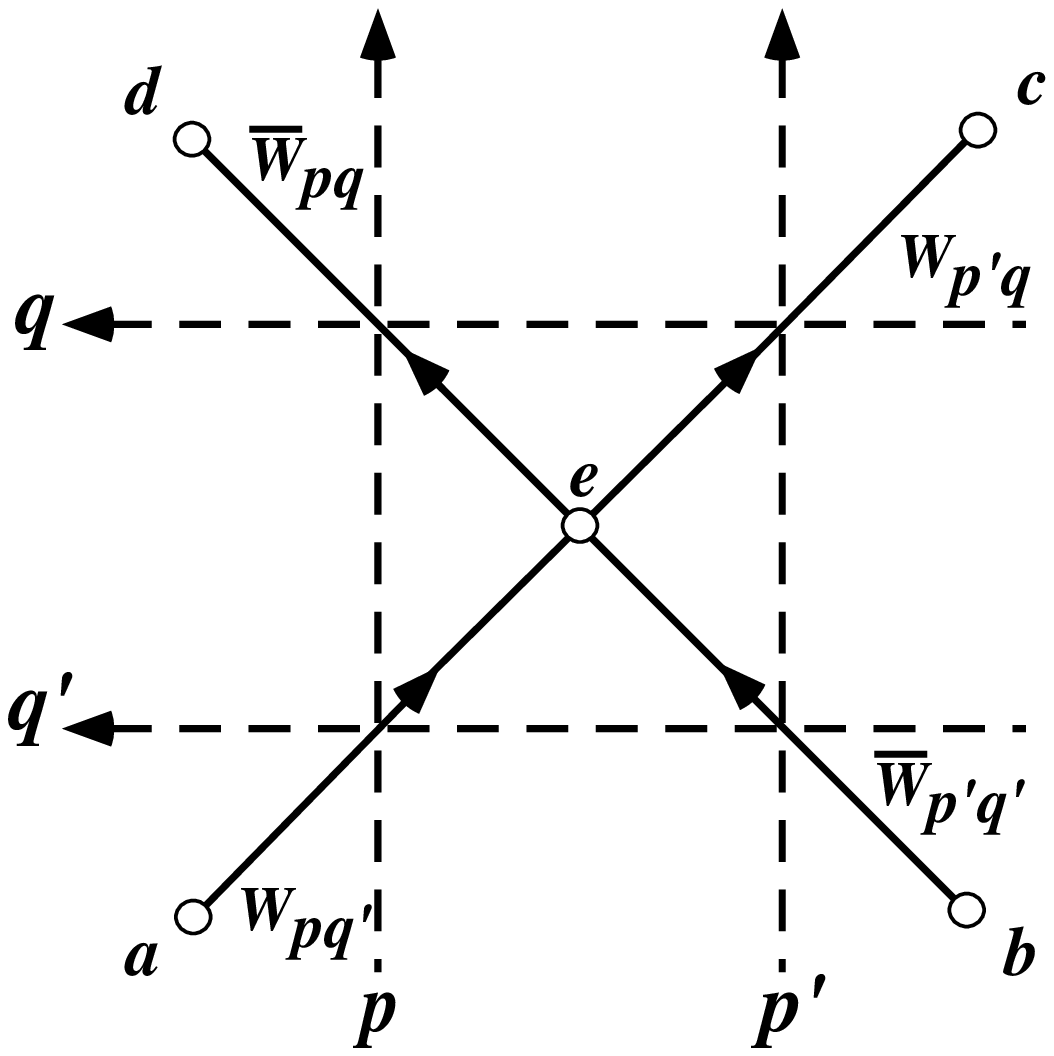} {3} {Gauge transformation at a particular site
$e$ adds gauge factors to each of the four Boltzmann weights associated with
the four bonds meeting in $e$. The total contributions add up to zero.}\par

\fig D {7cm} {6cm} {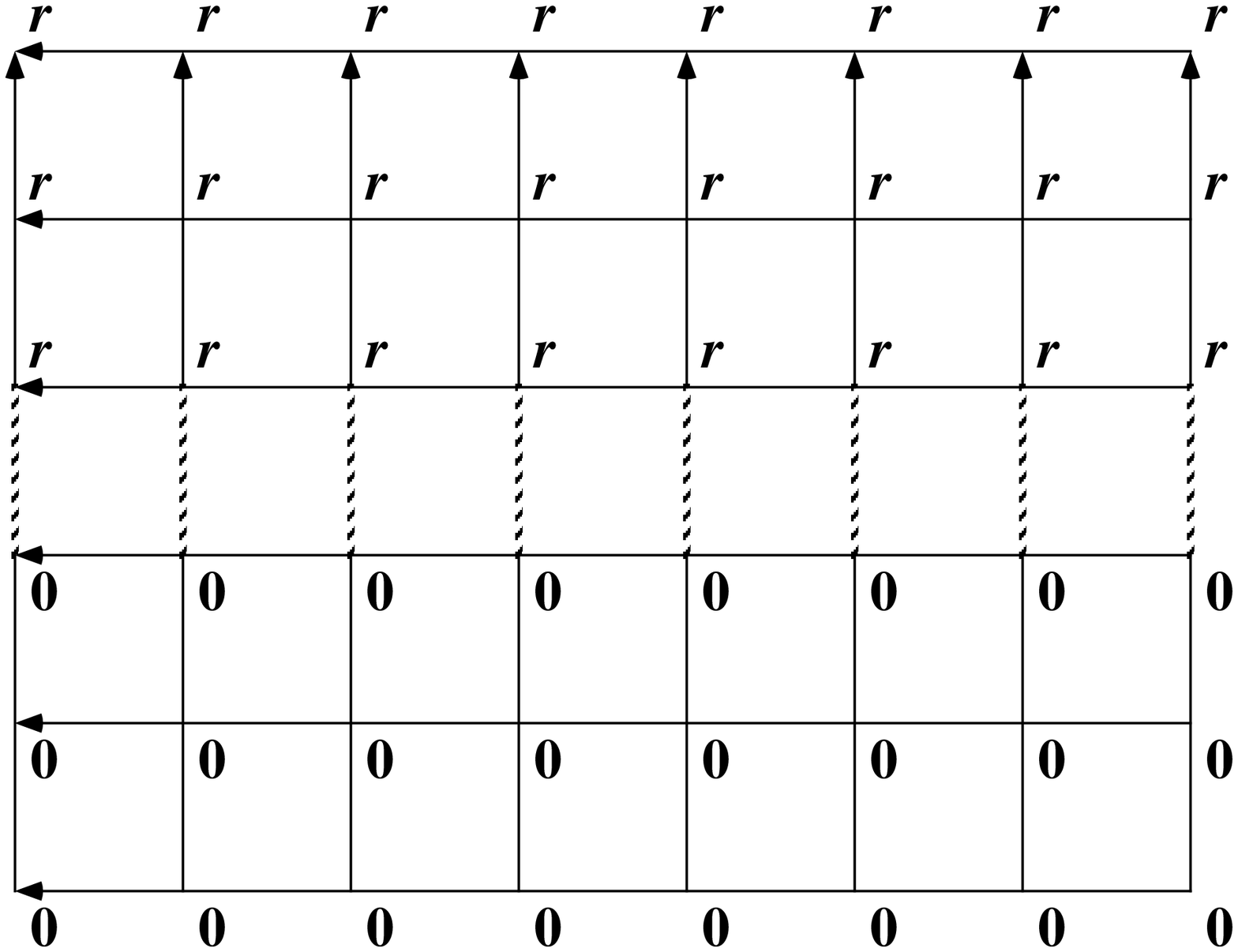} {4} {Interface at zero temperature in a
lattice with $\cL$ rows and $\cM$ columns. The interaction between a
vertical nearest-neighbor pair of spins, $\omega^n$ and $\omega^{n'}$, is
$\cE(n-n')$, while it is $\bCE(n-n')$ for a horizontal pair.}\par

\fig E {7cm} {6cm} {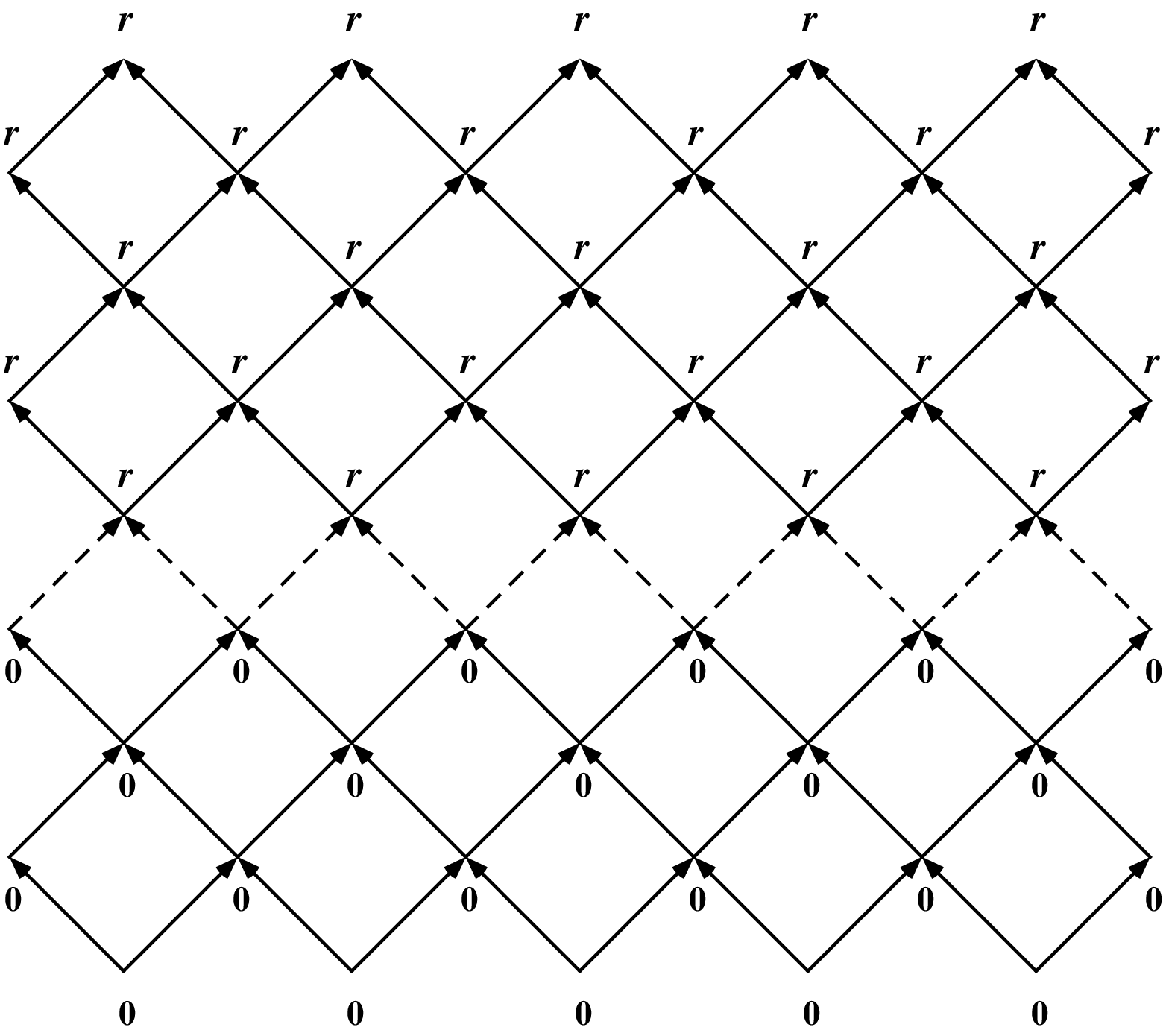} {5} {Interface at zero temperature for the
symmetric case with $W=\wb$ in a lattice with $M$ diagonal columns and $L$
diagonal rows, with skew boundary conditions imposed on top and bottom
rows and periodic boundary conditions on the utmost left and right
columns.}\par

\figth {8cm} {6cm} {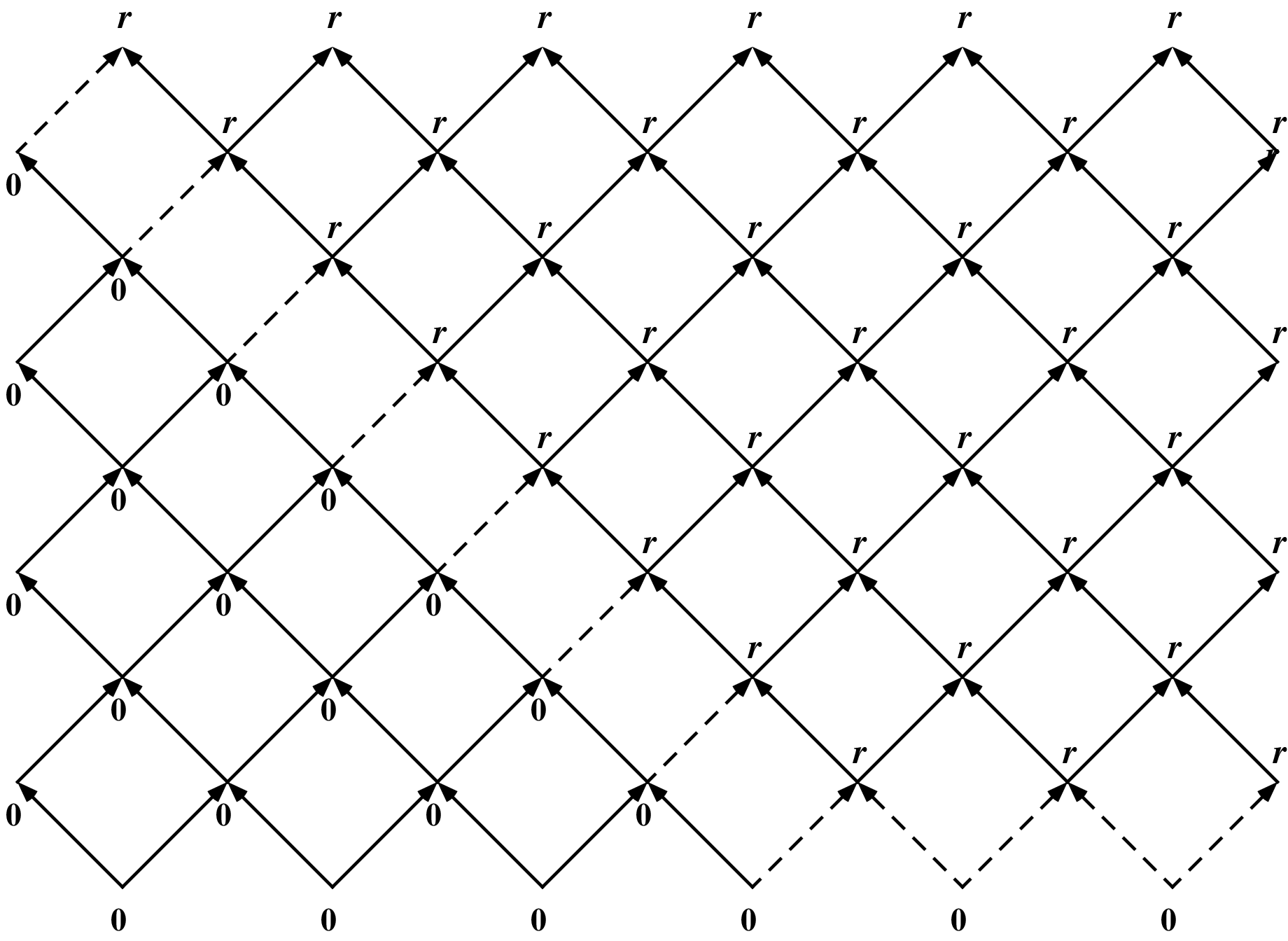} {6cm} {3cm} {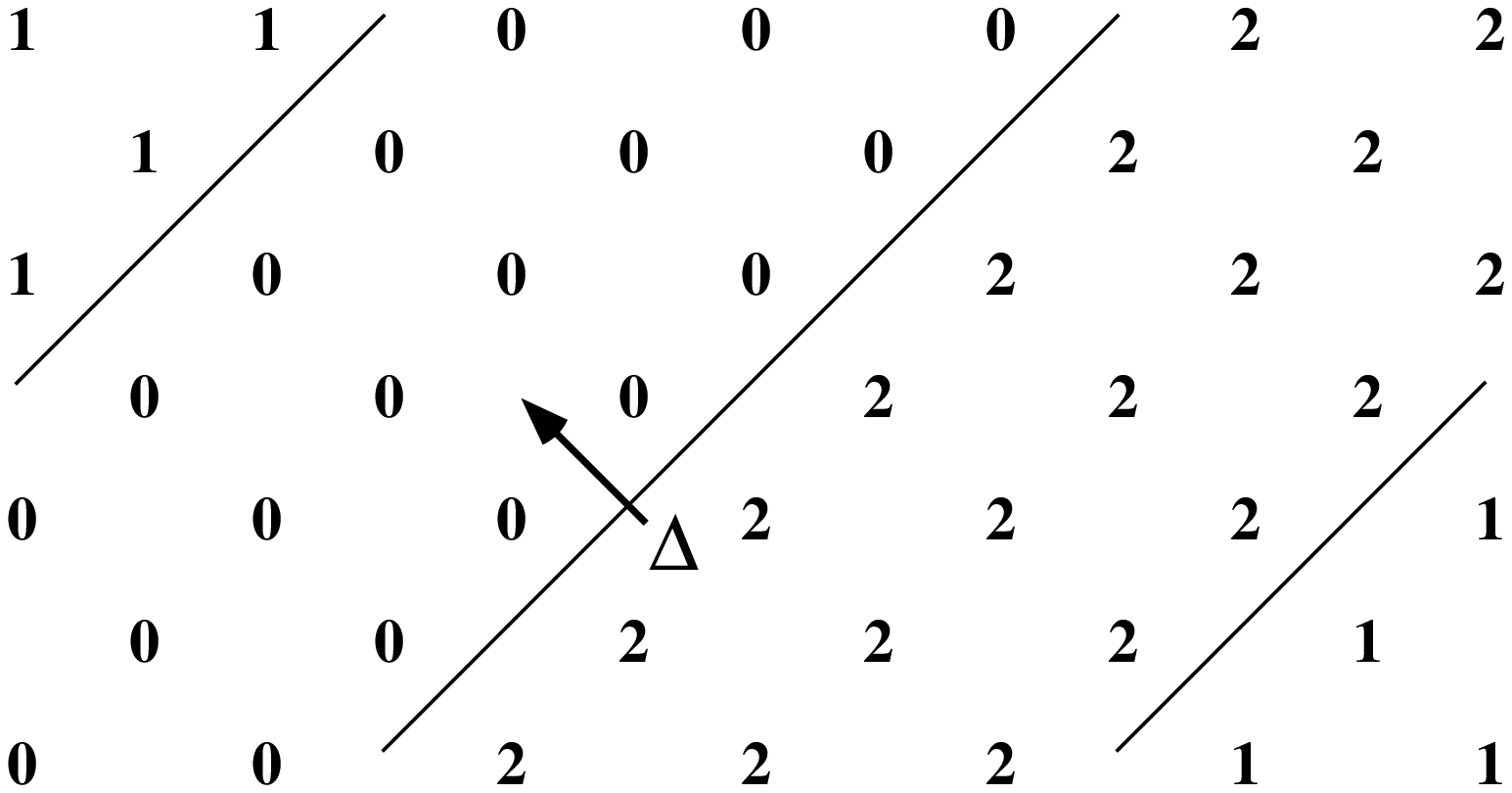} {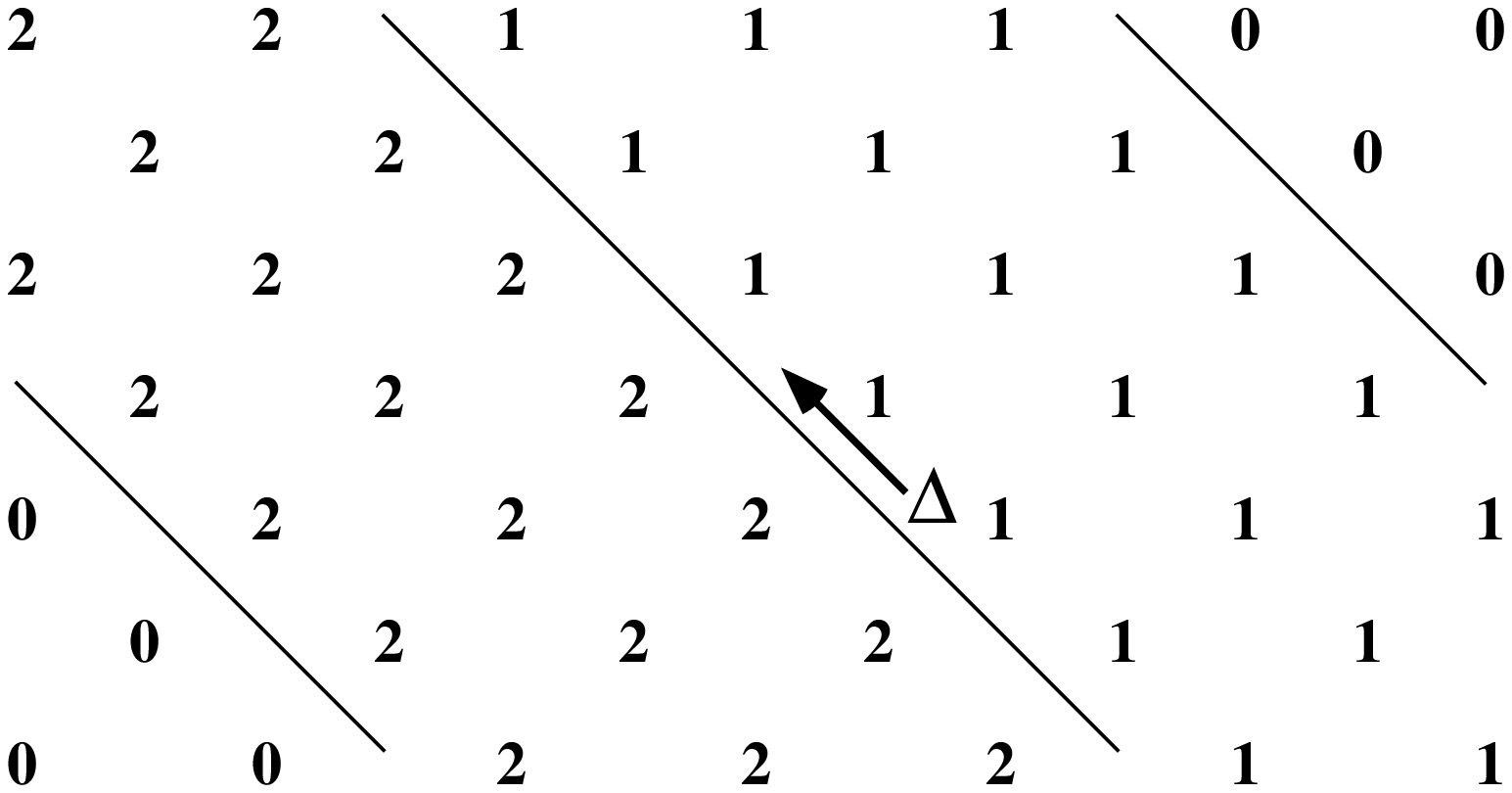} {6} %
{(a) For the case with skewed boundary conditions in both directions,
an interface is allowed to wind around the torus of vertical perimeter
$L$ and horizontal perimeter $M$, crossing the seams of modified bonds
several times. Thus for the Ostlund-Huse model, a single-step interface
with tension $\epsilon_1$ prefers to walk perpendicular to the
chiral field and then continue by crossing the seams.
(b) A double-step interface with tension $\epsilon_2$ tends to walk parallel
to the chiral field.
(c) If fixed boundary conditions are imposed on the top and bottom rows,
and free boundary conditions imposed on the outer columns, it is
energetically more favorable for the system to arrange itself into a
configuration with a mismatch. Therefore if we let $M\to\infty$, and fix the
ratio $M/L>1$, the excess in free energy at $T=0$ can have any arbitrary
value depending on $M/L$.}\par

\figth {8cm} {6cm} {figf.eps} {6cm} {3cm} {figfb.eps} {figfc.eps} {6} %
{(a) For the case with skewed boundary conditions in both directions,
an interface is allowed to wind around the torus of vertical perimeter
$L$ and horizontal perimeter $M$, crossing the seams of modified bonds
several times. Thus for the Ostlund-Huse model, a single-step interface
with tension $\epsilon_1$ prefers to walk perpendicular to the
chiral field and then continue by crossing the seams.\pagebreak
(b) A double-step interface with tension $\epsilon_2$ tends to walk parallel
to the chiral field.
(c) If fixed boundary conditions are imposed on the top and bottom rows,
and free boundary conditions imposed on the outer columns, it is
energetically more favorable for the system to arrange itself into a
configuration with a mismatch. Therefore if we let $M\to\infty$, and fix the
ratio $M/L>1$, the excess in free energy at $T=0$ can have any arbitrary
value depending on $M/L$.}\par

\figtw {8cm} {5.3cm} {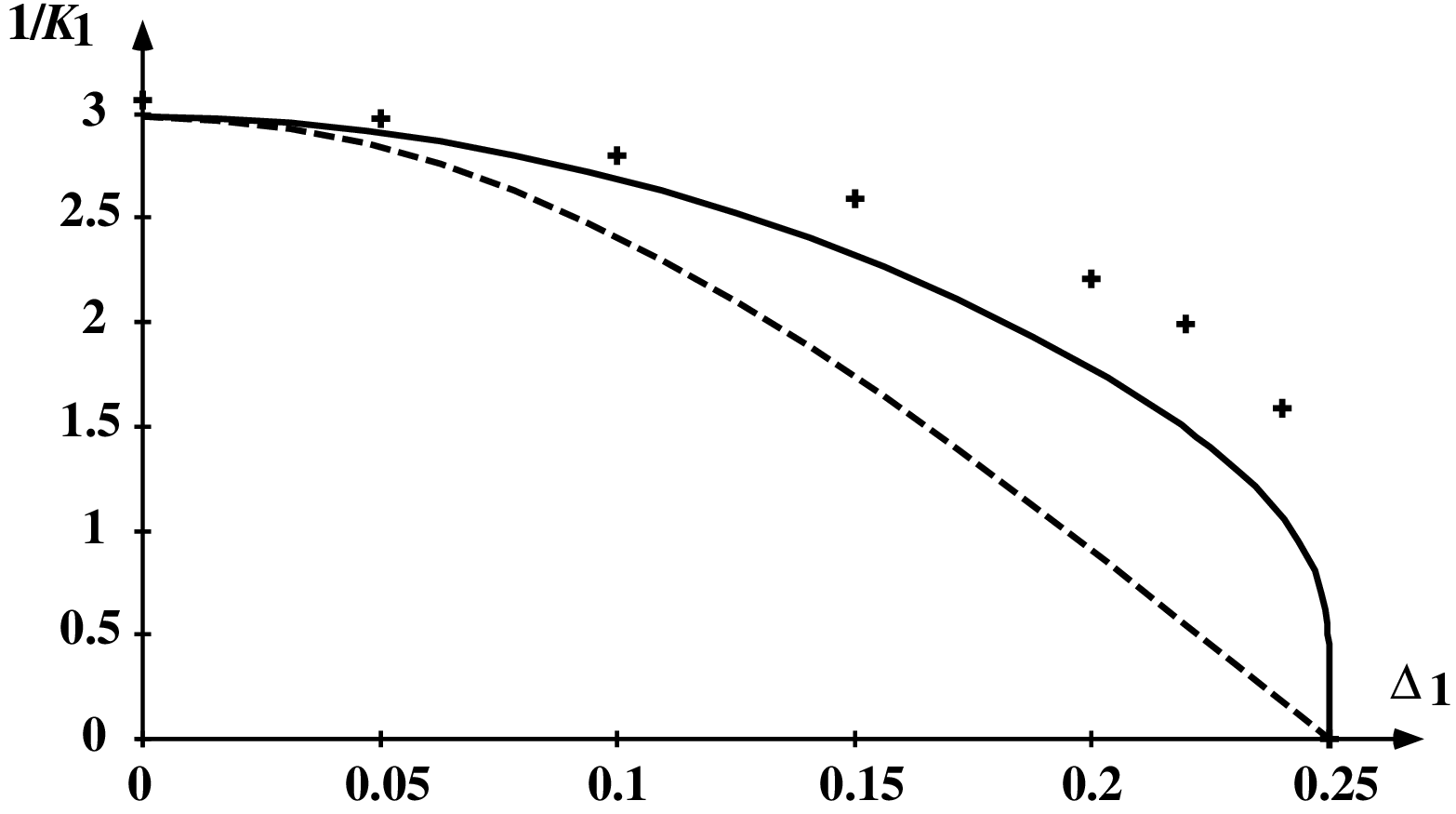} {8cm} {5.3cm} {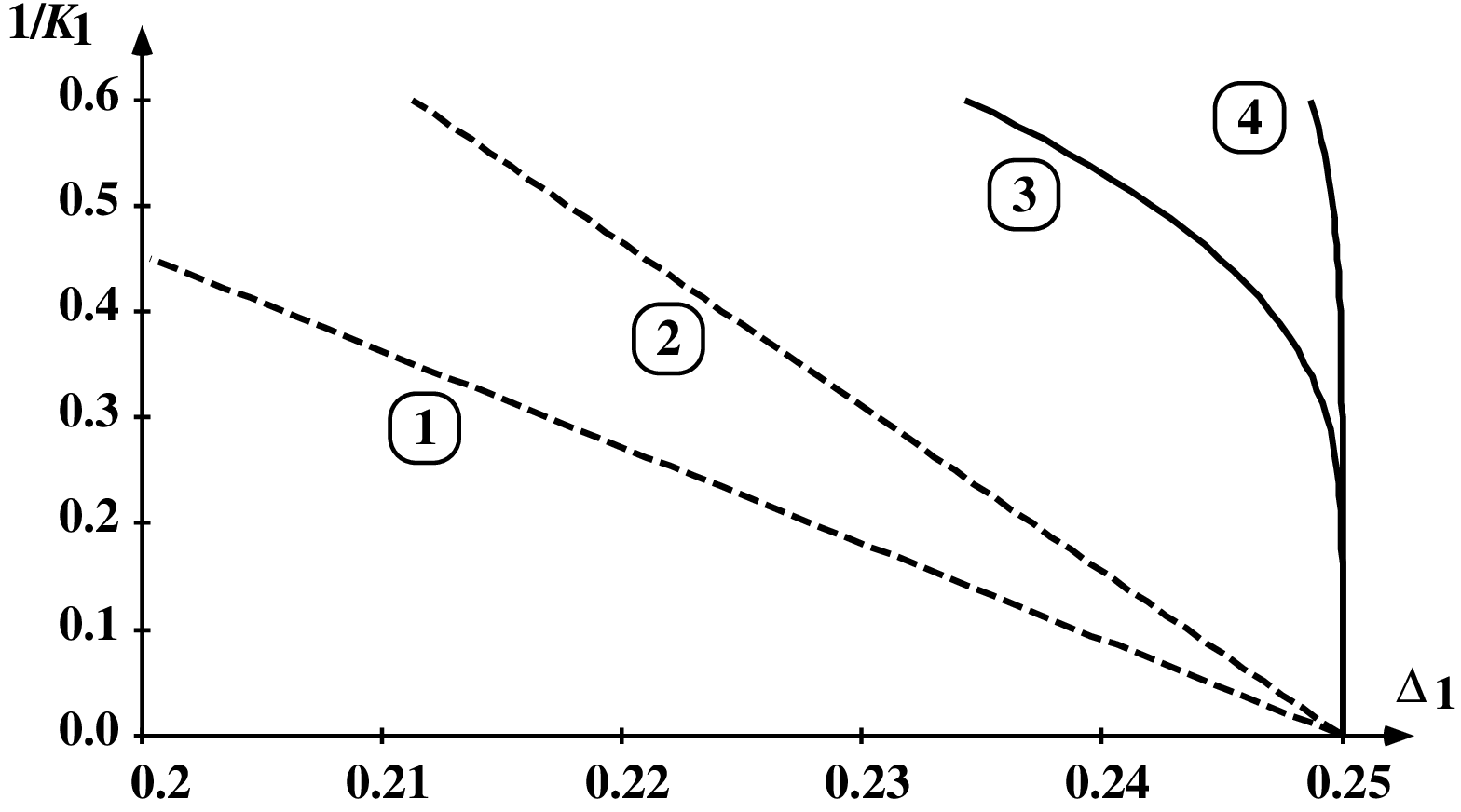} {7} {(a)
Integrable lines in the 3-state chiral clock model. The solid curve is
for the Ostlund-Huse fully asymmetric case and the dashed line is for the
symmetric case. The crosses are the numerical results of Stella et
al.\ for the fully asymmetric case. (b) We enlarge the part of the graph for
$T$ small and $\Delta_1$ near $\fract14$, with the integrable lines marked as
lines 1 and 3. We have also plotted the results of Huse et al.\ with line 2
denoting the wetting line of the symmetric lattice, and line 4 the one for
the fully asymmetric case. The graphs show that the integrable lines are
near but below the wetting lines, with the largest relative deviations
in the low-temperature regime.}\par

\fig H {8cm} {6cm} {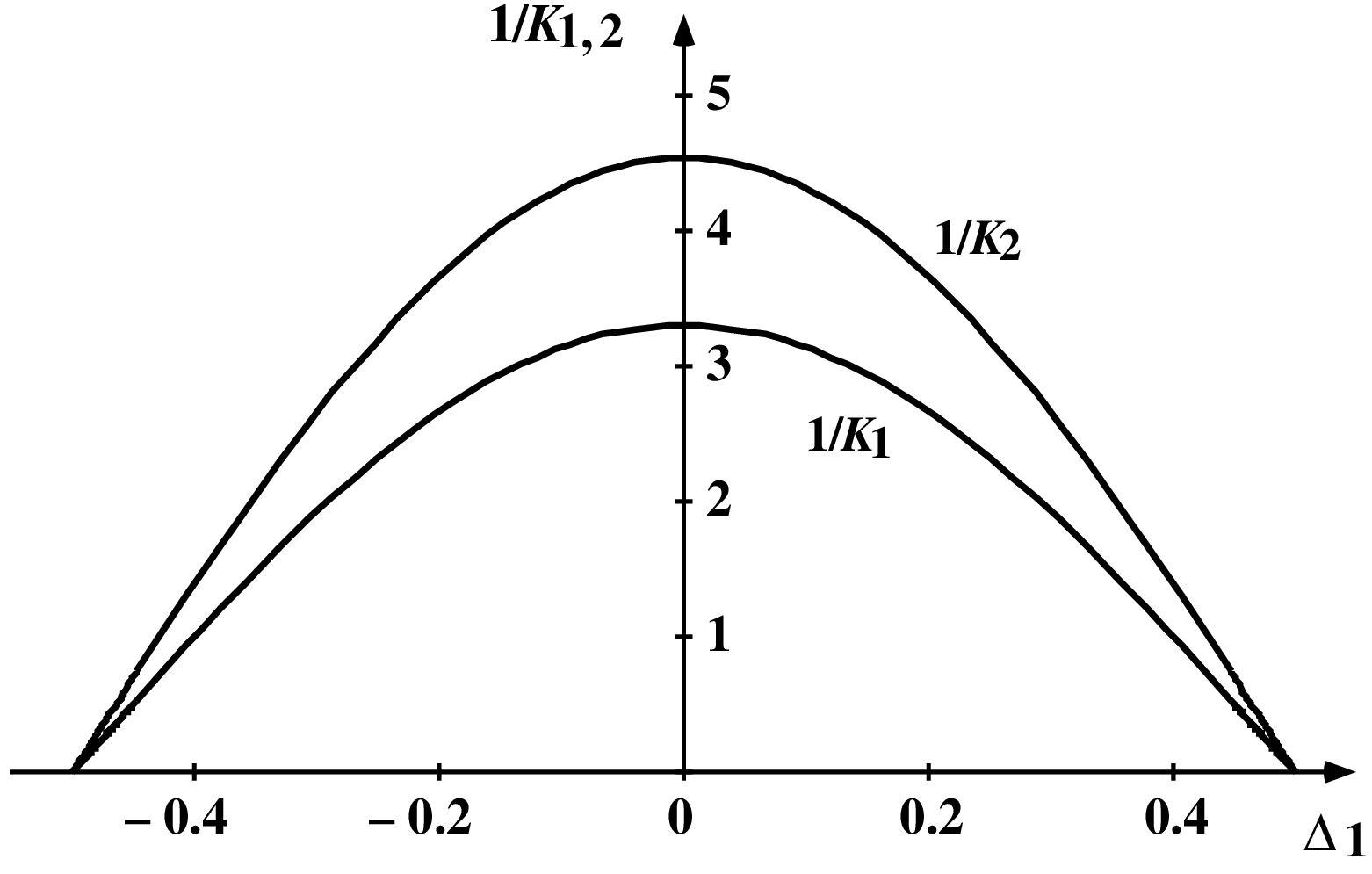} {8a} {The 4-state chiral Potts model:
$1/K_1$ and $1/K_2$ versus $\Delta_1$.}\par

\fig I {8cm} {6cm} {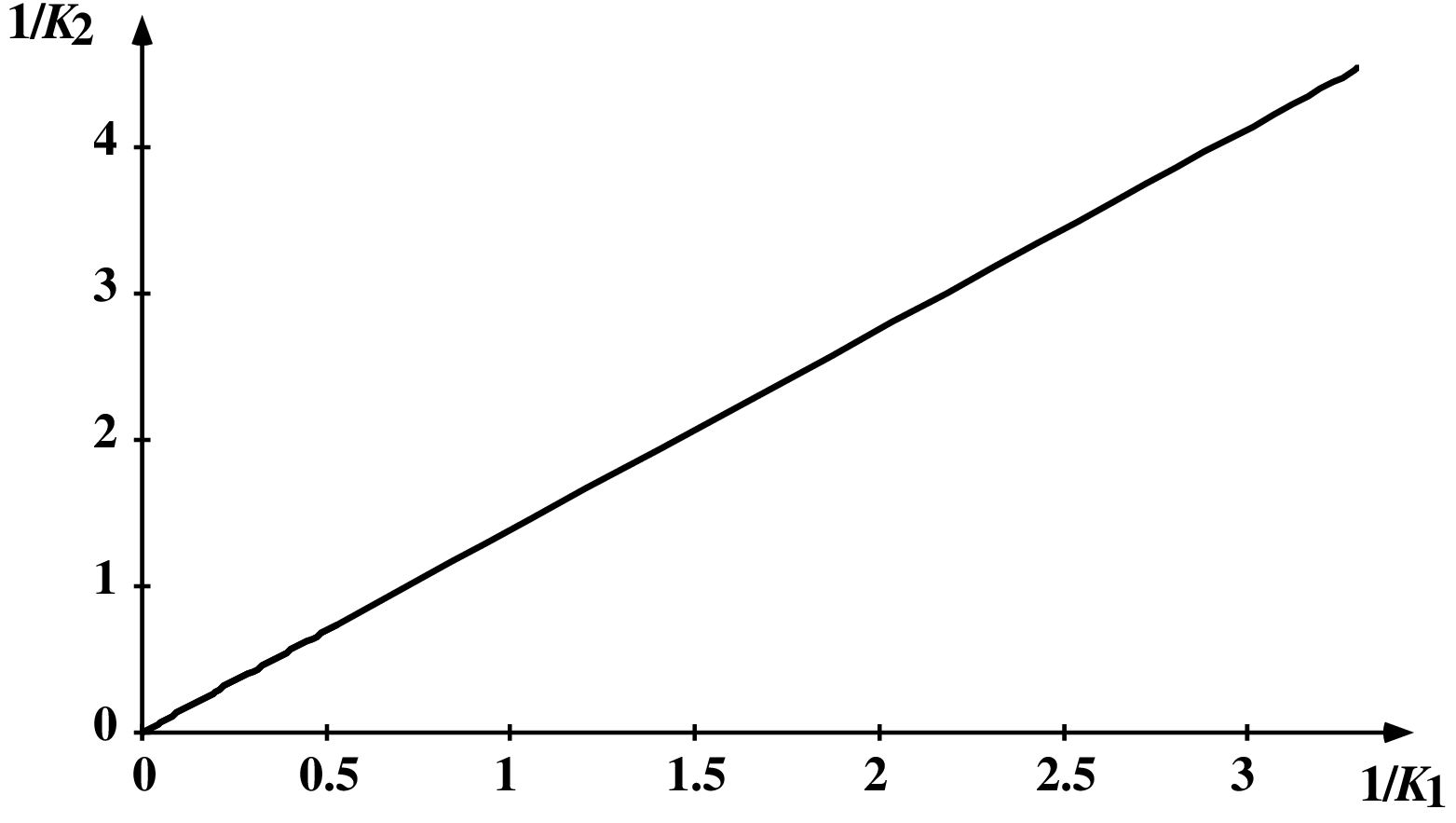} {8b} {$N=4$: $1/K_2$ versus $1/K_1$.}\par

\fig J {8cm} {6cm} {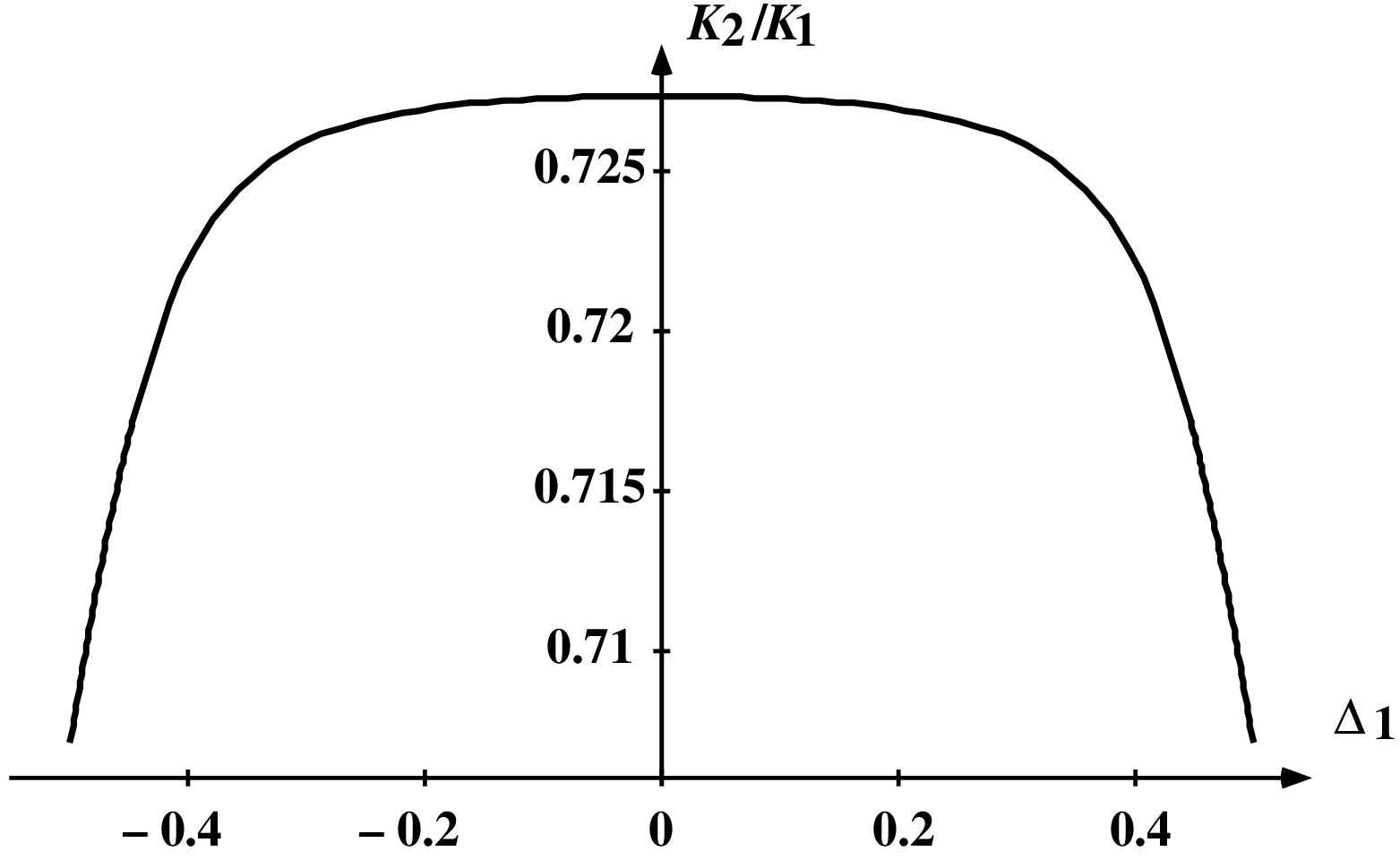} {8c} {$N=4$: $K_2/K_1$ versus
$\Delta_1$.}\par

\fig K {8cm} {6cm} {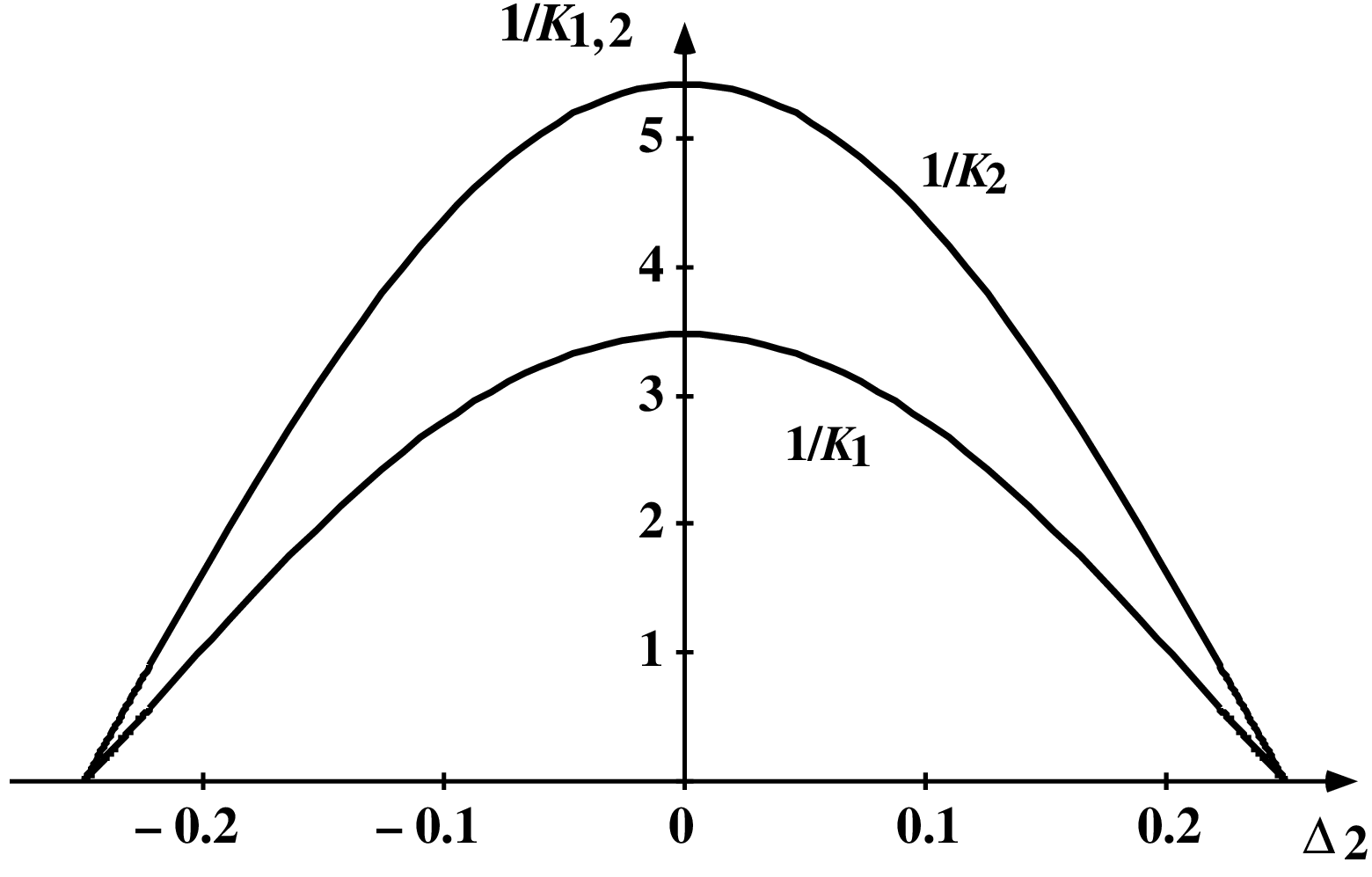} {9a} {The 5-state chiral Potts model:
$1/K_1$ and $1/K_2$ versus $\Delta_2$.}\par

\fig L {8cm} {6cm} {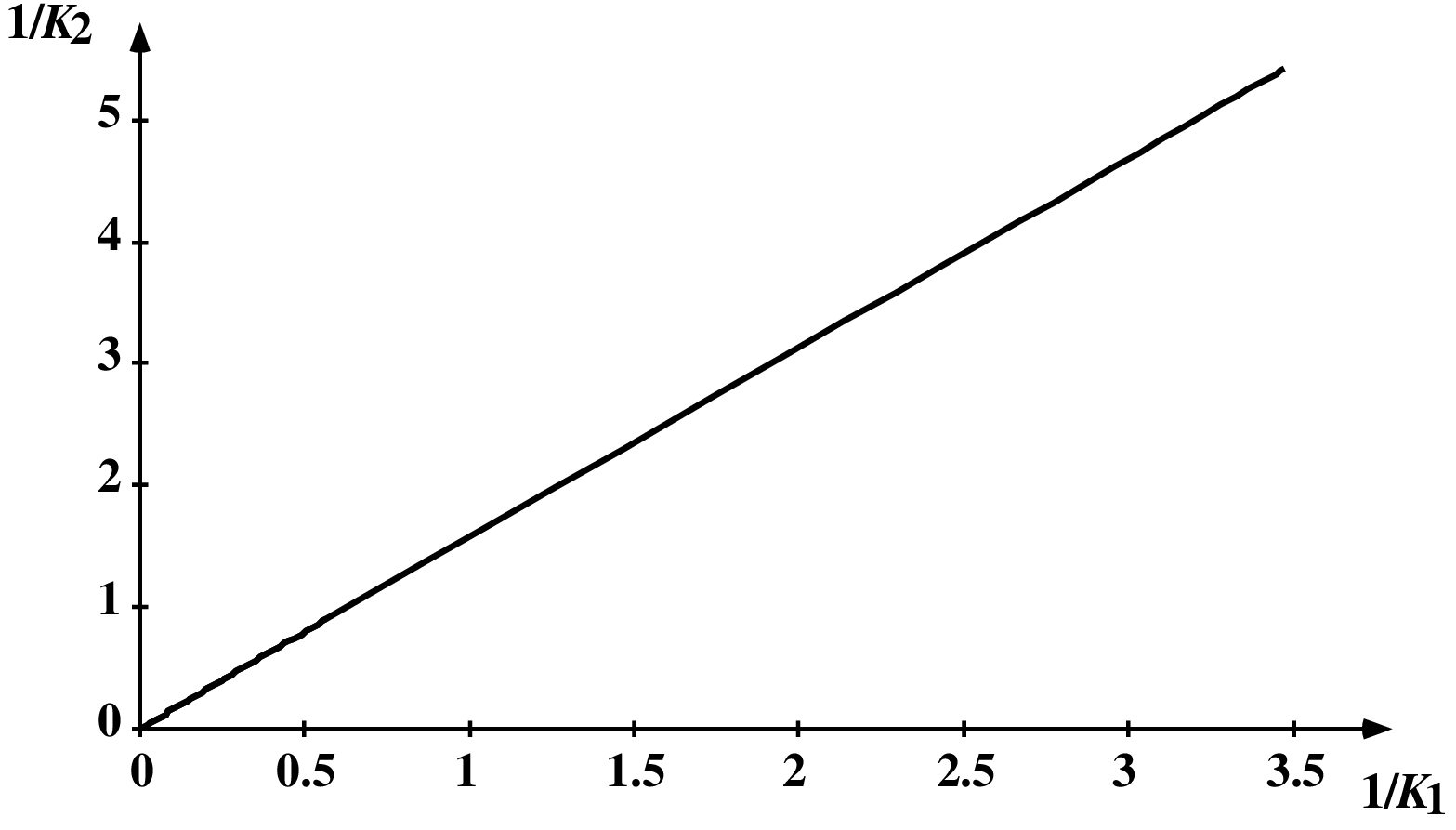} {9b} {$N=5$: $1/K_2$ versus $1/K_1$.}\par

\fig M {8cm} {6cm} {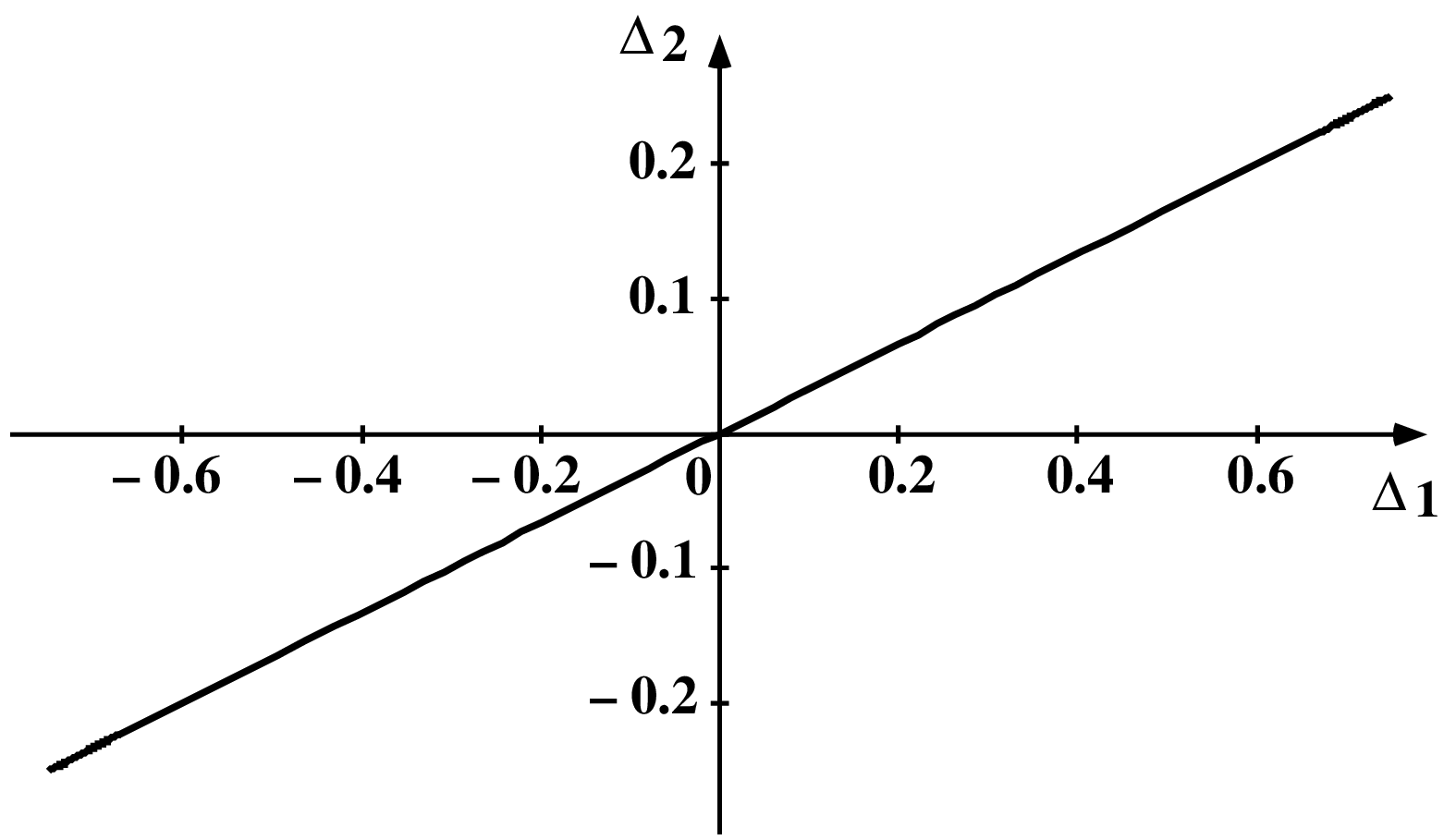} {9c} {$N=5$: $\Delta_2$ versus $\Delta_1$.}\par

\fig N {8cm} {6cm} {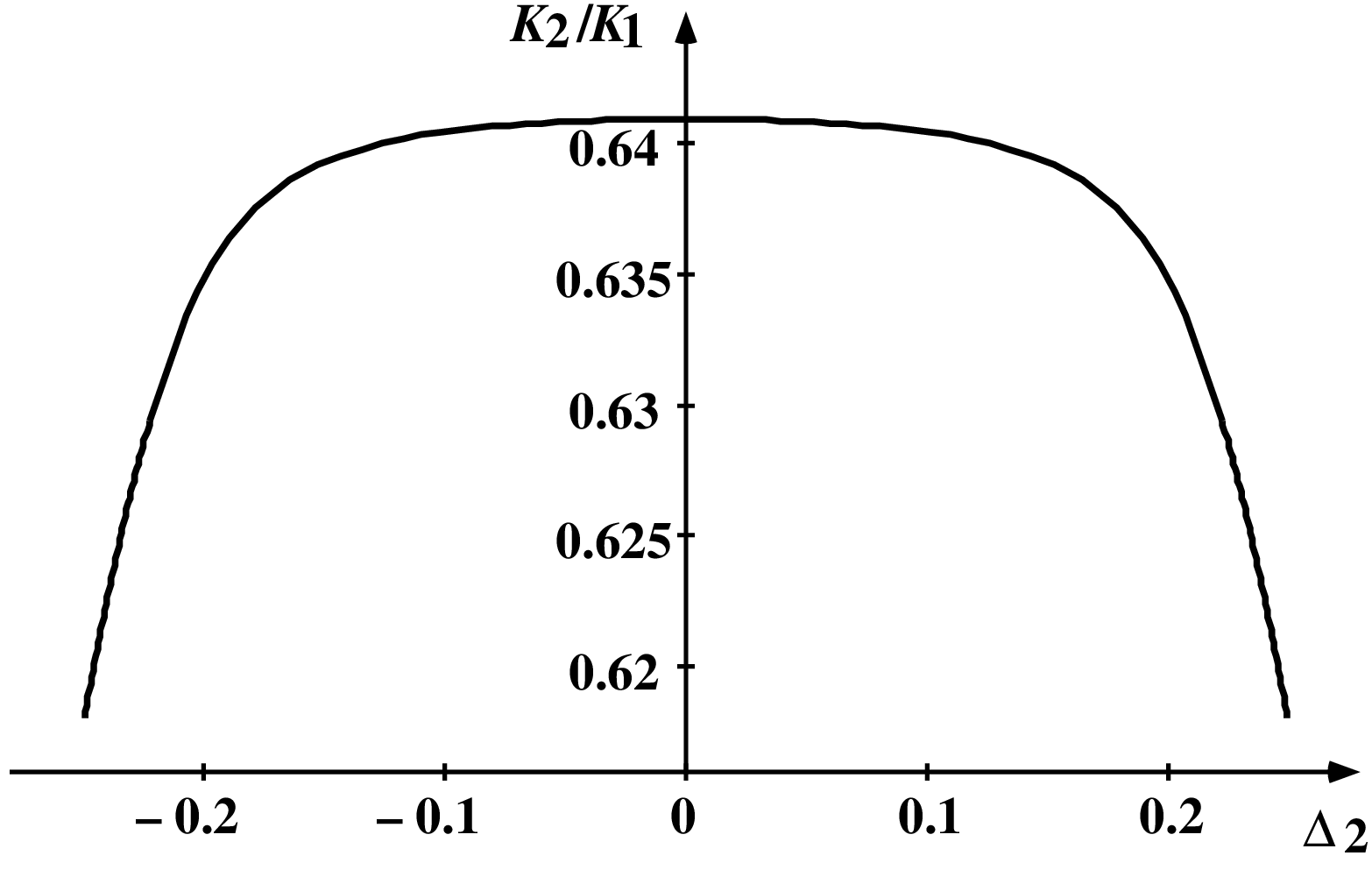} {9d} {$N=5$: $K_2/K_1$ versus
$\Delta_2$.}\par

\fig O {8cm} {6cm} {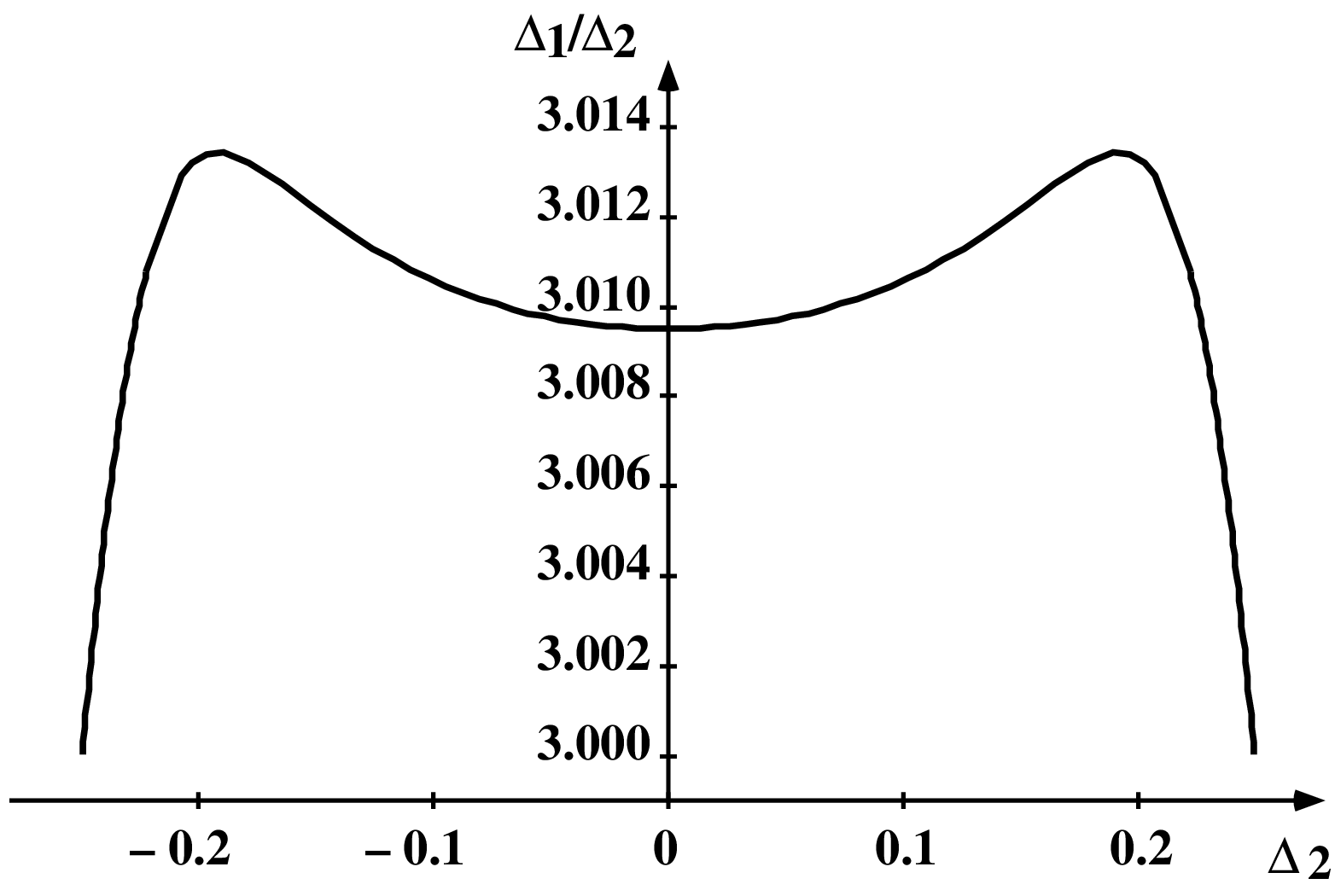} {9e} {$N=5$: $\Delta_1/\Delta_2$ versus
$\Delta_2$.}\par

\fig P {8cm} {6cm} {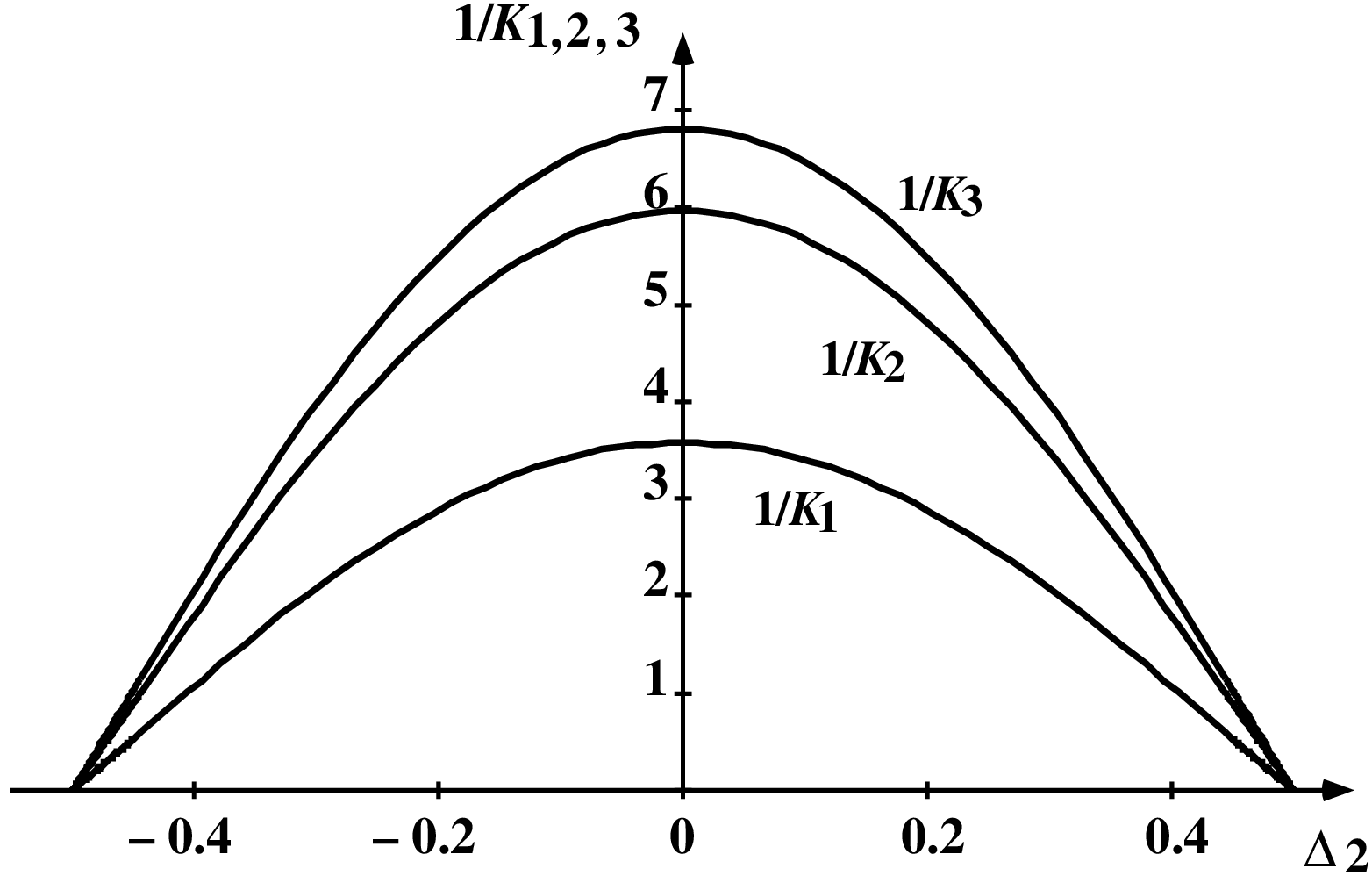} {10a} {The 6-state chiral Potts model:
$1/K_1$, $1/K_2$, and $1/K_3$ versus $\Delta_2$.}\par

\fig Q {8cm} {6cm} {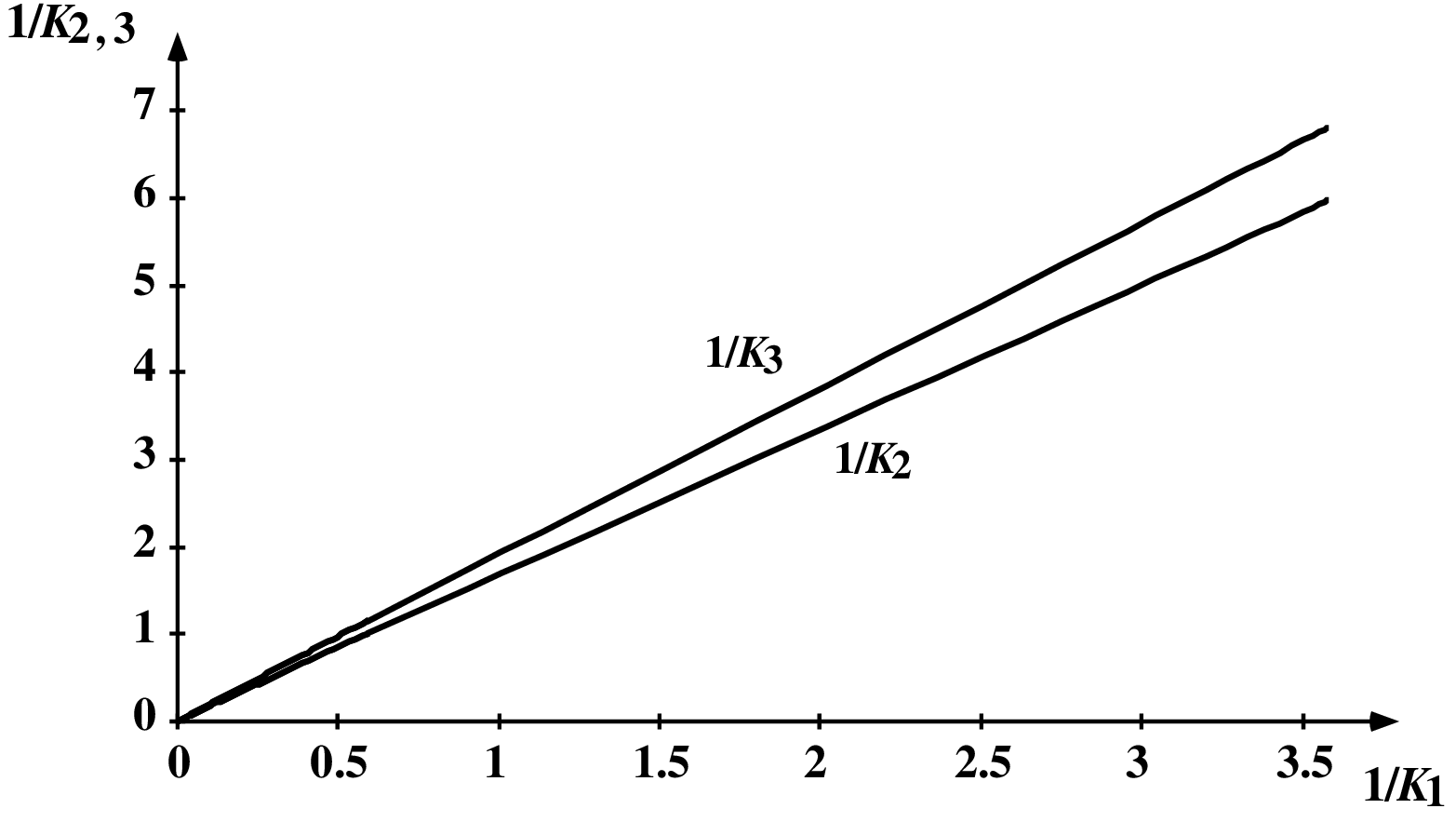} {10b} {$N=6$: $1/K_2$ and $1/K_3$ versus
$1/K_1$.}\par

\fig R {8cm} {6cm} {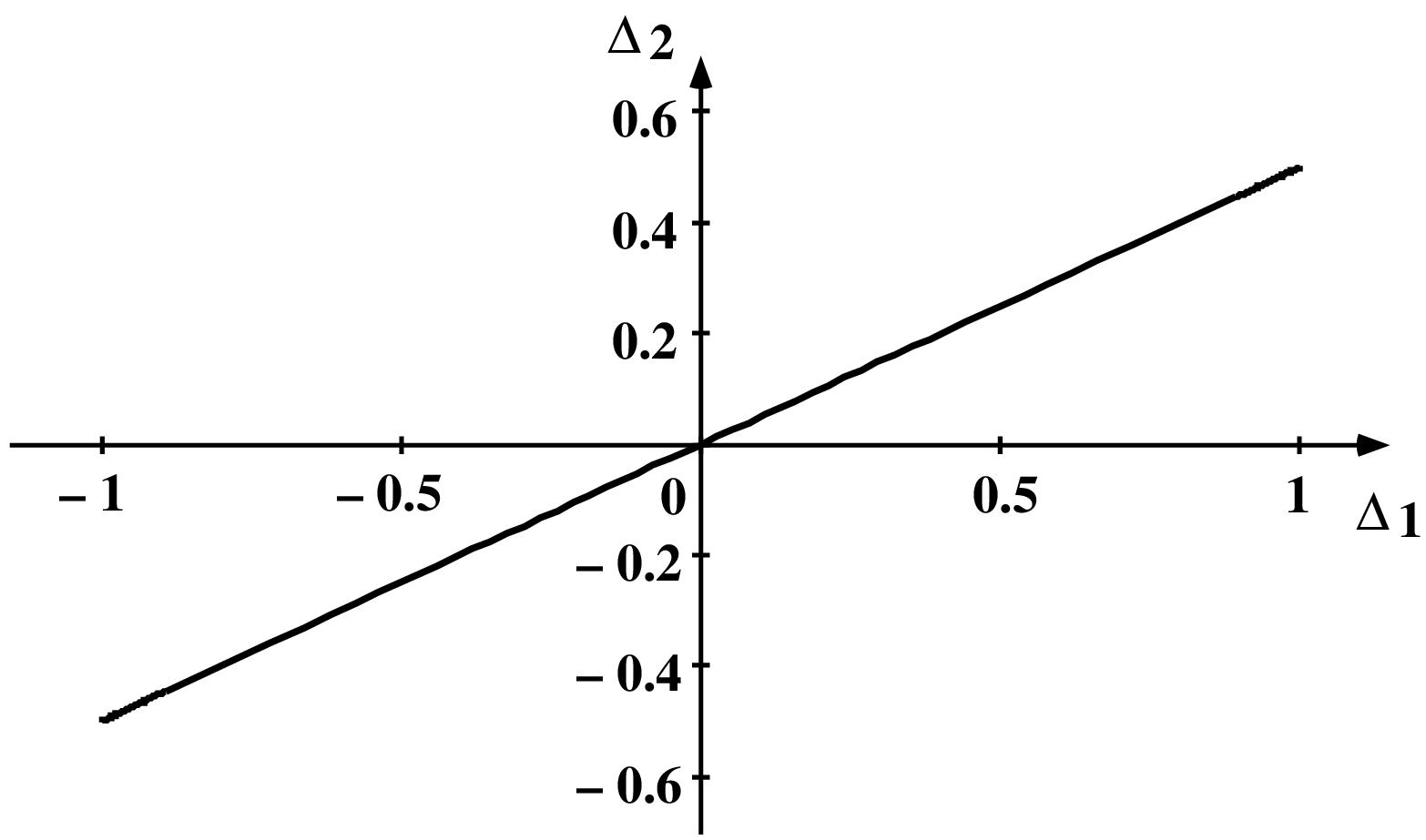} {10c} {$N=6$: $\Delta_2$ versus
$\Delta_1$.}\par

\fig S {8cm} {6cm} {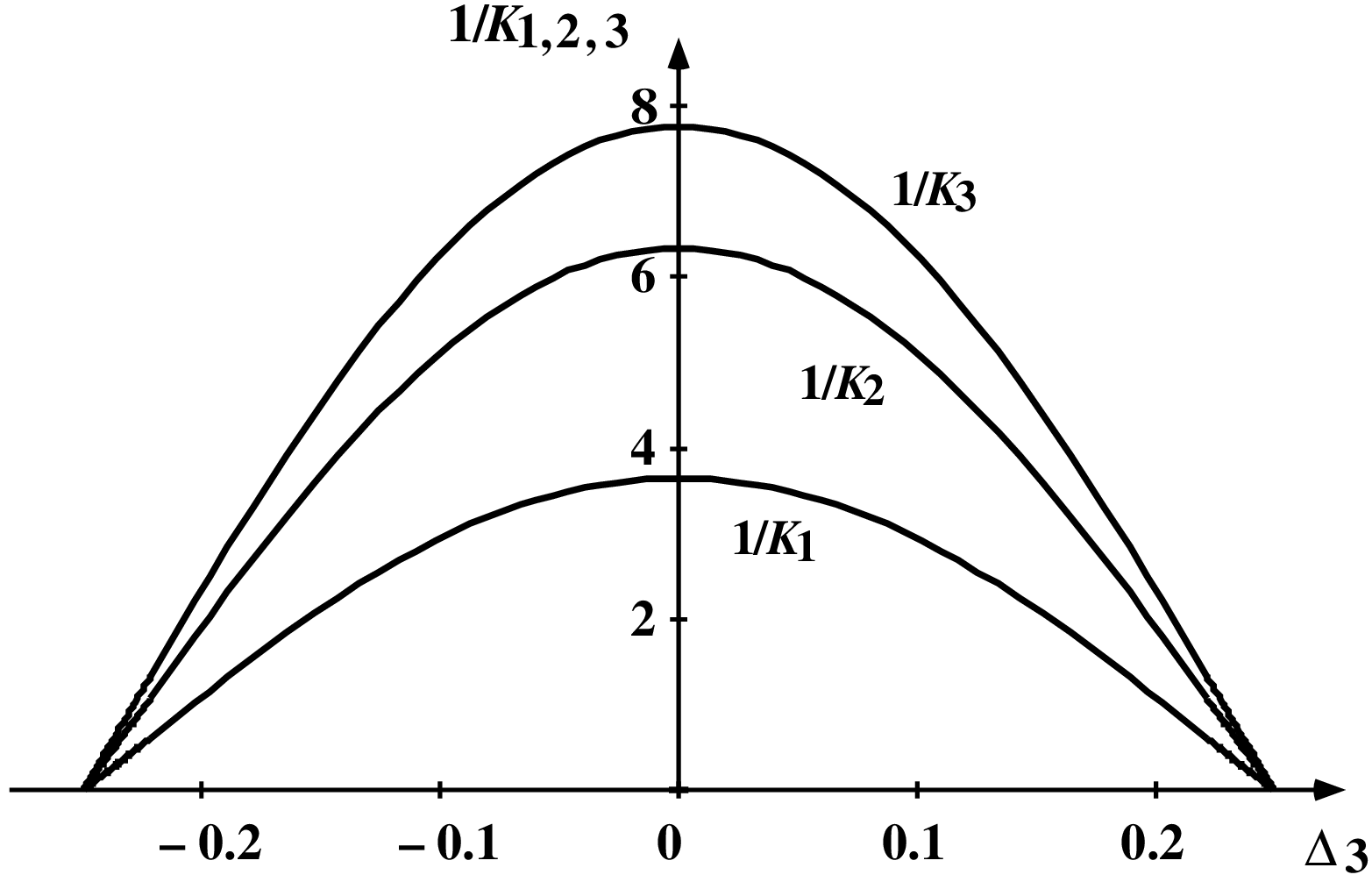} {11a} {The 7-state chiral Potts model:
$1/K_1$, $1/K_2$, and $1/K_3$ versus $\Delta_3$.}\par

\fig T {8cm} {6cm} {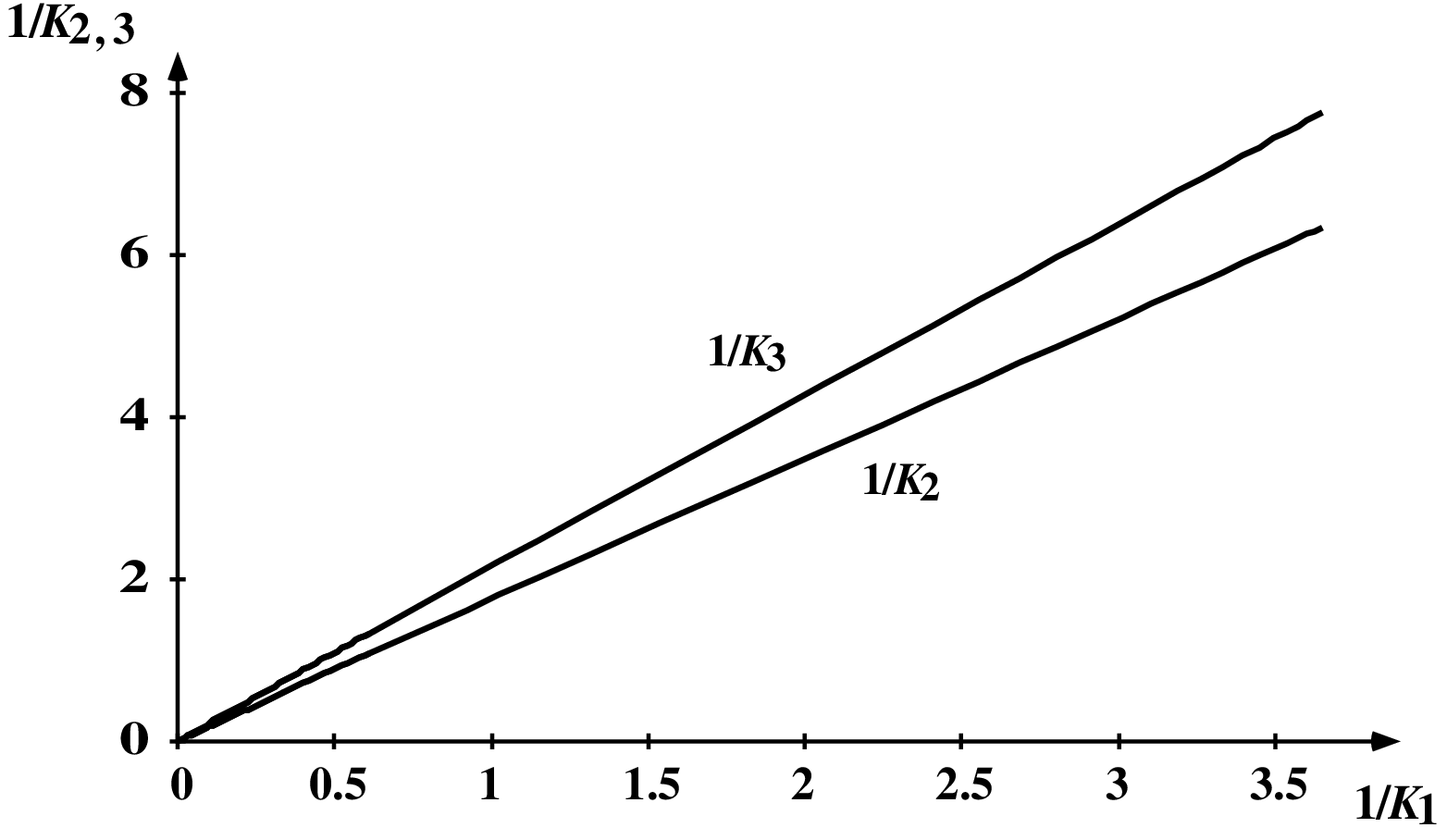} {11b} {$N=7$: $1/K_2$ and $1/K_3$ versus
$1/K_1$.}\par

\fig U {8cm} {6cm} {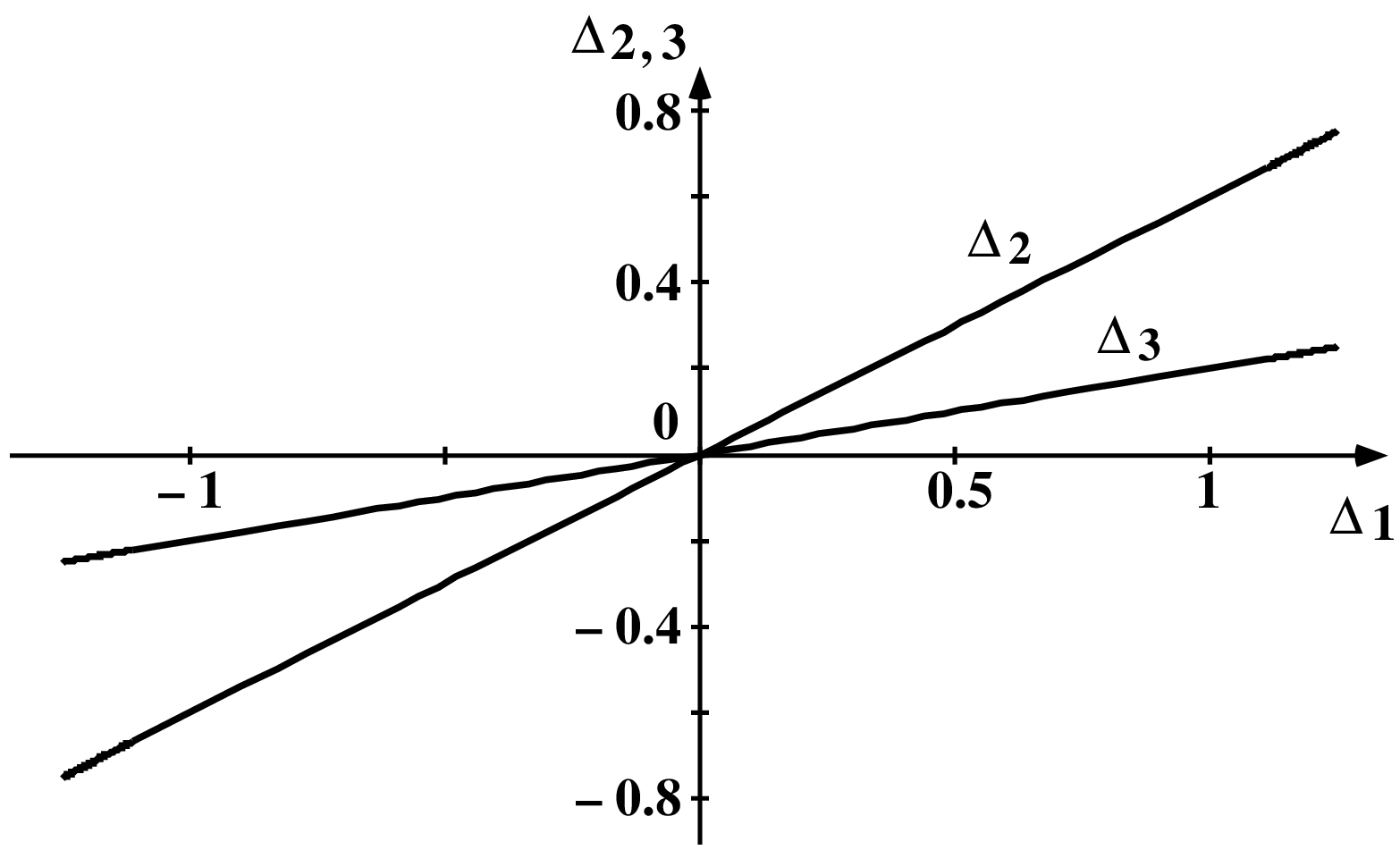} {11c} {$N=7$: $\Delta_2$ and $\Delta_3$
versus $\Delta_1$.}\par

\fig V {8cm} {6cm} {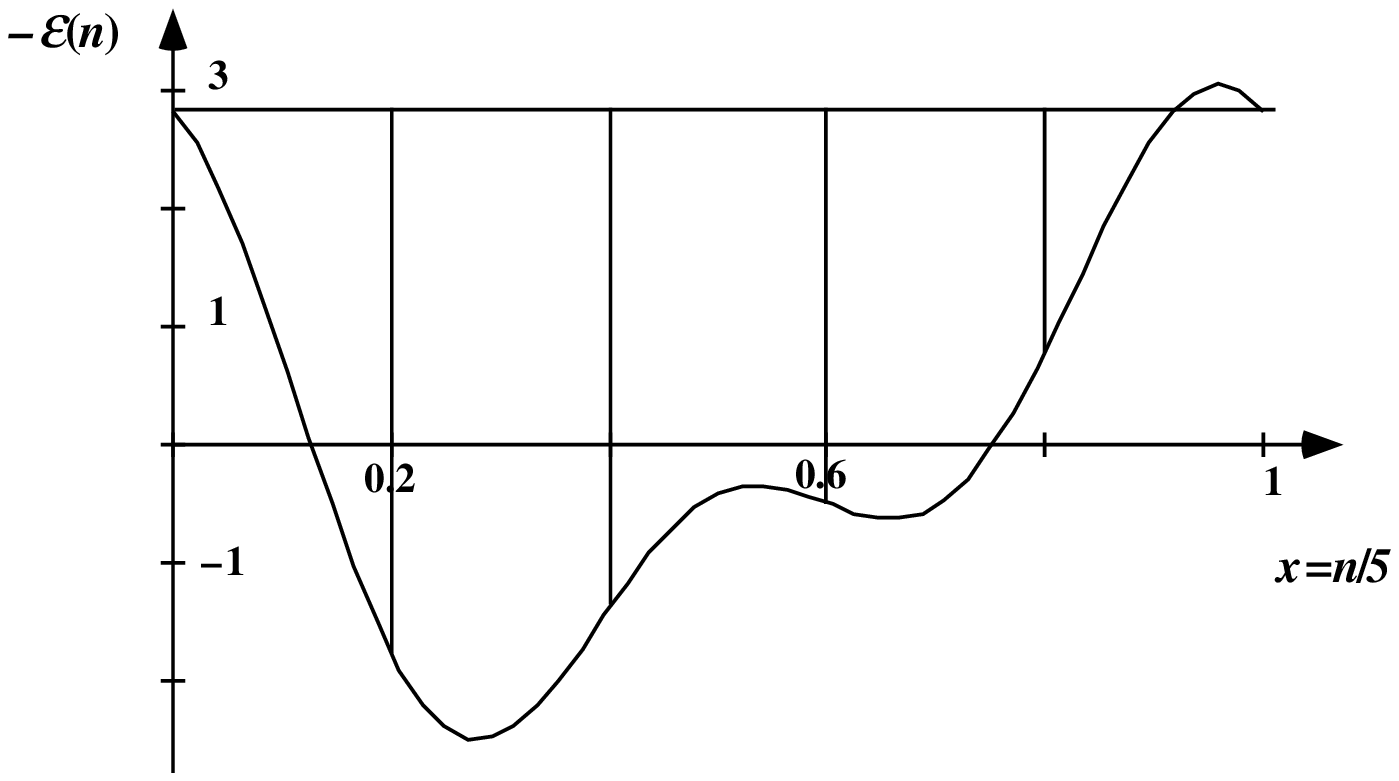} {12a} {Energy $-\cE(n)$ (or interfacial
tension at zero temperature) versus $x=n/N=n/5$ for the new model with $N=5$
and $\Delta=\fract16$: ground state in the ``dry phase" as no wetting can take
place.}\par

\fig W {8cm} {6cm} {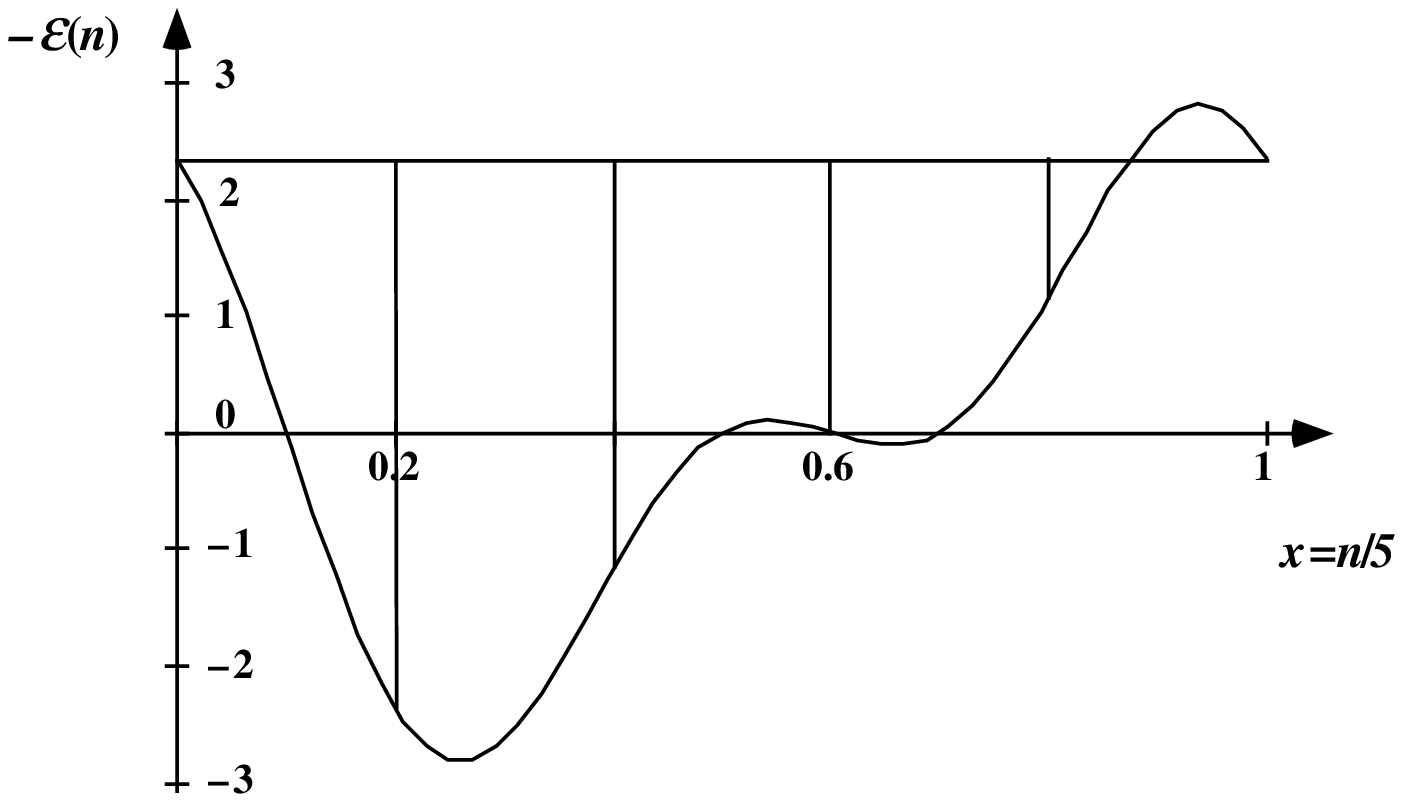} {12b} {Energy $-\cE(n)$ for $N=5$ and
$\Delta=\fract14$: the ground-state is at a (super)wetting transition.}\par

\fig X {8cm} {6cm} {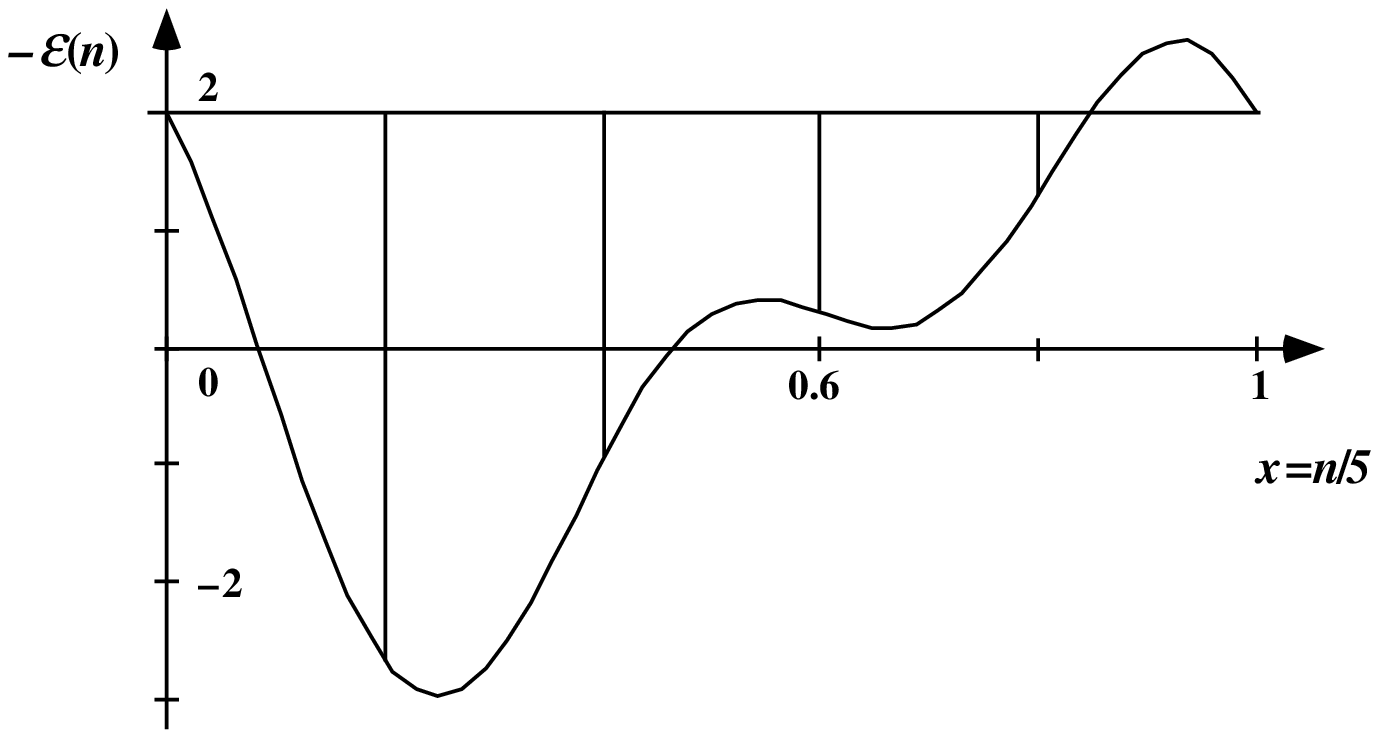} {12c} {Energy $-\cE(n)$ for $N=5$ and
$\Delta=\fract3{10}$: ground state in the wet phase.}\par

\fig Y {8cm} {6cm} {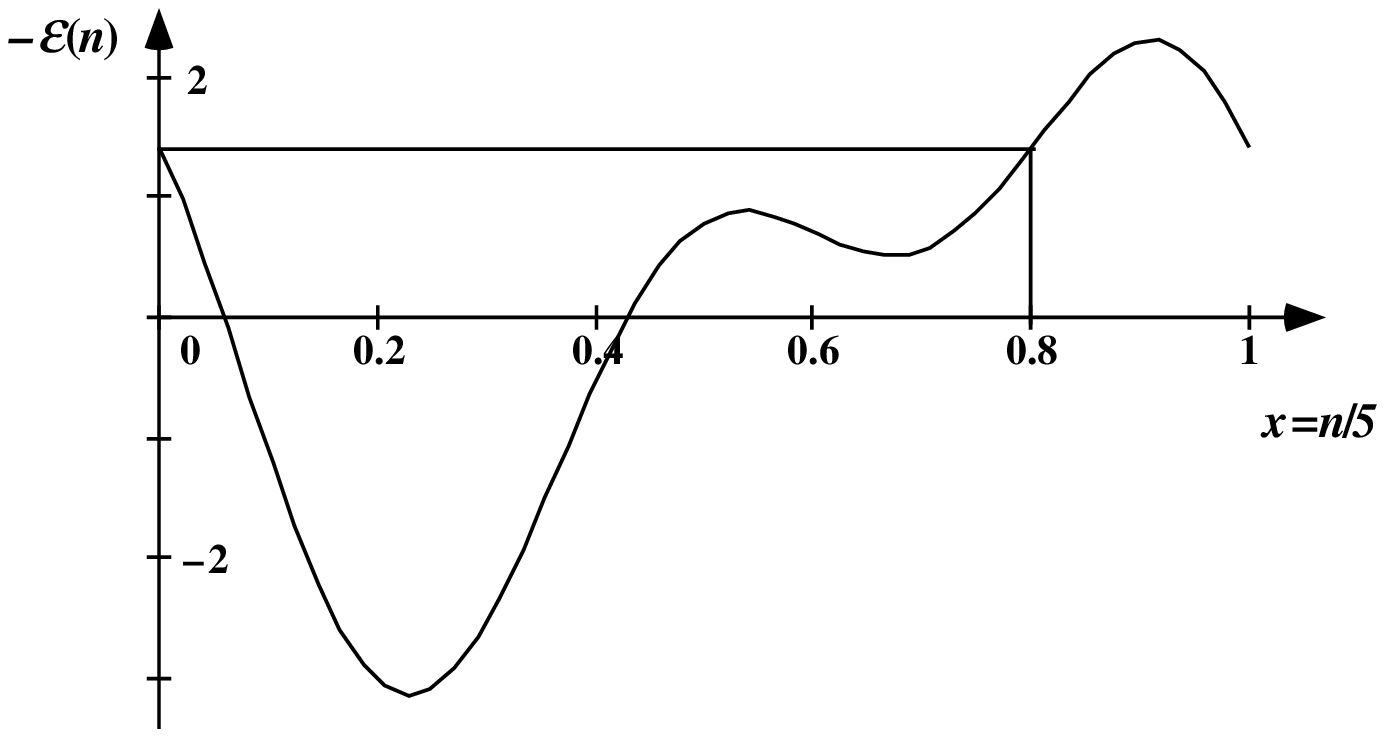} {13} {Energy $-\cE(n)$ for $N=5$ and
$\Delta=\fract38$: the ground state is highly degenerate as $W(0)=W(N-1)$
here.}\par

\fig Z {8cm} {6cm} {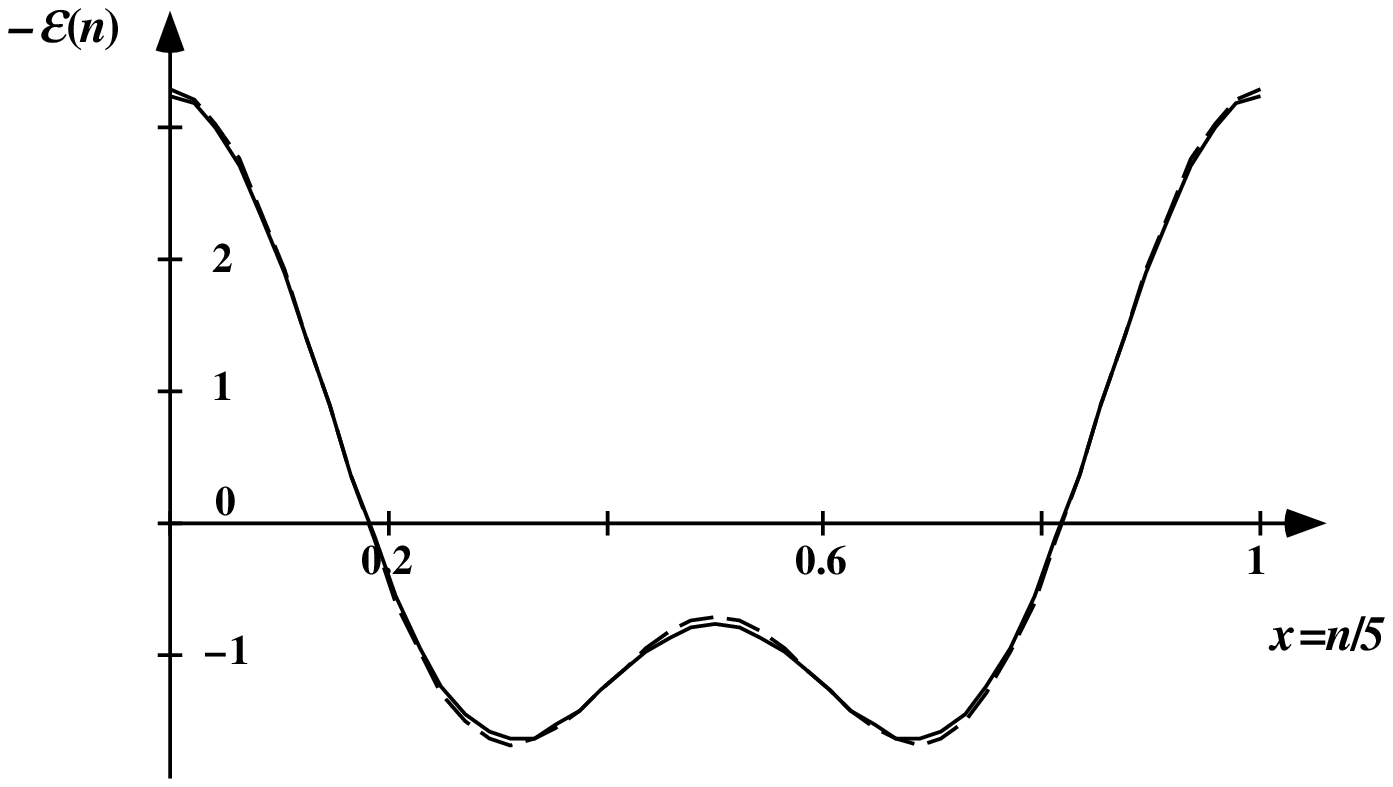} {14a} {Comparison of the energies $-\cE$ of
the self-dual integrable chiral Potts model and the new model with $\Delta=0$
for $N=5$; the two curves are almost on top of each other.}\par

\fig z {8cm} {6cm} {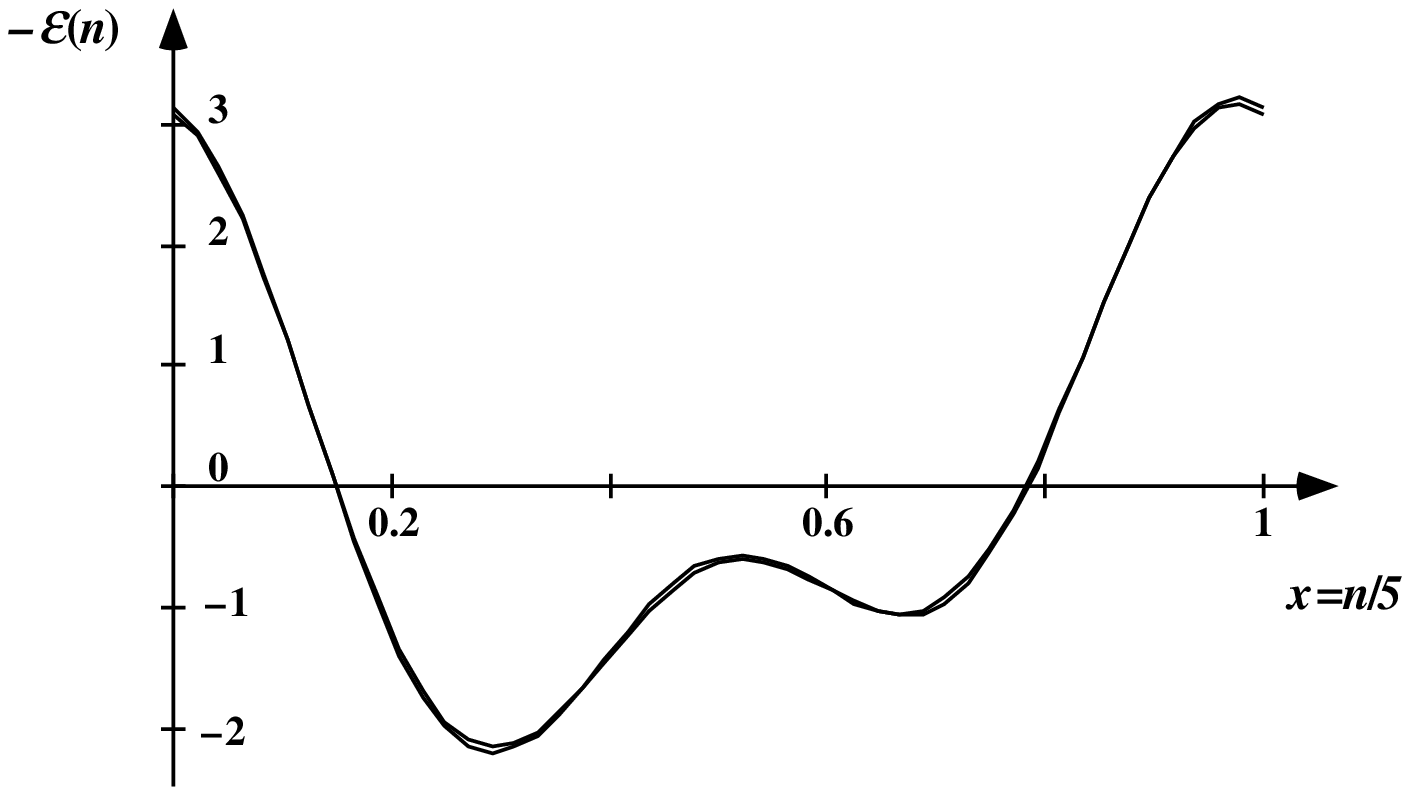} {14b} {Comparison of the energies $-\cE$
of the integrable chiral Potts and the new model at $\Delta=.0977875$ for
$N=5$; again the two curves are almost on top of each other.}\par

\input{psbox}

\begin{document}
\section*{The Chiral Potts Models Revisited\hfil}
\subsubsection*{Helen Au-Yang and Jacques~H.H.~Perk%
\footnote{Department of Physics, Oklahoma State University,
Stillwater, OK 74078-0444, USA.}$^{,}$%
\footnote{Supported in part by NSF Grants Nos.\ DMS 91-06521
and PHY 93-07816.}\hfil}
\vglue 0pt
{\leftskip=5em\par\noindent\footnotesize
{\it Abstract:}
\par\noindent\hrulefill\vglue 5pt
\par\noindent
In honor of Onsager's ninetieth birthday, we like to review some exact\B
results obtained so far in the chiral Potts models and to translate these
results into language more transparent to physicists, so that experts in
Monte Carlo\B calculations, high and low temperature expansions, and
various other\B methods, can use them.
\par
We shall pay special attention to the interfacial tension $\epsilon_r$ between
the $k$ state and the $k-r$ state. By examining the ground states, it is seen
that the integrable line ends at a superwetting point, on which the relation
$\epsilon_r=r\,\epsilon_1$ is satisfied, so that it is energetically neutral
to have one interface or more. We present also some partial results on the
meaning of the integrable line for low temperatures where it lives in the
non-wet regime. We make Baxter's exact results more explicit for the symmetric
case. By performing a Bethe Ansatz calculation with open boundary conditions
we confirm a dilogarithm identity for the low-temperature expansion which
may be new.
\par
We propose a new model for numerical studies. This model has only two
variables and exhibits commensurate and incommensurate phase transitions\B and
wetting transitions near zero temperature. It appears to be not\B integrable,
except at one point, and at each temperature there is a point, where it is
almost identical with the integrable chiral Potts model.
\par\noindent\hrulefill\vglue 5pt
\par\noindent{\bf KEY WORDS:} Chiral Potts model; chiral clock model;
star-triangle\B equations; Yang-Baxter equations; interfacial tension;
wetting; superwetting; scaling; corrections to scaling;
low-temperature expansions; dilogarithms; Bethe Ansatz.
\par\leftskip=0pt}
\vglue 0pt

\section{INTRODUCTION\label{sec.intro}}

When Onsager published his solution of the two-dimen\-sional Ising model\B
in 1944\cmyref{\ref{O1}} this was almost instantly recognized as a milestone
in the\B development of statistical mechanics. Many new developments
were inspired by his results. On the other hand, Onsager's techniques
were far ahead of his time and, when he announced his incredibly simple
result for the\B spontaneous magnetization as a comment to a conference
talk\cmyref{\ref{O2},\ref{O5}} his paper attained a mystical status for many
years after. That his student Kaufman simplified the solution through Clifford
algebras (i.e.\ fermions)\myref{\ref{O3},\ref{O4}} did little to change this.

Now, nearly fifty years later, we can appreciate Onsager's methods\B
much better. He was the first to introduce the star-triangle
equation\myref{\ref{Kennelly}} to\B statistical
mechanics\myref{\ref{O1},\ref{O5}} even though now the name Yang-Baxter
equation is commonly used\pmyref{\ref{McGuire}-\ref{AP1}} He also was the
first to introduce loop algebras\myref{\ref{VKac}} as a solvability
principle. Onsager and Kaufman have clear priority over Wick for the Wick
theorem\myref{\ref{O4}} and, together with Bethe\cmyref{\ref{Bethe}}
they scouted a new area of mathematics which is now called quantum
groups\pmyref{\ref{Kulish}-\ref{FaReTa}}

Onsager's (only partly published) work on two-point functions in the Ising
model got extended in 1966\cmyref{\ref{Wu}-\ref{FisherBurford}} but it was only
in 1973, when Wu, McCoy, Tracy, and Barouch\myref{\ref{Wuetal}-\ref{Perk2}}
announced the Painlev\'e equation for the scaled two-point correlation,
that the theory of the two-dimen\-sional Ising model went beyond Onsager's
level. Also, the first two-dimen\-sional models solved that were more
complicated than the Ising model were Lieb's ice
model\myref{\ref{Lieb},\ref{LiebWu}} of 1967 and Baxter's eight-vertex
model\myref{\ref{BaxterBook}} of 1973.

Onsager's loop-algebra solution method was generalized only in 1985 when Von
Gehlen and Rittenberg\myref{\ref{vGR}} solved the Dolan-Grady\myref{\ref{DoG}}
criterium within a one-dimensional generalization of the quantum Potts model.
The connection with Onsager's 1944 paper was noticed by
Perk\cmyref{\ref{Perk},\ref{Davies}} showing that the chiral Potts
model\myref{\ref{AMPTY},\ref{BPA},\ref{AP1}} is the first genuine
generalization of the Ising model, with its ``super\-integrable"
case\myref{\ref{AlMPT}} solvable for two reasons: star-triangle
integrability\myref{\ref{AMPTY},\ref{BPA},\ref{AP1}} and loop-algebra
integrability\pmyref{\ref{Perk},\ref{vGR}} The
chiral\mydouble{\eject\noindent} Potts model upgrades the fermions
of Kaufman\myref{\ref{O3}} to parafermions.\footnote{We note that the
parafermions here are of ``cyclic root-of-unity" type\cmyref{\ref{Morris}}
generalizing the Weyl algebra, not of the more usual ``highest-weight" type
introduced by Green\pmyref{\ref{Green}} Within the original Ising model and
its fermion approaches these two types are isomorphic, however.} It also
provides an infinite hierarchy of quantum groups at roots of unity, with Ising
as its first entry\pmyref{\ref{BazhanovS},\ref{dCK}} Thus the works of Onsager
are still at the center of attention.

This paper is organized as follows: In section \ref{sec.cpm} we define the
chiral Potts model and review some of the exact results obtained so
far\pmyref{\ref{Bax1}-\ref{AlMP}} We pay special attention to the interfacial
tensions $\epsilon_r$, giving several results that have not been presented
before as such.\footnote{We gratefully acknowledge several private
communications with Dr.\ Baxter.} We show that the solvable chiral Potts
models ``superwet," that is $\epsilon_r=r\epsilon_1$, at $T=0$ and
$T=T_{\rm c}$. However, Baxter\myref{\ref{Bax2},\ref{Bax3}} has shown that,
for $0<T<T_{\rm c}$, the interfacial tensions satisfy the inequalities
$\epsilon_r<\epsilon_{r-j}+\epsilon_j$, so that the integrable line is in the
not-wetted region. We use low-temperature expansions and the exact results of
Baxter to further analyze the effects of various boundary conditions. We also
discuss what the critical exponents obtained exactly mean for the scaling
function. In sections \ref{sec.con} to \ref{sec.sym}, we discuss the
relation between the integrable subcases, whose Boltzmann weights can be
represented by a product-form, and the\B generalized clock-model
representations of the pair-interaction energies.\B Special attention is given
to the symmetric case (with equal horizontal and vertical interactions) and
several numerical details and graphs are given. In section \ref{sec.bal}, we
outline a Bethe Ansatz calculation for the leading corrections to the
zero-temperature diagonal interfacial tensions. Finally, in section
\ref{sec.new}, we introduce a new model with only few parameters, which is
very close to the integrable model and may deserve detailed further study by
numerical means.

\section{CHIRAL POTTS MODEL\label{sec.cpm}}

\subsection{Integrable Chiral Potts Model\label{sus.icp}}

In our original paper\cmyref{\ref{AMPTY}} new exact-solution manifolds were
discovered within the chiral Potts model. To be more specific, the Potts
model\cmyref{\ref{Potts}-\ref{Baxbk2}} which is itself a generalization of the
two-dimensional Ising model solved by Onsager\myref{\ref{O1}} in 1944, and
whose interaction energy for the two spins on an edge is given by
\begin{equation}
\cE(\sigma,\sigma')=E\,\delta_{\sigma,\sigma'},
\tag {2.1}
\end{equation}
was generalized to the chiral Potts model (or {\mysym Z}$_q$-%
model\myref{\ref{WuRev}})
with interaction energy
\begin{equation}
\cE(\sigma,\sigma')=\cE(n-n')=\sum_{j=1}^{N-1}E_j\,\omega^{j(n-n')},
\tag {2.2}
\end{equation}
where
\begin{equation}
\omega=e^{2\pi i/N},\quad\sigma=\omega^n,\quad\sigma'=\omega^{n'}.
\tag {2.3}
\end{equation}
Clearly the interaction energy defined by (\ref{2.2}) satisfies the relation
$\cE(\sigma,\sigma')=\cE(n-n')=\cE(n-n'+N)$. For $E_j=E$, ($1\le j\le N-1$),
using
\begin{equation}
\sum_{j=1}^{N-1}\omega^{j(n-n')}=N\delta_{\sigma,\sigma'}-1,
\tag{2.4}
\end{equation}
we find that the interaction energy is identical to that of the Potts model
except for an overall constant. When $E_j=E_{N-j}$, it includes the
integrable Fateev-Zamolodchikov\myref{\ref{FZ}} self-dual ${\bbZ}_N$ model
as a special case.

The Boltzmann weight for an edge is
\begin{equation}
W(n-n')=e^{-\cE(n-n')/{k_{\rm B}T}}
\tag {2.5}
\end{equation}
with $k_{\rm B}$ the Boltzmann constant. Since duality transform is
equivalent\B to Fourier transform\myref{\ref{Potts},\ref{WuRev}} of these
weights, the weights are said to be\B ``self-dual," if they are equal
(or proportional) to their Fourier transforms. With a great deal of
effort\myref{\ref{AMPTY},\ref{BPTS},\ref{AMPT}} and some luck we were able to
find\B self-dual solutions to the star-triangle equations and found that
they can be written in product forms.

In Australia, Baxter and the two of us\myref{\ref{BPA},\ref{APTa}} --- mainly
through guessing\B and especially guided by Onsager's work\myref{\ref{APTa}}
--- found more general solutions\ of the star-triangle or ``checkerboard
Yang-Baxter" equation,
\begin{equation}
\sum^N_{d=1}\,\wb_{qr}(b-d)\,W_{pr}(a-d)\,\wb_{pq}(d-c)=
R_{pqr}\,W_{pq}(a-b)\,\wb_{pr}(b-c)\,W_{qr}(a-c).
\tag {2.6}
\end{equation}
They are also given in product form. To be more specific, we found
that in the chiral Potts model, the weights $W_{pq}$ or $\wb_{pq}$
depend on two line (or rapidity) variables denoted by $\hbox{\bf p}\equiv
(x_p,y_p,\mu_p)$ and $\hbox{\bf q}\equiv(x_q,y_q,\mu_q)$,
shown in Fig.\ 1; the weights are given as
\begin{equation}
{W_{pq}(n)\over W_{pq}(n-1)}=\biggl({\mu_p\over\mu_q}\biggr)
\biggl({y_q-x_p\omega^n\over y_p-x_q\omega^n}\biggr),
\qquad
{\wb_{pq}(n)\over\wb_{pq}(n-1)}=\bigl(\mu_p\mu_q\bigr)
\biggl({\omega x_p-x_q\omega^n\over y_q-y_p\omega^n}\biggr).
\tag {2.7}
\end{equation}

\myfA

Here, the parameters $\hbox{\bf p}$ and\vphantom{$W_{W_{W_W}}$}
$\hbox{\bf q}$ are restricted by the two periodicity\B requirements
$W_{pq}(N+n)=W_{pq}(n)$ and $\wb_{pq}(N+n)=\wb_{pq}(n)$, yielding the
conditions
\begin{equation}
\biggl({\mu_p\over\mu_q}\biggr)^N={y_p^N-x_q^N\over y_q^N-x_p^N},
\qquad(\mu_p\mu_q)^N={y_q^N-y_p^N\over x_p^N-x_q^N}.
\tag {2.8}
\end{equation}
These imply the existence of numbers $k$ and $k'$ related by
${k\vphantom{'}}^2+{k'}^2=1$ such that the equations
\begin{equation}
\mu_p^N=k'/(1-k x_p^N)=(1-k y_p^N)/k',\qquad x_p^N+y_p^N=k(1+x_p^N y_p^N)
\tag {2.9}
\end{equation}
\noindent hold for each $\hbox{\bf p}$. (This may require using the ambiguity
in defining the\B $x_p$, $x_q$, $y_p$, and $y_q$ in (\ref{2.7}) and rescaling
them with a common factor.) The equations (\ref{2.9}) describe a complex
curve, and the genus of this curve is\B $g=N^2(N-2)+1$.

\myfB

{}From Fig.\ 2, one can see that the star-triangle equations allow one to move
the rapidity line ${\bf p}$ through the vertex (the intersection of the other
two rapidity lines). Because of this, one can permute these rapidity
lines\B without changing the partition function, except possibly some
constant\B factors. Baxter called such lattice models
$Z$-invariant\pmyref{\ref{Bax-ZI}} This also means that transfer matrices
associated with different rapidity variables commute. We can see from
(\ref{2.7}) to (\ref{2.9}) that for given $k$, there is only one free\B
variable $x_p$, associated with each rapidity variable $\hbox{\bf p}$, and that
$y_p$ and $\mu_p$ can be determined from (\ref{2.9}). For the rectangular
lattice with just two\B rapidity variables $p$ and $q$, and therefore two
kinds of weights $W_{pq}$ and $\wb_{pq}$, there are three free variables.

By comparing with the Ising model ($N=2$)\cmyref{\ref{O1}} where $k$ and $k'$
are the elliptic modulus and its complementary modulus, we conclude that $k$
and $k'$ describe how far the system is from its critical point.

\subsection{Gauge Transformations\label{sus.gtr}}

Moreover, let
\begin{equation}
\eta_p=\eta(x_p,k),\quad
W_{pq}(n)'=\biggl({\eta_p\over\eta_q}\biggr)^n\,W_{pq}(n),\quad
\wb_{pq}(n)'=\bigl(\eta_p\eta_q\bigr)^n\,\wb_{pq}(n),
\tag {2.10}
\end{equation}
for any arbitrary function $\eta$. We can then replace the $W$ and $\wb$ in
(\ref{2.6}) by the $W'$ and $\wb'$ in (\ref{2.10}). We find that, if
$W_{pq}(n)$ and $\wb_{pq}(n)$ satisfy (\ref{2.6}), then $W_{pq}(n)'$ and
$\wb_{pq}(n)'$ also satisfy the star-triangle equations (\ref{2.6}).

\myfC

Furthermore, the transformation (\ref{2.10}) leaves the partition function for
a system with periodic boundary conditions invariant. This can be seen easily
by examining what happens at a particular site $e$ under such a transformation
in a checkerboard lattice with $p$, $p'$, $q$, and $q'$ as the rapidity
variables, as shown in Fig.\ 3. One finds that the additional factor
$\eta_q^{e}$ in $\wb_{pq}(e-d)'$ cancels out the factor $\eta_q^{-e}$ in
$W_{p'q}(e-c)'$, and the $\eta_p^e$ in $\wb_{pq}(e-d)'$ cancels out the
$\eta_p^{-e}$ in $W_{pq'}(a-e)'$, etc, leaving the net contribution at each
site unchanged.

Particularly, if we choose $\eta_p=\mu_p^{-1}$, then $W_{pq}(n)'$ and
$\wb_{pq}(n)'$ are no longer periodic, and they differ from (\ref{2.7}) by
dropping the $\mu_p$ and $\mu_q$ factors, making the weights and transfer
matrices depend rationally on $x_p$, $x_q$, $y_p$, and $y_q$ only, and
therefore more manageable. Indeed, as we are left only with the last equation
in (\ref{2.9}) the genus of the curve is then reduced to $g=(N-1)^2$.

On the other hand, if $\eta(x,k)$ in (\ref{2.10}) is a constant function, then
only $\wb$ picks up a factor. When we let $\eta_p=\eta_q=\omega^{\rho}$, then
$\wb$ remains periodic, whenever $\rho$ is an integer.

\subsection{Integrable Model and Zero-Temperature Limit\label{sus.izt}}

We now compare the weights of the integrable model with the weights given
by (\ref{2.5}) and (\ref{2.2}), which has $2(N-1)$ variables $E_j$ and $\bE_j$.
Hence for (\ref{2.2}) to be integrable, there must be $2N-5$ equations between
these $2N-2$ variables. We rewrite the $2(N-1)$ variables as
\begin{eqnarray}
-\,\betaE{E_j}=K_j\,\omega^{\Delta_j},&\qquad&
-\,\betaE{E_{N-j}}=K_j\,\omega^{-\Delta_j},\cr
&&\cr
-\,\betaE{\bE_j}=\bK_j\,\omega^{\bD_j},&\qquad&
-\,\betaE{\bE_{N-j}}=\bK_j\,\omega^{-\bD_j},
\tag {2.11}
\end{eqnarray}
for $1\le j\le [N/2]$, with $[x]$ denoting the integral part of $x$.
Therefore (\ref{2.2}) becomes
\begin{equation}
-\,\betaE{\cE(n)}=\left\{\begin{array}{ll}
\displaystyle\sum_{j=1}^{\halfs(N-1)}2K_j
\cos\biggl({2\pi\over N}(jn+\Delta_j)\biggr),
\qquad\mbox{$N$ odd},\vspace*{10pt}\cr
\,\,\displaystyle\sum_{j=1}^{\halfs N-1}2K_j
\cos\biggl({2\pi\over N}(jn+\Delta_j)\biggr)+K_{\halfs N}(-1)^n,
\qquad\mbox{$N$ even},\end{array}\right.
\tag {2.12}
\end{equation}
with similar equations for $\bCE$. For real $K_j$ and
$\Delta_j$, the Boltzmann weights are real and positive. When $N$ is odd, we
can think of the interactions as composed of $\half(N-1)$ chiral clock model
terms; particularly, for $N=3$, it is the three-state chiral clock model. For
even $N$ there is an additional Ising-like term; for example, for $N=4$, it is
composed of a four-state clock model with an Ising term.

The weights of the integrable models given in (\ref{2.7}) through (\ref{2.9})
can be rewritten in the form
\begin{equation}
{W(n)\over W(0)}={(\!(1,\alpha)\!)_{0,n}\over(\!(1,\beta)\!)_{0,n}},
\qquad
{\wb(n)\over\wb(0)}={(\!(1,\bA)\!)_{0,n}\over(\!(1,\bB)\!)_{0,n}},
\tag {2.13}
\end{equation}
where we have used the definitions
\begin{equation}
(\!(1,\alpha)\!)_{m,n}={(1,\alpha)_{m,n}\over\Delta(\alpha)^{n-m}},\quad
\Delta(\alpha)={(1-\alpha^N)}^{1/N},\quad
(1,\alpha)_{m,n}=\!\prod_{j=m+1}^n\!(1-\omega^j\alpha),
\tag {2.14}
\end{equation}
for $m<n$. It is easy to verify that
\begin{equation}
(\!(1,\alpha)\!)_{m,n}=(\!(1,\alpha)\!)_{m,k}
(\!(1,\alpha)\!)_{k,n}.
\tag {2.15}
\end{equation}
Since $(\!(1,\alpha)\!)_{m,m}=1$, we can extend the definition to
$m>n$ by
\begin{equation}
(\!(1,\alpha)\!)_{m,n}=1/
(\!(1,\alpha)\!)_{n,m}.
\tag {2.16}
\end{equation}
Moreover, because of the normalization factor $\Delta(\alpha)$, we have
\begin{equation}
(\!(1,\alpha)\!)_{m,m+N}=1.
\tag {2.17}
\end{equation}
Hence, the weights in (\ref{2.13}) are always periodic with period $N$.

In order that the weights (\ref{2.13}) satisfy the star-triangle equations
(\ref{2.6}), they have to satisfy only one necessary condition on the four
constants $\alpha$, $\beta$, $\bA$, and $\bB$, namely
\begin{equation}
{\alpha\over\beta}={\bB\over\omega\,\bA}.
\tag {2.18}
\end{equation}
This is consistent with (\ref{2.7}); it can also be derived directly.
Substituting $W(n)$, $\wb(n)$ (for $W_{pq}$ and $\wb_{pq}$), $W'(n)$, and
$\wb\s'(n)$ (for $W_{pr}$ and $\wb_{pr}$), as given by
(\ref{2.13}), into (\ref{2.6}), we can solve $W''(n)$, and $\wb\s''(n)$
(for $W_{qr}$ and $\wb_{qr}$), provided
\begin{equation}
{\alpha\over\,\bB\,}={\beta\over\omega\,\bA\,}=
{\alpha\s'\over\,\bB\s'}={\beta\s'\over\omega\,\bA\s'},\quad
{\Delta(\alpha)\Delta(\bA)\over\Delta(\beta)\Delta(\bB)}=
{\Delta(\alpha\s')\Delta(\bA\s')\over\Delta(\beta\s')\Delta(\bB\s')}.
\tag {e.20}
\end{equation}
{}From this, we precisely reproduce the $Z$-invariant periodic solutions
(\ref{2.7}) to (\ref{2.9}), up to possible gauge transformations as mentioned
in (\ref{2.10}). Other than that, the only other allowed variation on the
weights is that $\omega$ may be chosen to be any root of unity $\omega^N=1$ or
$\omega=e^{2\pi i j/N}$ for any integer $1\le j\le N-1$. Here $j$ and $N$ must
be relative prime, otherwise we would need to redefine $\Delta(\alpha)$ in
order to retain periodicity.
Letting
\begin{equation}
\alpha=e^{2i\theta},\quad \beta=e^{2i\phi},\quad
\bA=e^{2i\bT},\quad\bB=e^{2i\bP},
\tag{n.21}
\end{equation}
then the weights can be rewritten as
\begin{eqnarray}
{W(n)\over W(0)}\st&=&\st\left[{\sin(N\phi)\over\sin(N\theta)}\right]^{n/N}
\prod_{j=1}^n\,\left[{\sin(\theta+{\pi j/N})
\over\sin(\phi+{\pi j/ N})}\right],\cr&&\cr
{\wb(n)\over\wb(0)}\st&=&\st\left[{\sin(N\bP)\over\sin(N\bT)}\right]^{n/N}
\prod_{j=1}^n\,\left[{\sin(\bT+{\pi j/N})\over\sin(\bP+{\pi j/N})}\right],
\tag{n.22}
\end{eqnarray}
which are real as long as $\theta$, $\phi$, $\bT$, and $\bP$ are real. Now
(\ref{2.18}) becomes
\begin{equation}
\bP-\bT={\pi\over N}-\phi+\theta.
\tag {e.23}
\end{equation}
Therefore, the weights are functions of three independent variables. We can
relate the ``elliptic modulus" $k$ with these three variables as most of the
exact results are given in terms of this $k$. Comparing (\ref{2.7}) with
(\ref{2.13}), we find
\begin{equation}
\alpha=e^{2i\theta}=\frac{x_p}{y_q},\quad\beta=e^{2i\phi}=\frac{x_q}{y_p},
\quad\bA=e^{2i\bT}=\frac{x_q}{\omega x_p},\quad\bB=e^{2i\bP}=\frac{y_p}{y_q}.
\tag {e.24}
\end{equation}
Therefore
\begin{equation}
e^{2i(\theta-\phi)}=\frac{x_p y_p}{x_q y_q},\quad
e^{2i(\phi-\bT)}=\frac{\omega\,x_p}{y_p},\quad
e^{2i(\theta+\bT)}=\frac{x_q}{\omega\,y_q}.
\tag {e.25}
\end{equation}
We can then use (\ref{2.9}) to obtain\mydouble{\pagebreak}
\begin{eqnarray}
k^2\st&=&\st\sin^{-2}\Big(N(\theta-\phi)\Big)
\Big[\cos^2\Big(N(\theta+\bT)\Big)+\cos^2\Big(N(\phi-\bT)\Big)\cr&&\cr
&&\qquad-2\cos\Big(N(\theta+\bT)\Big)\cos\Big(N(\phi-\bT)\Big)
\cos\Big(N(\theta-\phi)\Big)\Big].
\tag {e.26}
\end{eqnarray}
This is also easily verified substituting (\ref{e.25}) in the right-hand
side of (\ref{e.26}) and eliminating $y_p^N$ and $y_q^N$ using the last
equality in (\ref{2.9}).

In section \ref{sec.con}, we express $\theta$ and $\phi$ in (\ref{n.21}) and
(\ref{n.22}) in terms of the $N-1$ variables $K_j$ and $\Delta_j$
of (\ref{2.12}). Clearly, these variables must satisfy $N-3$
consistency relations. The four variables $\theta$, $\phi$ and $\bT$,
$\bP$ can be easily rewritten in terms of $K_j$, $\Delta_j$, $\bK_j$,
$\bD_j$. The integrability condition (\ref{e.23}) gives another
condition relating them.

For the square lattice with $\wb=W$, (which is called the symmetric
case in the following), it then follows from (\ref{n.22}) that we must
have $\bT=\theta$ and $\bP=\phi$. Consequently, the integrability
condition becomes
\begin{equation}
\phi-\theta={\pi\over 2 N}\quad\mbox{or}\quad
\phi=\theta+{\pi\over 2 N}.
\tag {2.19}
\end{equation}
Now it is very easy to express $K_j=\bK_j$ and $\Delta_j=\bD_j$ in terms of
the single variable $\theta$, and plot graphs for different $N$. The details
are included in section \ref{sec.sym} and here we outline a few of the
conclusions. We find, with the energy unit convention $k_{\rm B}TK_1=1$ of
subsection \ref{sus.gra}, that the integrable curve ends at zero temperature at
\begin{eqnarray}
-\cE(n)\st&=&\st\sum_{j=1}^{N-1}\,{\sin(\pi/N)\over\sin(\pi j/N)}\,
\cos\left({2\pi nj\over N}\pm\Bigl({1\over2}-{j\over N}\Bigr)\pi\right)\cr\cr
\st&=&\st\sum_{l=1}^{N-1}\,{\sin(\pi/N)\over\sin(\pi l/N)}\,\,
\omega^{\,ln\,\mp\,(2l-N)/4}\cr\cr
\st&=&\st\left(N\pm2n-1+2N[\mp\fracd nN]\right)\sin{\pi\over N}.
\tag {2.20}
\end{eqnarray}
The last form is linear in $n$, periodically extended, with $[x]$ again
denoting the integral part of $x$. The first form in (\ref{2.20}) hides
the linearity with $n$, but will suggest an interesting generalization.
As the temperature $T$ increases, $\Delta_1$ decreases, and on
the integrable line, the ratios $K_j/K_1$ and $\Delta_j/\Delta_1$ remain
almost constant for $1\le j\le [\half (N-1)]$, namely
\begin{equation}
{K_j\over K_1}={\sin(\pi/N)\over\sin(\pi j/N)}+\kappa_j,
\quad{\Delta_j\over\Delta_1}={N-2j\over N-2}+\delta_j,
\tag {2.21}
\end{equation}
with $\kappa_j,\delta_j\lsim0.02$.
For $\Delta_j=0$, the self-dual and therefore critical case, we have
\begin{equation}
K_{j\rm c}=\sum_{m=1}^N{\sin\Big(\pi j(2m-1)/N\Big)\over2N\sin(\pi j/N)}\,
\log\left[{\sin\Big((m-\frac14)\pi/N\Big)\over
\sin\Big((m-\frac34)\pi/N\Big)}\right].
\tag {2.22}
\end{equation}
The curves of $1/K_j$ versus $\Delta_k$ are symmetric with respect to the
vertical axis.

\subsection{Boundary Conditions and Interfacial Tension at
{\mybmi T}\mysp=\mysp0\label{sus.bit}}

In the study of the interfacial tensions in model
systems\cmyref{\ref{Bax1}-\ref{Bax3},\ref{HSF}-\ref{SXEB}} various boundary
conditions and different orders of taking the limit are being used. We shall
here examine the resulting differences, using the chiral clock model as an
example.

In almost all of the numerical studies of the chiral clock model, fixed
boundary conditions are preferred. Specifically, in the work of Yeomans and
Derrida\cmyref{\ref{YD}} they consider a lattice with $\cL$ rows and
$\cM$ columns, as shown in Fig.\ 4, and demand that the spins on the
boundary rows have fixed values: $\sigma(m,0)=r$ and $\sigma(m,\cL)=0$;
but they impose periodic (or cyclic) boundary conditions on the utmost
left and right columns. They choose to have finite $\cL$, but $\cM\to\infty$;
that is, in the direction along the interface, the system is infinite from the
start.

\myfD

Huse et al., in their low-temperature analysis\myref{\ref{HSF}} of the wetting
transition for the symmetric case with $W=\wb$, consider a lattice oriented
diagonally with $L$ rows and $M$ columns,\footnote{We have made a trivial
reflection. They have a ``vertical" interface, whereas we have a ``horizontal"
one.} as shown in Fig.\ 5.
The spins in the top row have fixed values $r$ and in the bottom row they are
$0$. Moreover, they find it convenient to pin the diagonal interface in the
middle by further demanding the spins in the upper halves of the boundary
columns to have the fixed values $r$, and in the lower halves fixed values 0.
That is $\sigma(0,\ell)=\sigma(M,\ell)= r$ for $0\le\ell\le\half L$ and
$\sigma(0,\ell)=\sigma(M,\ell)= 0$ for $\half L<\ell\le L$. Their analysis is
done by taking $L\to\infty$ first, then $M\to\infty$; that is, in the direction
perpendicular to the interface, the system is infinite from the start.

\myfE

In the analytical works of Baxter\cmyref{\ref{Bax1}-\ref{Bax3}} such a
diagonally oriented\B lattice is being used for computational convenience,
because the diagonal\B transfer matrices form commuting families. Just
as in the Ising model, where Onsager obtained the interfacial
tension\myref{\ref{O1}} by imposing antiperiodic and\B periodic ($N\!=\!2$)
boundary conditions, here skew boundary conditions are imposed on the top
and bottom boundary ``rows," that is $\sigma_{m,L+1}=\sigma_{m,0}-r$,
($r\!=0,\ldots,N\!-\!1$), for
the two boundary spins in the same ``column," while cyclic boundary
conditions are imposed on the two utmost left and right boundary columns
with $\sigma(0,\ell)=\sigma(M,\ell)$. Due to such skew boundary conditions,
a ``horizontal" (actually, diagonal) interface occurs. The skew boundary
conditions do not affect the commutation properties\pmyref{\ref{AP1}}

\subsubsection{Interfacial Tension at {\mybmi T}\mysp=\mysp0\label{sss.itz}}

At zero temperature only ground states contribute to the partition function
$Z$, and to its logarithm, which is proportional to the free energy $F$. We
would typically expect a low-temperature behavior
\begin{equation}
-\log Z\equiv{F\over k_{\rm B}T}={E_{\rm g}\over k_{\rm B}T}-
\mysig_0+{\rm o}(T),
\tag {z.1}
\end{equation}
where $E_{\rm g}$ is the ground-state energy, $\mysig_0$ is an entropic term
related to the ground-state degeneracy, and ${\rm o}(T)$  stands for (usually
exponentially) small corrections. In this subsection, we shall concentrate
on the ground-state energy and surface or interface corrections to the
bulk behavior of it.

In the ferromagnetic case, $\cE(r)<\cE(0)$, $\bCE(r)<\bCE(0)$ for $r\ne0$,
and with the fixed boundary conditions mentioned above, the ground state
has to have a seam or interface. For the lattice shown in Fig.\ 4, the
excess free energy at $T=0$, or the increase in the ground-state energy
due to the mismatch of spins in two adjacent rows, divided by the interface
length $\cM$, is defined to be the horizontal interfacial tension
$\epsilon_r$. Hence the interfacial tensions at $T=0$ are given
by\footnote{Several statements in this subsection also apply to nonzero
temperatures, provided we replace excess energies by excess free energies per
``surface area" (length) $\area$. The resulting interfacial tensions or
surface tensions are independent of the choice of the ensemble and its
corresponding thermodynamic potential, as long as, with the ``surface volume"
(area) $V^{\rm s}$, each surface order parameter or its corresponding surface
field vanishes\pmyref{\ref{Griff}}}
\begin{equation}
\epsilon_r=\delta\cE(-r)\equiv\cE(-r)-\cE(0)>0,\quad\hbox{for }r=1,\ldots,N-1.
\tag {2.23}
\end{equation}
Now, from (\ref{2.20}), we find that the $N-1$ interfacial tensions of the
symmetric integrable chiral Potts model at $T=0$ satisfy
\begin{equation}
\epsilon_r=2r\sin\frac{\pi}N=r\epsilon_1=\epsilon_{r-1}+\epsilon_1,
\tag {2.24}
\end{equation}
which is a consequence of the special form of the $W$ weights.

Hence, at zero temperature, this not only signifies a wetting transition, but
for $N>3$ a more special phenomenon is taking place which we shall call
``superwetting," with the maximal amount of interface degeneracy as each
interface of type $r$ is free to break up into two interfaces of types $j$ and
$r-j$, for any $j$ between $1$ and $r-1$.\footnote{Interfacial wetting can
occur only in systems with three or more bulk phases ($N>2$), contrary to
surface wetting which can occur in the Ising model
($N\ge2$)\pmyref{\ref{AbrahamKoS}} However, ``surface superwetting" also
requires $N>2$ and interfacial superwetting $N>3$.} It is as if we have energy
levels given by spin operators $S^z$ for $2S+1=N$, similar to what happens in
the superintegrable chiral Potts model\pmyref{\ref{vGR},\ref{AlMPT}}

It should be obvious that the horizontal couplings $\bCE$ have no role to play
at zero temperature. Particularly, for the three-state chiral clock model,
there is no difference in the horizontal interfacial tension between the
symmetric case with $\cE(n)=\bCE(n)$ and the Ostlund-Huse asymmetric case
with the same $\Delta_1\ne0$ but $\bD_1=0$.

On the other hand, if we interchange $\cE(n)$ and $\bCE(n)$, then the
incremental free energy due to the mismatch is now, for the Ostlund-Huse
$N=3$ case,
\begin{equation}
\delta\bCE(-1)=\delta\bCE(1)=3,\quad\delta\bCE(r)\equiv\bCE(r)-\bCE(0),
\tag {n.28}
\end{equation}
different from
\begin{equation}
\delta\cE(\mp1)=3\cos\Bigl({2\pi\over3}\Delta_1\Bigr)\mp
\sqrt3\sin\Bigl({2\pi\over3}\Delta_1\Bigr)
=2\sqrt3\sin\Bigl({\pi\over3}(1\mp2\Delta_1)\Bigr).
\tag {n.29}
\end{equation}
Thus the excess in free energies at $T=0$ calculated for the asymmetric
case are different in the two directions. This means that the interfacial
tensions are anisotropic. We denote the angle-dependence by letting
$\epsilon_r=\epsilon_r(\varphi)$, whenever confusion may occur, with
$\epsilon_r(0)$ denoting the horizontal, $\epsilon_r({\half\pi})$ the
vertical, and $\epsilon_r({\quarter\pi})$ the diagonal interfacial tension.

The diagonal interfacial tension at zero temperature can also be calculated
by considering the incremental ground-state energy due to the mismatch of
bonds in a lattice, for example as shown by dashed lines in Fig.\ 5.
Following Baxter\cmyref{\ref{Bax2},\ref{Bax3}} we calculate the incremental
energy per horizontal and vertical bond pair. For the symmetric case with
$W=\wb$, we find
\begin{equation}
\epsilon_r=2\delta\cE(-r)=4r\sin(\pi/N),
\tag {2.25}
\end{equation}
which is double the amount in (\ref{2.24}) and the same whether cyclic or
free boundary conditions are imposed on the left and right boundary
columns. Again this shows that the integrable model is at a superwetting
transition at zero temperature.

For the asymmetric cases, however, the diagonal interfacial tensions at $T=0$
are given by
\begin{equation}
\epsilon_r({\quarter\pi})=\delta\cE(-r)+\delta\bCE(-r).
\tag {2.26}
\end{equation}
For the three-state Ostlund-Huse model at $\Delta_1=\quarter$, we have
\begin{equation}
\delta\cE(-1)=\sqrt3,\quad\delta\cE(-2)=2\sqrt3,\quad
\delta\bCE(-1)=\delta\bCE(-2)=3.
\tag {2.28}
\end{equation}
Substituting these into (\ref{2.26}), we find that
${1\over 2}\epsilon_1({\quarter\pi})$ is greater than $\epsilon_1(0)$
given in (\ref{2.24}) and ${1\over 2}\epsilon_2({\quarter\pi})$ is smaller
than $\epsilon_2(\half\pi)$ in (\ref{2.24}). Hence this diagonal
interfacial tension, as well as the vertical interfacial tensions, in the
Ostlund-Huse asymmetric case with $W\ne\wb$, do not satisfy the superwetting
condition even at $T=0$.

At zero temperature, we can also easily find the interfacial tensions
for general angle $0\le\varphi\le\half\pi$, as
\begin{equation}
\epsilon_r(\varphi)={{\cal N}(\varphi)}^{-1}
\Bigl(\delta\cE(-r)\cos\varphi+\delta\bCE(-r)\sin\varphi\Bigr),
\tag {z.2}
\end{equation}
which can be normalized\footnote{For the excess energy per unit length we
would have to use ${\cal N}(\varphi)=1$.} per horizontal (or vertical) bond
using
\begin{equation}
{\cal N}(\varphi)=\max\Bigl(\cos(\varphi),\sin(\varphi)\Bigr).
\tag {z.3}
\end{equation}
Such a ground state is highly degenerate as there are many configurations
of the interface with a given number of horizontal and vertical bonds. For
ground-state wetting we need that either a horizontal or a vertical piece
of the interface wets. So, for $N=3$ the system is not wet when both
$0\le\Delta_1,\bD_1<\quarter$, and it is wet when at least one of the
$\Delta_1,\bD_1\ge\quarter$ within the interval $0\le\Delta_1,\bD_1\le\half$ .
A ground-state wetting transition occurs for
\begin{equation}
\Delta_1=\quarter,\quad 0\le\bD_1\le\quarter\quad {\em or}\quad
0\le\Delta_1\le\quarter,\quad \bD_1=\quarter,
\tag {z.4}
\end{equation}
which is the boundary of these two regimes.

At $\Delta_1=\half$, we find from (\ref{n.28}) and (\ref{n.29}) that
\begin{equation}
\delta\cE(-1)=0,\quad \delta\cE(-2)=3,\quad
\delta\bCE(-1)=\delta\bCE(-2)=3.
\tag {z.5}
\end{equation}
Now we can use (\ref{2.26}) to find that the diagonal interfacial tensions do
satisfy the condition for the onset of wetting: $\epsilon_2=2\epsilon_1$.
This means that the wetting transition of the diagonal interface
of the Ostlund-Huse model occurs at $T=0$ and $\Delta_1=\half$.
It is interesting to note that the chiral melting line starts at the
same point. The finite-strip calculations of Yeomans and
Derrida\myref{\ref{YD}} show that even the vertical interface (which is
parallel to the chiral field) is wet at this point. It is interesting to
investigate whether the wetting curve of the diagonal interface is
identical to the chiral melting curve.

We shall now consider some other subtleties of the interfacial tensions, as
they relate to different boundary conditions and different orders of taking
the thermodynamic limit. We shall restrict ourselves to diagonally oriented
lattices only.

\subsubsection{The Limit
{\mybmi M}\mysp{\mybsy\char'041}\mysp{\mybsy\char'061}, then
{\mybmi L}\mysp{\mybsy\char'041}\mysp{\mybsy\char'061}\label{sss.lml}}

If one lets the number of columns $M\to\infty$ first and the number of rows
$L\to\infty$ afterwards, then the boundary condition imposed on the columns
should not have any effect. This can also be seen by considering the
elements\B of the column transfer matrix $T^{(0r)}$ with finite $L$ ---
dividing by the largest eigenvalue of $T^{(00)}$ (the bulk term). At non-zero
temperature, these elements are all positive and thus the Perron-Frobenius
theorem holds. Consequently, the largest eigenvalue is nondegenerate. In the
limit $M\to\infty$, only the largest eigenvalue survives. This shows that the
resulting interfacial tension is independent of the boundary conditions
imposed on the columns for $T>0$.

Nevertheless, in this limit, the interfacial tensions depend heavily on the
boundary conditions imposed on the top and bottom rows. We may consider the
interface (or domain wall) as a random walker\cmyref{\ref{MEF}} who tends to
walk in the direction of higher probability. For the skewed boundary condition,
this allows an interface winding around the cylinder of length $L$ and
perimeter $M$ crossing the seam of modified bonds several times. Thus for the
Ostlund-Huse model, with\myref{\ref{HSF}}
\begin{eqnarray}
&\xr\equiv w_1\equiv W(-1)/W(0),\quad\yr\equiv w_2\equiv W(-2)/W(0),\cr\cr
&\zr\equiv\wo_1\equiv\wb(-1)/\wb(0)=\wo_2\equiv\wb(-2)/\wb(0),\cr\cr
&\xr<\zr<\yr,
\tag{os.2}
\end{eqnarray}
the $\epsilon_1$ interface prefers to walk perpendicular to the chiral field
and then continue by crossing the seam, while the $\epsilon_2$ interface
tends to walk parallel to the chiral field, as shown in Figs.\ 6a and 6b. Thus
one does obtain the horizontal interfacial tension $\epsilon_1(0)$ and the
vertical interfacial tension $\epsilon_2({\half\pi})$ using a diagonally
oriented lattice with skewed boundary conditions. For fixed boundary
conditions on the boundary rows, one obtains instead the diagonal interfacial
tensions $\epsilon_r({\quarter\pi})$. Specifically, we
find\mydouble{\pagebreak}
\begin{eqnarray}
&\mbox{skewed b.c.:}&\qquad
{\displaystyle\lim_{L\to\infty}\lim_{M\to\infty}\,k_{\rm B}T\log(Z_r/Z_0)=
\min_\theta\,\epsilon_r(\theta)},\cr&&\cr
&\mbox{fixed b.c.:}&\qquad
{\displaystyle\lim_{L\to\infty}\lim_{M\to\infty}\,k_{\rm B}T
\log(Z_{0r}/Z_{00}) =\epsilon_r({\quarter\pi})},
\tag{bc.1}
\end{eqnarray}
where $Z_r$ and $Z_{0r}$ denote the corresponding partition functions.

\mysingle{\myfF}

\mydouble{\myfFa}

\mydouble{\myfFb}

In the latter case, with fixed boundary conditions for the boundary spins, we
have free open boundary conditions for the domain walls, which can touch but
not cross the boundaries. On the other hand, a seam due to skewed boundary
conditions can be moved to any place using gauge transformations. In spite of
the Perron-Frobenius theorem, these two cases are very different, even in the
limit $M\to\infty$, when the size of the row transfer matrix becomes infinite.

\subsubsection{The Limit
{\mybmi L}\mysp{\mybsy\char'041}\mysp{\mybsy\char'061}, then
{\mybmi M}\mysp{\mybsy\char'041}\mysp{\mybsy\char'061}\label{sss.llm}}

If we take the limit that the number of rows $L\to\infty$ first, then
it is necessary to pin the interface roughly at the middle\footnote{This is
equivalent to the subtraction procedure used by Baxter\cmyref{\ref{Bax2}}
who needs to omit an $L^2Z_1^2$ term from $Z_2$ in order to obtain
$\epsilon_2$, see his (5.27), (A9), and surrounding text.} in order
to get a true $\epsilon_r$ interface. Otherwise, because there are $L$
positions to place the $\epsilon_r$ interface, while there are
({\scriptsize$\!\!\begin{array}{c}L\\r\end{array}\!\!$}) possible positions
to place the $r$ $\epsilon_1$ interfaces, the term with $r$ $\epsilon_1$
interfaces dominates the partition function, in this limit.

Now it is easy to see that the boundary conditions on the top and bottom
rows can have no impact, as an interface of finite length $M$ could not
possibly reach the boundary rows which are infinitely far away. On the other
hand, with one end of the interface being pinned to the middle, cyclic
boundary conditions imposed on the boundary columns force the interface
to come back giving the diagonal interfacial tension, while free boundary
conditions on the boundary columns allow the interface to wander and
settle to its lowest energy configuration, yielding the minimum of the
interfacial tensions.

\subsubsection{The Limit
{\mybmi L},\mysp{\mybmi M}\mysp{\mybsy\char'041}\mysp{\mybsy\char'061},
with fixed {\mybmi M}\mysp{\mybmi\char'075}\mysp{\mybmi L}\label{sss.lmf}}

In exact calculations, however, it is cumbersome to pin the interface
in the middle. As Baxter\myref{\ref{Bax3}} chooses to let $M,L\to\infty$
simultaneously, we shall examine the case when they are proportional to
one another. Since the Perron-Frobenius theorem may not hold
on infinite matrices, a difference in boundary conditions could play
an important role.

The full complexity seems to arise when $M/L>1$. It is most interesting to
consider the case when fixed boundary conditions are imposed on the top and
bottom rows and free boundary conditions on the columns. Then it is easily
seen that it is energetically more favorable for the system to arrange itself
into a configuration with a mismatch as shown in Fig.\ 6c. Therefore, if we
let $M\to\infty$ and fix the ratio $M/L>1$, the excess in free energy at $T=0$
can have any arbitrary value depending on $M/L$. A similar problem arises,
when skewed boundary conditions are imposed on the top and bottom rows. For if
the ratio $M/L$ is an integer, or free boundary conditions are imposed on the
boundary columns, we will obtain the minimum of the two interfacial tensions
shown in Figs.\ 6a and 6b; otherwise, the interfacial tension will be a
function of $M/L$.

For $M/L<1$, however, this complexity disappears, and cyclic boundary
conditions on the columns give the diagonal interfacial tensions, whereas
free boundary conditions give the minimum of the interfacial tensions
over all directions. Thus by taking the limits in an appropriate way, we may
calculate horizontal or vertical interfacial tensions using a diagonally
oriented lattice.

\subsubsection{Single Interface at Low Temperatures\label{sss.sil}}

Let us now discuss the effect of boundary conditions on the restricted\B
partition functions with one domain wall. As before, $M$ denotes the size of
the system in the direction of the interface, and $L$ the size  perpendicular
to it. Besides the partition function without the domain wall $Z_0(L,M)$, we
can introduce various partition functions with a single domain wall. First, we
can introduce $Z_{1,\rm pin}(L,M|s)$ for the partition function with the
interface pinned in the middle on one side and pinned $s$ steps from the
middle on the other side. We can also introduce $Z_{1,\rm per}(L,M)$ for the
periodic case, where the interface starts and ends at the same, but otherwise
free, position, and $Z_{1,\rm free}(L,M)$ for the case where the interface is
left free on both sides. These three partition functions are closely related
when $L$ becomes large, namely
\begin{eqnarray}
&&\frac{Z_{1,\rm per}(L,M)}{Z_0(L,M)}\approx
L\,\lim_{L'\to\infty}\,\frac{Z_{1,\rm pin}(L\,',M|0)}{Z_0(L\,',M)},\cr&&\cr
&&\frac{Z_{1,\rm free}(L,M)}{Z_0(L,M)}\approx
L\,\sum_{s=-\infty}^{+\infty}\,\lim_{L'\to\infty}\,
\frac{Z_{1,\rm pin}(L\,',M|s)}{Z_0(L\,',M)}.
\tag{si.1}
\end{eqnarray}
Therefore, these quantities are easily evaluated at low temperatures, for
which overhangs can be ignored, using e.g.\ random walks, transfer matrix
techniques, or Szeg\"o's theorems for Toeplitz matrices.

For the horizontal interface as in Fig.\ 4 and using (\ref{os.2}), we find
\begin{equation}
\lim_{\cM\to\infty}\lim_{\cL\to\infty}\frac1{\cM}\log
\frac{Z_{1,\rm free}(\cL,\cM)}{\cL\,Z_0(\cL,\cM)}\approx
\log\frac{(1-\xr\yr)\,\xrb}{(1-\xr)(1-\yr)}
\tag{si.2}
\end{equation}
and\mysingle{\pagebreak}
\begin{eqnarray}
&&\lim_{\cM\to\infty}\lim_{\cL\to\infty}\frac1{\cM}\log
\frac{Z_{1,\rm per}(\cL,\cM)}{\cL\,Z_0(\cL,\cM)}\cr&&\cr
&&\qquad\approx\lim_{\cL\to\infty}\lim_{\cM\to\infty}\frac1{\cM}\log
\frac{Z_{1,\rm fix}(\cL,\cM)}{Z_{0,\rm fix}(\cL,\cM)}\approx
\log\frac{(1-\xr\yr)\,\xrb}{{(1-\sqrt{\xr\yr})}^2},
\tag{si.3}
\end{eqnarray}
where ``fix" stands for various boundary conditions which keep the interface
continuing within a horizontal strip of width $\cL$. The results
differ only by exponentially small terms in the temperature $T$, following
a leading term of order $1/T$, showing that ``kinks" are important.

However, for the diagonal interface as in Fig.\ 5, we find
\begin{equation}
\lim_{M\to\infty}\lim_{L\to\infty}\frac1M\log
\frac{Z_{1,\rm free}(L,M)}{L\,Z_0(L,M)}\approx
2\log(\xr+\xrb)
\tag{si.4}
\end{equation}
and
\begin{eqnarray}
&&\lim_{M\to\infty}\lim_{L\to\infty}\frac1M\log
\frac{Z_{1,\rm per}(L,M)}{L\,Z_0(L,M)}\cr&&\cr
&&\qquad\approx\lim_{L\to\infty}\lim_{M\to\infty}\frac1M\log
\frac{Z_{1,\rm fix}(L,M)}{Z_{0,\rm fix}(L,M)}\approx
2\log(2\sqrt{\xr\xrb}),
\tag{si.5}
\end{eqnarray}
where the factor 2 is to have agreement with Baxter's convention. Now the
differences occur in the order 1, and the corrections are entropic in nature.

{}From the results (\ref{si.2}) to (\ref{si.5}), we see that the effects of
the boundary disappear when the system is reflection symmetric with respect to
an axis perpendicular to the interface, so that $\xr=\yr$ in (\ref{si.2}) and
(\ref{si.3}), or $\xr=\xrb$ in (\ref{si.4}) and (\ref{si.5}). In those two
cases, for which the interface is perpendicular to the chiral direction, we
can use ref. \ref{HSF} to obtain further detail.

\subsection{Wetting Transitions\label{sus.wet}}

Wetting transitions can occur when there are different types of domains
and domain walls. More precisely, the wetting transition is defined as
``the\B bubbles of $B$ domain absorbed on the $A|\!|C$ interface merge
into an essentially\B macroscopic layer of $B$ domain which wets the
entire interface," (see e.g.\ page 377 of ref.\ {\ref{HSF}}). We need
to compare a configuration with a domain wall $A|\!|C$ with a
configuration with two domain walls $A|B$ and $B|C$. Hence, the wetting
temperature $T_{\rm w}$ is defined by
\begin{equation}
T_{\rm w}=\min_\theta\,T_{\rm w}(\theta),\hbox{ with }T_{\rm w}(\theta)
\hbox{ from }\epsilon_2(\theta,T)=2\,\epsilon_1(\theta,T).
\end{equation}
As the interfaces can be oriented differently, the wetting transition can\B
occur at different temperatures $T_{\rm w}(\theta)$, but its minimum is the
true wetting temperature for given chirality $\Delta_1$.

We note that we need to compare the interfacial tensions $\epsilon_2(\theta)$
and $2\epsilon_1(\theta)$ for the same angle $\theta$, as is implicit in all
the calculations mentioned above\pmyref{\ref{HSF}-\ref{SXEB}} It is often
energetically more favorable to wet than to turn the interface through an angle
for which it is necessary to flip macroscopically many spins. We should be
careful not to calculate $T_{\rm w}$ from
$\min\epsilon_2(\theta,T)=2\,\min\epsilon_1(\theta,T)$,
as a calculation with skewed boundary conditions in both directions could
lead to and which would cause us to overestimate $\Delta_{\rm wet}(T)$.

It seems that an interface perpendicular to the ``chiral field" $\Delta_1$
wets first. In fact, Huse et al.\ calculate the wetting curve by
considering diagonal interfaces for the symmetric case, with
$\Delta_{\rm ver}=\Delta_{\rm hor}$, and horizontal interfaces for the
Ostlund-Huse model with $\Delta_{\rm ver}\ne 0$ and $\Delta_{\rm hor}=0$.
Particularly, for the diagonal interface of the Ostlund-Huse
model, its wetting line would have to start at zero temperature at
$\Delta_1=\half$, above the wetting curve of the horizontal interface,
which wets at $\Delta_1=\quarter$ and $T=0$. We note that a diagonal
interface is a superposition of many allowed walks with a given number
of horizontal and vertical bonds. So with a chiral field in the vertical
direction, horizontal segments of the interface will wet at $\quarter$,
while vertical segments will not wet, lowering $\Delta_{\rm wet}(T\!=\!0)$
from $\half$ to $\quarter$.

It is easy to extend the low-temperature analysis of Huse et
al\pmyref{\ref{HSF}} to the case with $K_1\ne\bK_1$ and one finds that, at low
temperature, the wetting line for the Ostlund-Huse model,
($\xrb=\yrb\equiv\zrb$), is given by\footnote{There is a small discrepancy
with a factor 3 given by Huse et al\pmyref{\ref{HSF}} instead of 2 as given
below, which we believe to be due to a misprint.}
\begin{equation}
\frac{\yr}{\xr^2}=
\exp\left[6K_1\sin\left(\fracd16\pi(1-4\Delta_1)\right)\right]=
1+2\zrb^2+{\rm O}(\zrb^3),
\quad\zrb=\exp(-6\bK_1),
\tag{os.3}
\end{equation}
while the integrable line is given by
\begin{equation}
\frac{\yr}{\xr^2}=1+2\zrb+{\rm O}(\zrb^2).
\tag{os.4}
\end{equation}
The integrable line, denoted by line 3, and the wetting line, denoted by
line 4, are shown in Fig.\ 7b. They rise faster than any power due to the
importance of kinks, see also the text below (\ref{si.3}).

\mysingle{\myfG}

\mydouble{\myfGa}

\mydouble{\myfGb}

For the symmetric lattice with $W=\wb$, ($\xrb=\xr,\yrb=\yr$), however, we
find that the diagonal interface wets first, with the wetting line given by
Huse et al.\ as
\begin{equation}
\frac{\yr}{\xr^2}=\fracd32+\cdots,
\tag{dia.1}
\end{equation}
whereas the integrable line is given by
\begin{equation}
\frac{\yr}{\xr^2}=2+\cdots.
\tag{dia.2}
\end{equation}
These are also plotted in Fig.\ 7b as lines 2 and 1 respectively.
The wetting curve for the horizontal or vertical interface in the
symmetric case also starts at $\Delta_1=\quarter$ and $T=0$, but rises
in a much steeper way. Since all the interfaces wet on this curve,
it would be interesting to compare the behaviors near this curve and near
the chiral melting curve of the Ostlund-Huse model.

We show in Fig.\ 7a the integrable line for the $3$-state chiral
clock\myref{\ref{HSF}-\ref{SXEB}} model. The solid curve is for the
Ostlund-Huse asymmetric case, while the dashed line is for the symmetric case,
with equal horizontal and vertical interactions. We can see that both curves
end up at the same point for zero temperature. The numerical results of
Stella et al\pmyref{\ref{SXEB}} for the wetting line in the Ostlund-Huse
model are represented here by crosses.

In Fig.\ 7b, we enlarge the graphs for $T$ small and $\Delta_1$ near
$\quarter$, with the integrable lines marked as lines 1 and 3 added to the
wetting line results of Huse et al.\ with line 2 for the symmetric lattice and
line 4 for the fully asymmetric case. The curves show that the integrable
lines are near but below the wetting lines. For $0<T<T_{\rm c}$, therefore,
contrary to what we have conjectured before\cmyref{\ref{APTa}} and in
agreement with Baxter\cmyref{\ref{Bax2},\ref{Bax3}} we find that the
integrable model is in the non-wetted region.

\subsection{Low-Temperature Regimes\label{sus.ltr}}

As suggested to us by Baxter,\footnote{Private communications.} we
can approximate the Boltzmann weights of the integrable chiral Potts model,
\begin{eqnarray}
w_n\st&\equiv&\st{W(-n)\over W(0)}={1\over\Lambda\strut^{n\,}}
\,\prod_{j=0}^{n-1}\,\left[{\sin(j\lambda-\phi)
\over\sin(j\lambda-\theta)}\right],
\cr&&\cr
\wo_n\st&\equiv&\st{\wb(-n)\over\wb(0)}={1\over\bL^{\,n}}
\,\prod_{j=0}^{n-1}\,\left[{\sin(j\lambda-\bP)
\over\sin(j\lambda-\bT)}\right],
\tag{b.12}
\end{eqnarray}
where
\begin{equation}
\lambda\equiv{\pi\over N}=\phi+\bP-\theta-\bT,\quad
\Lambda\strut^N\equiv{\sin(N\phi)\over\sin(N\theta)},
\quad\bL^{\,N}\equiv{\sin(N\bP)\over\sin(N\bT)},
\tag{b.13}
\end{equation}
see also (\ref{2.7}), (\ref{n.22}), and section \ref{sec.thr},
by replacing $\theta$, $\bT$, $\phi$, $\bP$ inside the products in
(\ref{b.12}) by their zero-temperature limiting values. For the
symmetric case, with
\begin{equation}
-\lambda<\theta\equiv\bT<-\threequarter\lambda,\quad
\phi\equiv\bP=\theta+\half\lambda,
\tag{b.14}
\end{equation}
we can express $\theta$ and $\phi$ in terms of $k$ using (\ref{e.26}). We
find successively
\begin{eqnarray}
&&k=\cos(2N\theta),\quad\sin(N\theta)=-\sqrt{\half(1-k)},\quad
\sin(N\phi)=-\sqrt{\half(1+k)},\cr&&\cr
&&\Lambda\strut^N=\bL^{\,N}=\sqrt{1+k\over1-k}={1+k\over k'},\quad
\phi=\bP=-{1\over N}\,\arcsin\sqrt{\half(1+k)},\cr&&\cr
&&\theta=\bT=-{1\over N}\,\Bigl(\pi-\arcsin\sqrt{\half(1-k)}\Bigr)=
\phi-{\pi\over2N}.
\tag{b.15}
\end{eqnarray}
The zero-temperature limit then corresponds to
\begin{equation}
\theta\equiv\bT\to-\lambda,\quad\phi\equiv\bP\to-\half\lambda,\quad
\Lambda\equiv\bL\equiv{\Big({1+k\over k'}\Big)}^{1/N}\to\infty,
\tag{b.16}
\end{equation}
and we find the asymptotic low-temperature formula for the weights
\begin{equation}
w_n=\wo_n={1\over\Lambda\strut^n}\,\prod_{j=1}^n\,
{\sin\Big((j-\half)\lambda\Big)\over\sin(j\lambda)}.
\tag{b.17}
\end{equation}
Here we can allow $\bL\ne\Lambda$, as this corresponds to a gauge
transformation.

Baxter applied a Bethe Ansatz method with skew-periodic boundary\B
conditions\cmyref{\ref{Bax2}} shown in Fig.\ 5, together with a complicated
subtraction\B method, in order to obtain the first two diagonal interfacial
tensions in the low-temperature regime. Indeed, substituting
\begin{equation}
f=g=\tan^2(\half\lambda),\quad\chi_0=0,
\tag{b.18}
\end{equation}
corresponding to the symmetric case, his results (5.36) and (A12) reduce
to\pagebreak
\begin{eqnarray}
&&{\epsilon_1\over k_{\rm B}T}=-\log(4w_1\wo_1)=
2\log\Big(\Lambda\cos(\half\lambda)\Big),\cr&&\cr
&&{\epsilon_2\over k_{\rm B}T}-2{\epsilon_1\over k_{\rm B}T}=
\log\Big({(1-f^2)}^2\Big)=2\log\Big(1-\tan^4(\half\lambda)\Big),
\tag{b.19}
\end{eqnarray}
or, equivalently,
\begin{eqnarray}
{\epsilon_1\over k_{\rm B}T}\st&=&\st
2\log\Lambda\,+\,\log\Big(1+\tan^2(\half\lambda)\Big),\cr&&\cr
{\epsilon_2\over k_{\rm B}T}\st&=&\st
4\log\Lambda\,+\,2\log\Big(1-\tan^2(\half\lambda)\Big).
\tag{b.20}
\end{eqnarray}
These results are useful to verify that the exact
calculations\myref{\ref{Bax3}} based on analytic continuation from the
superintegrable case are correct.

For the symmetric case, there is an easier Bethe Ansatz calculation than
the one done by Baxter\cmyref{\ref{Bax2}} namely one with free boundary
conditions. This gives the desired results directly, without a subtraction
procedure. As we do not only have to consider ``collisions" of interfaces,
but also reflections at the boundary, we use an extension of the Bethe
Ansatz\myref{\ref{Bethe}} similar to the ones introduced by
Gaudin\pmyref{\ref{Gaudin}} But as this requires further consistency
conditions, our method does not apply to asymmetric cases that differ from
(\ref{b.17}) by more than changing $\Lambda$ to $\bL$ for $\wo$.
We shall outline our Bethe Ansatz calculations in section \ref{sec.bal}.
These calculations show some special features on the integrable line,
which may be related to some pre-wetting phenomena.

We note that the Bethe Ansatz method to calculate interfacial tensions in
the low-temperature limit is similar to the M\"uller-Hartmann--Zittartz
approximation\myref{\ref{SP}} in ignoring overhangs. Moreover, that method can
be also used for non-symmetric and non-integrable cases, with the condition
\begin{equation}
\yr/\xr^2=e^{6K_1\sin\bigl(\pi(1-4\Delta_1)/6\bigr)},\quad
\yrb/\xrb^2=e^{6\bK_1\sin\bigl(\pi(1-4\bD_1)/6\bigr)}\quad\mbox{finite}.
\tag{bac.1}
\end{equation}
However, for more than two interfaces the additional condition
$\yr/\xr^2=\yrb/\xrb^2$ is needed for the Bethe Ansatz to work.

The above procedure (\ref{b.16}) to obtain the low-temperature expansion also
works for asymmetric cases with
\begin{equation}
\theta,\bT\to-\lambda,\quad k\to1,\quad\bP=-\lambda-\phi,\quad
\mbox{with}\quad\phi\not\to-\lambda\hbox{ or }0.
\tag{bb.1}
\end{equation}
Later on in section \ref{sec.thr} we shall see that this represents just one
part of the integrable manifold in the phase diagram for the 3-state chiral
clock model. So we shall call models with (\ref{bb.1}) ``quasi-symmetric"
cases. It is a regime which does not include, nor even border, fully
asymmetric cases as the Ostlund-Huse model, for which the vertical weights
satisfy the condition $\wb(n)=\wb(-n)$ while the horizontal weights are
chiral, $W(n)\ne W(-n)$.

As we shall show for $N=3$ later in section \ref{sec.thr}, to have
$\wb(n)=\wb(-n)$, we must require
\begin{equation}
\bT=-\lambda-\bP,\quad\mbox{with}\quad\bL=1.
\tag{bb.2}
\end{equation}
Together with the integrability condition (\ref{b.13}), we find then that
there can be only two free variables, say $\theta$ and $\phi$, such that
\begin{equation}
\bT=-\lambda-\half(\theta-\phi),\quad\bP=\half(\theta-\phi),
\tag{bb.3}
\end{equation}
Because of that, in the low-temperature limit given by
$\theta,\,\bT\to-\lambda$, we must also have $\phi\to-\lambda$. In order to
see whether the integrable model is or is not in the wetted region, it is
necessary to include higher order terms in the low-temperature expansion. In
fact, for the three-state Ostlund-Huse model, the\B condition $K=\bK$
reduces the free variables to just one, and in the low-temperature limit,
$\phi\to-\frac13\pi+\delta\phi$, we find $\theta=-\frac13\pi+\delta\phi^\rho$
with $\rho=(1+\sqrt 3)$. (The details are delegated to section \ref{sec.thr}).
Therefore, we have
\begin{equation}
\frac{w_2}{{w_1}^2}=\frac{\yr}{\xr^2}=
{\rm O}(\delta\phi)\sim\wo_1=\wo_2=\zr={{\sin[\frac12(\phi-\theta)]}\over
{\sin[\frac13\pi-\frac12(\phi-\theta)]}}.
\tag{bb.9}
\end{equation}
Because these are of the same order of magnitude, we have to take kinks in the
interface into account\cmyref{\ref{HSF}} as also noted below (\ref{si.3}).

It seems that the largest asymmetric integrable regime is  described by
$\bP\to0$, $\theta,\,\bT,\,\phi\to\-\lambda$. For $N=3$, this corresponds to
$\bK=K$ and $\bD=p\Delta$, ($p\!\ne\!0,1$), which again reduces the parameters
to just one. We find
\begin{equation}
\phi=-\frac13\pi+\delta\phi,\quad\theta=-\frac13\pi+\delta\theta,\quad
\bP=-\delta\bP,\quad\bT=-\frac13\pi+\delta\bT,
\tag{bb.10}
\end{equation}
with
\begin{equation}
\delta\theta\approx\delta\phi\,{\left(\frac2{\sqrt3}\,
\delta\phi\right)}^{\sqrt{b^2+b+1}},\qquad
\delta\bP\approx\delta\phi,\qquad
\delta\bT\approx\delta\phi\,{\left(\frac2{\sqrt3}\,\delta\phi\right)}^{b-1},
\tag{bb.11}
\end{equation}
where
\begin{equation}
b={\sin\left(\fracd16(1+p)\pi\right)}
\left/{\sin\left(\fracd16(1-p)\pi\right)}\right..
\tag{bb.12}
\end{equation}
For limits as in (\ref{bb.1}), we shall show in section \ref{sec.thr} that
\begin{equation}
\Delta_1=\frac14-\frac{\kappa}{K_1}+\cdots,
\qquad\bD_1=\frac14-\frac{\overline\kappa}{\bK_1}+\cdots,
\tag{bb.13}
\end{equation}
with $\kappa$ and $\overline\kappa$ some positive constants. This shows that
both $\Delta$ and $\bD\to\fracd14$ as $T\to0$, which is rather restricted.

\subsection{Baxter's Exact Interfacial Tension Results, Symmetric
Case\label{sus.ber}}

In two recent papers\cmyref{\ref{Bax2},\ref{Bax3}} Baxter obtained several
new exact results for the interfacial tensions in the integrable chiral Potts
model. At this moment, it is not clear to us if or how we can extract from his
results the interfacial tensions for the fully asymmetric Ostlund-Huse case.
Here we shall work out his results in explicit detail only for the fully
symmetric case, leaving the results for the more general case implicit for
now.

In equation (42) of Baxter's second paper, we can
substitute\myref{\ref{Bax3}}
\begin{eqnarray}
u_q\st&=&\st v_q={\pi\over2N}=\half\lambda,\quad
\Lambda_q=\sqrt{1+k\over1-k},\quad\eta_p=1,\quad\eta_q=\eta,\cr&&\cr
\eta^{N/2}\st&=&\!\!{\Lambda_q-1\over\Lambda_q+1}=\sqrt{1-k'\over1+k'},
\quad k'\equiv\sqrt{1-k^2},\quad m=m_0=1,
\tag{b.1}
\end{eqnarray}
resulting in the very explicit formula for the interfacial tensions
\begin{equation}
{\epsilon_r\over k_{\rm B}T}=2v_r(m_0)=
{8\over\pi}\,\int_0^{\eta}dy\,{\sin({\pi r\over N})\over
1+2y\cos({\pi r\over N})+y^2}\,
\hbox{artanh}\sqrt{\eta^N-y^N\over
1-\eta^N y^N},
\tag{b.2}
\end{equation}
in the symmetric case. Here $\eta$ is a temperature-like variable on the
integrable curve, defined in (\ref{b.1}).

In the low-temperature region,
\begin{equation}
k'\approx\half N(1-\eta)\to0,
\tag{b.3}
\end{equation}
we can expand (\ref{b.2}) as
\begin{equation}
{\epsilon_r\over k_{\rm B}T}=
-{2r\over N}\,\log(\half k')-\mub_{r,N}+{\rm O}(k'),
\tag{b.4}
\end{equation}
where the constant is given by a dilogarithm integral, which seems
different from the ones studied recently by Kirillov and many
others\cmyref{\ref{Kir},\ref{NRT}}
\begin{eqnarray}
\mub_{r,N}\st&\equiv&\st
{4\over\pi}\,\int_0^1dx\,{\sin({\pi r\over N})\over
1+2x\cos({\pi r\over N})+x^2}\,
\log\biggl({1+x^N\over1-x^N}\biggr)\cr&&\cr
\st&=&\st\log\biggl({1\over\cos^2(\half r\lambda)}
\prod_{j=1}^{[\halfs r]}\,{\cos^4\Big(\half(r-2j+1)\lambda\Big)
\over\cos^4\Big(\half(r-2j)\lambda\Big)}\biggr).
\tag{b.5}
\end{eqnarray}
This also satisfies the sum rule
\begin{equation}
\sum_{j=1}^{N-1}\mub_{j,N}=\log N.
\tag{b.6}
\end{equation}
We have not found a direct way yet to prove the last equality in (\ref{b.5}),
except for the cases $N=2$ (Ising) and $N=3$. In section \ref{sec.bal}, we
shall outline an indirect proof using a Bethe Ansatz method. We originally
guessed the formula by performing numerical integration to 100 places
and expanding the exponentials of the integrals in periodic continued
fractions, which gave us
\begin{eqnarray}
\mub_{1,2}\st&=&\st\log2,\quad
\mub_{1,3}=\log{4\over3},\quad
\mub_{2,3}=\log{9\over4},\tag{b.7}\\
\mub_{1,4}\st&=&\st\log\Bigl(2(2-\sqrt2)\Bigr),\quad
\mub_{2,4}=-\log\Bigl(2{(2-\sqrt2)}^2\Bigr),{\strut}_{\strut}\cr
\mub_{3,4}\st&=&\st\log\Bigl(4(2-\sqrt2)\Bigr),\tag{b.8}\\
\mub_{1,5}\st&=&\st\log\Bigl(2(5-\sqrt5)/5\Bigr),\quad
\mub_{2,5}=\log(5/4),{\strut}_{\strut}\cr
\mub_{3,5}\st&=&\st\log\Bigl({32/{(5-\sqrt5)}^3}\Bigr),\quad
\mub_{4,5}=\log\Bigl(5{(5-\sqrt5)}^2/16\Bigr).
\tag{b.9}
\end{eqnarray}
Combining these results with Baxter's two Bethe Ansatz
results\myref{\ref{Bax2}} for $r=1$ and 2, see also (\ref{b.20}) in the
previous subsection, we originally guessed (\ref{b.5}).

For the critical region,
\begin{equation}
\eta^{\halfs N}\approx\half k\to0,
\tag{b.10}
\end{equation}
we can expand (\ref{b.2}) by first expanding the $\hbox{artanh}$ in a Taylor
series of its argument. Then, it is straightforward to rewrite the integrand
in (\ref{b.2}) as a power series of $y$ and $\eta$, multiplied with
$\sqrt{\eta^N-y^N}$. Each term of the series leads to a Beta function
integral, $B(x,y)=\Gamma(x)\Gamma(y)/\Gamma(x+y)$. We find
\begin{equation}
{\epsilon_r\over k_{\rm B}T}=
{8\sin({\pi r\over N})B({1\over N},\half)\over\pi(N+2)}\,
\eta^{\halfs N+1}-
{8\sin({2\pi r\over N})B({2\over N},\half)\over\pi(N+4)}\,
\eta^{\halfs N+2}+
{\rm O}(\eta^{\halfs N+3}),
\tag{b.11}
\end{equation}
giving both the leading term\myref{\ref{Bax3}} and the first
correction term. Coefficients of further terms $\eta^{\halfs N+j}$,
for $j=3,4,\ldots$, can also be obtained easily.

\subsection{Other Exact Results and Further Speculations\label{sus.exa}}

A great deal of progress on the chiral Potts models has been achieved. The
exponent ${\alpha}$ for the specific heat --- not to be confused with the
independent\B variables for the integrable weights given in (\ref{2.13}) ---
has been obtained by Baxter\myref{\ref{Bax4}} for the case with real positive
Boltzmann weights and the\B exponents ${\beta}_j$ of the one-point
functions $<\sigma^j>$ have been conjectured by Albertini et
al\pmyref{\ref{AlMPT}} They are
\begin{equation}
{\alpha}=1-\frac2N,\qquad
{\beta}_j=\frac{j(N-j)}{2N^2}.
\tag {2.30}
\end{equation}
In fact, Albertini et al\pmyref{\ref{AlMPT}} also conjectured the exact
low-temperature formula
\begin{equation}
<\sigma^j>={(1-{k'}^2)}^{\beta_j}=k^{2\beta_j},
\tag {s.0}
\end{equation}
which generalizes the Onsager result\myref{\ref{O2},\ref{O5}} for the
spontaneous magnetization of the Ising model.

For $N=3$, we find ${\alpha}=\fract13$, which is identical to the three-state
Potts model result. It is generally believed that the chiral field $\Delta_1$
is a relevant variable, and the free energy in the scaling region can be
written as
\begin{equation}
F(T,\Delta_1,\bD_1,H)=F(T,\Delta_1,p\Delta_1,H)=|t|^{2-\alpha}
X(g\Delta_1/|t|^{\phi},hH/|t|^{\psi}),
\tag {s.1}
\end{equation}
for some crossover exponents $\phi$ and $\psi$ and with $t=T/T_{\rm c}-1$,
where $T_{\rm c}$ is the Potts critical point. The ratio of the two chiral
fields $\bD_1/\Delta_1$ is believed to be an irrelevant variable, which may
only change the two constants $g$ and $h$.

The exact calculation of Baxter\myref{\ref{Bax4}} gives
\begin{equation}
F(T,\Delta_1,\bD_1,0)\sim {|k^2|}^{2-\alpha},
\tag {bs.1}
\end{equation}
where $k$ is the elliptic modulus given by (\ref{e.24}), with $k=0$ for
$T=T_{\rm c}$ and $\Delta_1=\bD_1=0$.

For $k\sim 0$, we find the integrable line lies on the curve given by
\begin{equation}
t=-9.959616982\,(1+p^2)\Delta_1^2+{\rm O}(t^2,t\Delta_1^2,\Delta_1^4),
\quad p=\bD_1/\Delta_1,
\tag{cs.1}
\end{equation}
and
\begin{equation}
k^2\sim \frac{108\,(2+\sqrt3)\,K_{1\rm c}^2}
{1+K_{1\rm c}\,(7\sqrt3+12)}\,t=-4.981050242\,t,
\tag{cs.0}
\end{equation}
with $K_{1\rm c}=J_1/k_{\rm B}T_{\rm c}=\fract13\log(1+\sqrt3)$, see also
eq.\ (\ref{oh.25}). In the above equation, $p=1$ corresponds to the symmetric
case and $p=0$ to the Ostlund-Huse asymmetric case. The analytic expansion
(\ref{cs.1}) for the integrable line is similar in form to the usual
expression for the nonlinear thermal scaling field\myref{\ref{HSF}}
\begin{equation}
{\tilde t}=t+c_2\Delta_1^2+\cdots,
\tag{cs.2}
\end{equation}
in terms of which the free energy can be rewritten as
\begin{equation}
F(T,\Delta_1,\bD_1,H)=|{\tilde t}|^{2-\alpha}
X(g\Delta_1/|{\tilde t}|^{\phi},hH/|{\tilde t}|^{\psi}).
\tag {s.3}
\end{equation}
But to say more about this requires more detail from exact results than we
currently have available.

{}From the most recent results in (\ref{b.11}), we conclude that we have the
following asymptotic expansions for the surface tensions on the integrable
line,
\begin{equation}
\epsilon_r=|t|^{\frac12+\frac1N}\,D_r(|t|^{\frac1N}),
\tag{ce.3}
\end{equation}
which we must compare with the forms required by scaling
\begin{equation}
\epsilon_r=|t|^{\mu}\,D_r(\Delta_1/|t|^{\phi}).
\tag{ce.4}
\end{equation}
Now, as the temperature is an analytic function of $\Delta_1$ on the
integrable line, we need
\begin{equation}
\Delta_1\sim|t|^{\frac12},
\tag{ce.5}
\end{equation}
and we find
\begin{equation}
\mu=\frac12+\frac1N,\qquad \phi=\frac12-\frac1N.
\tag{ce.6}
\end{equation}

The critical exponents are most conveniently expressed in terms of the scaling
(conformal) dimensions $x_j$ and $y_j$, indicating how a local density (or
order parameter) $m_j$ and its corresponding field scale with a typical length
scale $R$ in the problem,
\begin{equation}
m_j\sim R^{-x_j},\qquad h_j\sim R^{-y_j},\qquad x_j+y_j=2.
\tag{ce.1}
\end{equation}
In our model, the subscript $j$ takes the values
${\rm T}\hbox{ (thermal)},1,\ldots,N\!-\!1,\Delta$ (chiral), with
$h_{\rm T}\equiv t\equiv(T-T_{\rm c})/T_{\rm c}$, $m_j\equiv<\sigma^j>$ for
$j=1,\ldots,N\!-\!1$, and $h_{\Delta}\equiv\Delta_1$.

In terms of these, the usual critical exponents are
\begin{eqnarray}
&&\nu=\mu=\frac1{y_{\rm T}},\qquad \eta_j=2x_j,\qquad
2-\alpha=\frac2{y_{\rm T}},\qquad \psi_j=\frac{x_j}{y_{\rm T}}\cr&&\cr
&&\beta_j=\frac{x_j}{y_{\rm T}},\qquad
\gamma_{ij}=\frac{y_i-x_j}{y_{\rm T}}=\frac{2-x_i-x_j}{y_{\rm T}},\qquad
\delta_j=\frac{y_j}{x_j},
\tag{ce.2}
\end{eqnarray}
where we use $\psi_j$ for the gap exponents (or crossover exponents), as
the more conventional notation $\Delta$ would lead to confusion. We are
writing $\phi\equiv\psi_{\Delta}$ for the chiral crossover exponent
in (\ref{s.1}).

This leads to the scaling dimensions
\begin{eqnarray}
&&x_{\rm T}=\frac4{N+2},\qquad x_j=\frac{j(N-j)}{N(N+2)},\qquad
x_{\Delta}=\frac{N+6}{N+2},\cr&&\cr
&&y_{\rm T}=\frac{2N}{N+2},\qquad y_j=2-x_j,\qquad
y_{\Delta}=\frac{N-2}{N+2}=y_{\rm T}-1.
\tag{ce.7}
\end{eqnarray}
For $N=3$, we find $\phi=\frac16$, $x_{\Delta}=\frac95$, in agreement with
earlier predictions\pmyref{\ref{dN},\ref{SXEB}} For general $N$,
$y_{\Delta}=y_{\rm T}-1$ provides the chiral exponents of the
Fateev-Zamolodchikov model\pmyref{\ref{FZ}}

The most detailed results have been obtained for the superintegrable model
with weights given by
\begin{equation}
{W(n)\over W(0)}={(\!(1,\alpha)\!)_{0,n}\over(\!(1,\beta)\!)_{0,n}},
\qquad
{\wb(n)\over\wb(0)}={(\!(1,\beta/\omega)\!)_{0,n}\over
(\!(1,\alpha)\!)_{0,n}}.
\tag {2.31}
\end{equation}
For the corresponding hermitian quantum spin chain, Albertini et
al\pmyref{\ref{AlMP}} have calculated results for the excitation spectrum,
demonstrating the special\B role of level crossings, as the Perron-Frobenius
theorem does not apply to this case. Baxter\myref{\ref{Bax1}} has calculated
the bulk and surface free energies, horizontal\B and vertical interfacial
tensions and finite size corrections for the lattice shown in Fig.\ 6 with
free or fixed spin configurations on the left and right columns. He has found
that the critical exponents for these physical\B quantities are given by
\begin{eqnarray}
&{\alpha}=1-\fracd2N,\qquad&\qquad
{\alpha}_{\rm s}=2-\fracd2N,\cr
&\mu_{\rm hor}=\nu_{\rm hor}=\fracd2N,&\qquad\mu_{\rm ver}=
\nu_{\rm ver}=1,\cr
&&\hskip-2.5em\nu_{\rm hor}+\nu_{\rm ver}=2-{\alpha}.
\tag {2.32}
\end{eqnarray}

These are extremely interesting and puzzling results, deserving a great
deal of attention. Even though the weights are asymmetric, one should not
associate $\nu_{\rm ver}$ to be in the direction of the vertical weight,
because the lattice is oriented diagonally. The two different correlation
lengths are obtained from the decay of the finite size effects of such
diagonally oriented lattices. Since the diagonal lattice is symmetric under
$90^\circ$ rotation, one is forced to conclude the different boundary
conditions give rise to the two different exponents for correlation
lengths.

Yet it is remarkable that the bulk properties should be influenced by the
boundary, particularly as the partition functions in the superintegrable case
are real. In the most recent work of Baxter\cmyref{\ref{Bax2},\ref{Bax3}} he
imposes cyclic and skew boundary conditions on the two directions, finding
that $2\mu=2-{\alpha}=1+2/N$.\B Assuming $\mu=\nu$, then the usual scaling
relation again holds and, with such boundary conditions, we would have that
the vertical and horizontal\B correlation lengths have the same exponents
with $\nu_{\rm ver}=\nu_{\rm hor}=1/2+1/N$.

In the superintegrable case, the weights are complex. By a gauge\B transform,
we can make the weights real, but we cannot make them all\B positive.
Therefore the Perron-Frobenius theorem does not hold. This may be the only
plausible reason for the above strange behavior. Because the bulk properties
can be shown to be boundary dependent for non-positive weights --- complex
weights are typical within the integrable chiral Potts model --- the
combination of using $Z$-invariance properties and analytic continuation, as
used by Baxter, may or may not lead to valid results for the positive-real
Boltzmann weight case. For this reason, we find it  necessary to verify some
of the calculations of Baxter by low-temperature expansions.

For real and positive weights, such boundary-dependent behavior is not
conceivable. Furthermore, because the anisotropy $p=\bD_1/\Delta_1$ is an
irrelevant variable, the exponent $\nu$ cannot be a function of $p$ either,
therefore we must have $\nu_{\rm hor}=\nu_{\rm ver}=\nu_{\rm dia}=\nu$.
Baxter\myref{\ref{Bax2},\ref{Bax3}} found that $\mu_{\rm dia}=1/2+1/N$. Hence,
assuming the scaling relation $\mu=\nu$, one finds that the scaling relation
$2\nu=2-{\alpha}=1+2/N$ again holds.

\subsection{A New Model\label{sus.nmo}}

In \ref{sus.gra} we shall present plots, for $N=7$, of $1/K_2$ and $1/K_3$
versus $1/K_1$, and of $\Delta_2$ and $\Delta_3$ versus $\Delta_1$. The curves
look like straight lines. Plots for $N=5,6$ also give the same impression. Yet
an explicit calculation shows that the ratios $K_j/K_1$ and $\Delta_j/\Delta_1$
are almost but not quite constant. We therefore propose a new model with
constant ratios, given by
\begin{equation}
-\,\betaE{\cE(n)}=K\sum_{j=1}^{N-1}\,
{\sin(\pi/N)\over\sin(\pi j/N)}\,
\cos\left({2\pi\over N}\Bigl(nj+(N-2j)\Delta\Bigr)\right),
\tag {2.29}
\end{equation}
which has only two parameters. This may be good for further numerical
studies. For $\Delta=\quarter$ and $K=-E/k_{\rm B} T$, (\ref{2.29}) and
(\ref{2.20}) are identical. Hence at zero temperature, it goes through a
superwetting transition at $\Delta=\quarter$. In section \ref{sec.new},
we shall describe some of the symmetries of this model and show that for
$\Delta=\quarter(N+1)/(N-1)$ we have $\cE(0)=\cE(-1)$. Hence, its
ground state is highly degenerate, as it is in the chiral clock
model\smyref{\ref{SY},\ref{ANNNI}} thus this model, like the chiral clock
model, can also be used to describe commensurate-incommensurate transitions.

For $T>0$, we shall show that for a certain value of $\Delta$, this model is
not very different from the integrable chiral Potts model. We can use the
universality hypothesis to argue that this model in two dimensions, having
the same ${\bbZ}_N$ symmetry as the integrable chiral Potts model, must have
the same exponents. Hopefully, the exact results obtained for the chiral Potts
model can be used to gauge the accuracy of the numerical studies.

\setcounter{equation}{0}
\section{CONSISTENCY EQUATIONS AND INVERSE PROBLEM\label{sec.con}}

In this section we derive the consistency equations that the weights must
satisfy in order to be put in product forms as in (\ref{2.13}). We also
express the variables $\alpha$ and $\beta$ in terms of these weights.

\subsection{Consistency Conditions for General {\mybmi N}\label{sus.ccn}}

Let us introduce the notation
\begin{equation}
f(n)={W(n)W(N-1)\over W(n-1)W(0)}.
\tag {3.1}
\end{equation}
Then substituting (\ref{2.13}) and (\ref{2.14}) into it, we find that the
$\Delta(x)$ terms cancel out leaving
\begin{equation}
f(n)={(1-\alpha\omega^n)(1-\beta)\over
(1-\beta\omega^n)(1-\alpha)},\qquad1\le n\le N-1.
\tag {3.2}
\end{equation}
Now we use (\ref{n.21}) to rewrite
\begin{eqnarray}
&&\qquad\cot(\pi n/N)=-i\,{1+\omega^n\over1-\omega^n},\cr\cr
&&\cot\theta=-i\,{1+\alpha\over1-\alpha},\quad
\cot\phi=-i\,{1+\beta\over1-\beta}.
\tag{3.3}
\end{eqnarray}
Consequently, (\ref{3.2}) become $N-1$ linear equations for $\cot\theta$
and $\cot\phi$, i.e.
\begin{equation}
\cot\theta-f(n)\,\cot\phi=\Bigl(f(n)-1\Bigr)\cot(\pi n/N),\qquad1\le n\le N-1.
\tag {3.4}
\end{equation}
The determinant of any three of these equations must vanish in order to
have nontrivial solutions. This gives the consistency equations
\begin{equation}
\sin\left({(n-2)\pi\over N}\right)\,\Bigl(f(n)-1\Bigr)\Bigl(f(2)-f(1)\Bigr)=
\sin\left({n\pi\over N}\right)\,\Bigl(f(1)-1\Bigr)\Bigl(f(n)-f(2)\Bigr),
\tag {3.5}
\end{equation}
for $3\le n\le N-1$. Thus if the weights $W(n)$ and $\wb(n)$ satisfy these
$N-3$ consistency equations, then they can be put in the product form.

We note that (\ref{3.5}) is equivalent to the more general equation
\begin{equation}
{\Bigl({f(n_1)-f(n_2)}\Bigr)\Bigl({f(n_3)-f(n_4)}\Bigr)\over
\Bigl({f(n_1)-f(n_4)}\Bigr)\Bigl({f(n_2)-f(n_3)}\Bigr)}=
{\sin\Bigl(\fracd{\pi}N(n_1-n_2)\Bigr)
\sin\Bigl(\fracd{\pi}N(n_3-n_4)\Bigr)\over
\sin\Bigl(\fracd{\pi}N(n_1-n_4)\Bigr)
\sin\Bigl(\fracd{\pi}N(n_2-n_3)\Bigr)}
\tag {e.1}
\end{equation}
which is solved by any expression of the form
\begin{equation}
f(n)\to{x_1\omega^n+x_2\over x_3\omega^n+x_4}.
\tag {e.2}
\end{equation}

One can solve any two of the equations (\ref{3.5}), yielding
\begin{eqnarray}
&&\cot\theta={f(n_2)\Bigl(f(n_1)-1\Bigr)\cot(\pi n_1/N)
-f(n_1)\Bigl(f(n_2)-1\Bigr)\cot(\pi n_2/N)\over\Bigl(f(n_2)-f(n_1)\Bigr)},\cr
\cr
&&\cot\phi={\Bigl(f(n_1)-1\Bigr)\cot(\pi n_1/N)
-\Bigl(f(n_2)-1\Bigr)\cot(\pi n_2/N)\over\Bigl(f(n_2)-f(n_1)\Bigr)},
\tag {3.6}
\end{eqnarray}
which expresses $\theta$ and $\phi$ in terms of the weights $W$.

Using the integrability condition (\ref{e.23}) we find
\begin{equation}
\Omega\,\bO=1,
\tag {3.8}
\end{equation}
with
\begin{equation}
\Omega^{-1}=\sin(\pi/N)\Bigl(\cot(\phi-\theta)-\cot(\pi/N)\Bigr)
\tag {3.9}
\end{equation}
and a similar equation for $\bO$. Since we have expressed $\theta$
and $\phi$ in terms of the ratios of weights $f(n)$ and similarly
$\bT$ and $\bP$ in terms of the $\bF(n)$, we can substitute
(\ref{3.6}) into (\ref{3.9}); then (\ref{3.8}) gives a relation relating
the $W$ and $\wb$ weights.

For $N=3$, (\ref{3.8}) and (\ref{3.9}) are the same equations that we presented
elsewhere without derivation\cmyref{\ref{AP1}} except for a slight change of
notation, as $K$ and $\Delta$ in the earlier work are changed to $2K_1$ and
$\Delta_1$ here. These equations are useful to check whether a particular model
is integrable or not. Yet even in the $N=3$ case, we find it more convenient to
use the parameters $\theta$ and $\phi$ instead. Moreover, even though we can in
principle use the equations (\ref{3.5}) to calculate the $K_j$ and $\Delta_j$
for $j\ge2$ in terms of $K_1$ and $\Delta_1$ and (\ref{3.8}) and (\ref{3.9})
for the relations between the $K_j$ and $\Delta_j$ and the $\bK_j$ and $\bD_j$,
we have found out that these algebraic equations become more and more complex
to solve, as $N$ increases. In the next subsections, we shall present an
alternative way.

\subsection{Interaction Energy Parameters of Chiral Potts Model\label{sus.iep}}

We may rewrite (\ref{2.13}) in terms of the variables $\theta$ and $\phi$ as
\begin{equation}
{W(n)\over W(n-1)}=\left({\sin(N\phi)\over\sin(N\theta)}\right)^{1/N}
{\sin(\theta+\pi n/N)\over\sin(\phi+\pi n/N)}.
\tag {4.1}
\end{equation}
Together with (\ref{2.2}) and (\ref{2.5}), we find
\begin{equation}
-\,{\cE(n)-\cE(n-1)\over k_{\rm B}T}=
\log{W(n)\over W(n-1)}=A+B_n=\sum_{j=1}^{N-1}r_j\,\omega^{jn},
\tag {4.2}
\end{equation}
with
\begin{eqnarray}
&\displaystyle{A={1\over N}\log{\sin(N\phi)\over\sin(N\theta)}},\qquad
\displaystyle{B_n=\log{\sin(\theta+\pi n/N)\over\sin(\phi+\pi n/N)}},
\tag{4.3}\\
\cr&r_j=-\,\betaE{E_j}\,(1-\omega^{-j}).
\tag{4.4}
\end{eqnarray}
The Fourier coefficients $r_j$ in (\ref{4.2}) can now be written as
\begin{equation}
Nr_0=NA+\sum_{n=1}^NB_n=0,\qquad r_j=N^{-1}\sum_{n=1}^N\omega^{-nj}B_n.
\tag {4.5}
\end{equation}
Consequently, we find from (\ref{4.4}) and (\ref{2.11}) that
\begin{equation}
K_j\,\omega^{\Delta_j}={r_j\over1-\omega^{-j}}=
-\,{S_j+iC_j\over2N\sin({\pi j/N})},
\tag {4.6}
\end{equation}
where
\begin{equation}
S_j=\sum_{n=1}^N\,B_n\,\sin\biggl((2n\!-\!1)j{\pi\over N}\biggr),\qquad
C_j=\sum_{n=1}^N\,B_n\,\cos\biggl((2n\!-\!1)j{\pi\over N}\biggr).
\tag {4.7}
\end{equation}
Using (\ref{4.6}) we obtain
\begin{equation}
\Delta_j={N\over2\pi}\,\arctan{C_j\over S_j},\qquad
K_j={(S_j^2+C_j^2)^{1/2}\over2N\sin({\pi j/ N})}.
\tag {4.8}
\end{equation}
To have $\Delta_j=0$, we must choose $\theta$ and $\phi$ such that $C_j=0$.
{}From (\ref{4.7}) and (\ref{4.3}), we find if $\theta+\phi=-\pi/N$, then
$B_n=-B_{N-n+1}$ and $C_j,\Delta_j=0$.

\subsection{ Alternative Approach\label{sus.alt}}

Even though the results in the previous subsection suffice, for future
reference it may be useful to provide more explicit formulae using a
normalization first used by Baxter\pmyref{\ref{Bax1}}

If the average energy for a bond is to be zero, the natural normalization is
\begin{equation}
\prod_{n=0}^{N-1}\,W(n)=1.
\tag {e.3}
\end{equation}
We may then rewrite the $W(n)$ in (\ref{2.13}) in terms of the variables
$\theta$ and $\phi$ given in (\ref{3.3}) explicitly as
\begin{equation}
W(n)=\prod_{m=0}^{N-1}{f_m}\!^{\alpha_{m,n}},
\tag {e.4}
\end{equation}
where
\begin{equation}
f_m\equiv\displaystyle{{\sin(\theta+\fracd{m\pi}N)\over
\sin(\phi+\fracd{m\pi}N)}}=\exp(B_m),
\tag {e.5}
\end{equation}
\begin{equation}
\alpha_{m,n}\equiv\half\,\hbox{sign}(n+\half-m)-{(n+\half-m)\over N},
\tag {e.6}
\end{equation}
\begin{equation}
\log W(n)=\sum_{m=0}^{N-1}\,\alpha_{m,n}\,\log f_m\equiv-\betaE{\cE(n)},
\tag {e.7}
\end{equation}
also using the identity
\begin{equation}
\sin(N\theta)=2^{N-1}\,\prod_{n=0}^{N-1}\,
\sin\biggl(\theta+{n\pi\over N}\biggr).
\tag {e.19}
\end{equation}

It is easy to show, replacing $\omega$ by $x$ and then taking the limit
$x\to\omega$, that
\begin{equation}
\sum_{n=0}^{N-1}\,\alpha_{m,n}\,\omega^{np}=
\left\{\begin{array}{ll}
{\displaystyle0,\quad\hbox{if }p=0,}\cr\cr
{\displaystyle{\omega^{mp}\over{1-\omega^{p}}},\quad\hbox{if }p\ne0},
\end{array}\right.
\tag {e.8}
\end{equation}
and its inverse is
\begin{equation}
\sum_{p=1}^{N-1}\,{\omega^{-np}\over N}\,{\omega^{mp}\over{1-\omega^{p}}}
=\alpha_{m,n}.
\tag {e.9}
\end{equation}
We can then rewrite
\begin{equation}
\cE(n)=-k_{\rm B}T\,\log W(n)=
-{1\over\beta}\,\sum_{m=0}^{N-1}\,\alpha_{m,n}\,\log f_m
=\sum_{l=1}^{N-1}\,E_l\,\omega^{ln},
\tag {e.10}
\end{equation}
with
\begin{equation}
E_l={1\over N}\sum_{n=0}^{N-1}\,\cE(n)\,\omega^{-ln},\quad E_0\equiv0,
\tag {e.11}
\end{equation}
see also (\ref{2.2}), together with the inverse results
\begin{equation}
E_l=-\,{1\over\beta N}\,\sum_{m=0}^{N-1}\,{\omega^{-ml}\over{1-\omega^{-l}}}
\,\log f_m=-\,{\omega^{(2l-N)/4}\over2\beta N\sin\fracd{\pi l}N}\,\,
\sum_{m=0}^{N-1}\,\omega^{-ml}\,\log f_m,
\tag {e.12}
\end{equation}
\begin{equation}
E_{N-l}={E_l}^{\raise2pt\hbox{$\ast$}},\quad E_N\equiv E_0=0.
\tag {e.13}
\end{equation}
These results also agree with (\ref{4.6}) of the previous subsection.

\subsection{Superwetting in the Ground State\label{sus.swg}}

Let us say that the interface is superwet at $T=0$, if we can find a suitable
relabeling of the state differences $n$ such that
\begin{equation}
\cE(n)-\cE(0)=n\Bigl(\cE(1)-\cE(0)\Bigr)>0,\quad n=0,\ldots,N-1.
\tag {e.14}
\end{equation}
This is equivalent to the second-order difference equation
\begin{equation}
\cE(m+1)-2\cE(m)+\cE(m-1)=0,\quad m=1,\ldots,N-2.
\tag {e.15}
\end{equation}
Imposing the condition, see also (\ref{e.3}),
\begin{equation}
\sum_{n=0}^{N-1}\,\cE(n)=0,
\tag {e.16}
\end{equation}
we can solve these conditions, in terms of one constant $C$, by
\begin{equation}
\cE(n)=-2NC\,\alpha_{0,n}=-C(N-2n-1),\quad C>0,
\tag {e.18}
\end{equation}
together with its Fourier coefficients
\begin{equation}
E_l=\,-C\,{\omega^{(2l-N)/4}\over\sin(\pi l/N)}=
E_{N-l}^{\,\,\raise2pt\hbox{$\ast$}},\quad l=1,\ldots,N-1,
\tag {e.17}
\end{equation}
where we have used (\ref{e.6}) and (\ref{e.8}). We note that the results
(\ref{e.18}) and (\ref{e.17}) correspond also to the zero-temperature
super\-integrable\myref{\ref{AlMPT}} chiral Potts model and its corresponding
quantum chain\pmyref{\ref{vGR}}

\setcounter{equation}{0}
\section{INTEGRABLE THREE-STATE CHIRAL CLOCK MODEL\label{sec.thr}}
In this section, we take a closer look at the 3-state integrable
chiral clock models. For $N=3$, we use the chiral clock
representation (\ref{2.12}) to write the Boltzmann weights as
\begin{eqnarray}
&&W(n)=\exp\left[{2K_1\cos\left({2\pi\over3}(n+\Delta_1)\right)}\right],
\cr&&\cr
&&\wb(n)=\exp\left[{2\bK_1\cos\left({2\pi\over3}(n+\bD_1)\right)}\right].
\tag {oh.1}
\end{eqnarray}
Letting
\begin{equation}
w_n=W(-n)/W(0),\qquad \wo_n=\wb(-n)/\wb(0).
\tag{oh.2}
\end{equation}
and writing $K=2K_1$ and $\bK=2\bK_1$, we find from (\ref{oh.1}),
\begin{equation}
w_2/w_1=\exp[-\sqrt 3 K\sin (2\pi\Delta/3)],\quad
w_2 w_1=\exp[-3 K\cos(2\pi\Delta/3)],
\tag{oh.3}
\end{equation}
where we have dropped the subscript $1$ in $\Delta$. Consequently, we find
\begin{equation}
K=2K_1=\frac13\sqrt{\log^2(w_1w_2)+3\,\log^2(w_2/w_1)}
\tag{oh.4}
\end{equation}
and
\begin{equation}
\Delta=\Delta_1=\frac3{2\pi}\,\arctan\left({\sqrt3\log(w_2/w_1)\over
\log(w_1w_2)}\right).
\tag{oh.5}
\end{equation}
{}From (\ref{n.22}), we find
\begin{eqnarray}
&&w_1\equiv w_1(\theta,\phi)=
{\left[\frac{\sin^2(\phi)\sin(\theta+\frac13\pi)\sin(\theta+\frac23\pi)}
{\sin^2(\theta)\sin(\phi+\frac13\pi)\sin(\phi+\frac23\pi)}\right]}^{1/3},
\cr&&\cr
&&w_2\equiv w_2(\theta,\phi)=
{\left[\frac{\sin(\phi)\sin^2(\theta+\frac13\pi)\sin(\phi+\frac23\pi)}
{\sin(\theta)\sin^2(\phi+\frac13\pi)\sin(\theta+\frac23\pi)}\right]}^{1/3}.
\tag{oh.6}
\end{eqnarray}
Similar relations hold for $\bK$, $\bD$, $\wo_1$, and $\wo_2$ in terms of
$\bT$ and $\bP$.

\subsection{Ferromagnetic regions\label{sus.fer}}

We want to determine for which values of $\theta$ and $\phi$
the (relative) Boltzmann weights are positive real and in the ferromagnetic
regime. This means that we have
to look for a suitable fundamental domain as the relative Boltzmann
weights have a few simple symmetries: They are periodic modulo $\pi$ in
$\theta$ and $\phi$,
\begin{eqnarray}
&&w_1(\theta+\pi,\phi)=w_1(\theta,\phi+\pi)=w_1(\theta,\phi),\cr&&\cr
&&w_2(\theta+\pi,\phi)=w_2(\theta,\phi+\pi)=w_2(\theta,\phi),
\tag{oh.7}
\end{eqnarray}
they invert under the interchange of $\theta$ and $\phi$,
\begin{equation}
w_1(\phi,\theta)\,w_1(\theta,\phi)=1,\qquad
w_2(\phi,\theta)\,w_2(\theta,\phi)=1,
\tag{oh.8}
\end{equation}
and there is a transformation interchanging $w_1$ and $w_2$,
\begin{equation}
w_1(\fracd23\pi-\phi,\fracd23\pi-\theta)=w_2(\theta,\phi),\qquad
w_2(\fracd23\pi-\phi,\fracd23\pi-\theta)=w_1(\theta,\phi).
\tag{oh.9}
\end{equation}
After some work, one finds that modulo $\pi$ the ferromagnetic regimes
are given by
\begin{eqnarray}
0\le  w_2\le  w_1\le1\quad&\hbox{for}&\quad
-\fracd13\pi\le\theta\le\phi\le-\fracd13\pi-\theta,\cr&&\cr
&&\quad\hbox{or}\quad\pi-\theta\le\phi\le\theta\le\fracd23\pi,
\tag{oh.10}\\&&\cr
0\le  w_1\le  w_2\le1\quad&\hbox{for}&\quad
-\fracd13\pi-\phi\le\theta\le\phi\le0,\cr&&\cr
&&\quad\hbox{or}\quad0\le\phi\le\theta\le\fracd13\pi-\phi.
\tag{oh.10a}
\end{eqnarray}
Here, we can ignore the latter two choices (\ref{oh.10a}) as they reduce
to the first two choices (\ref{oh.10}) under the interchange transformation
(\ref{oh.9}).

For the integrable 3-state chiral Potts model, we also need to require
the integrability condition, see (\ref{e.23}),
\begin{equation}
\phi+\bP=\theta+\bT+\fracd13\pi,\quad(\hbox{modulo }\pi).
\tag{oh.11}
\end{equation}
Combining this with (\ref{oh.10}) above, we arrive at a unique fundamental
domain
\begin{equation}
-\fracd13\pi\le\theta\le\phi\le-\fracd13\pi-\theta,\qquad
-\fracd13\pi\le\bT\le\bP\le-\fracd13\pi-\bT,
\tag{oh.12}
\end{equation}
where\pagebreak
\begin{eqnarray}
&&0\le  w_2=w_2(\theta,\phi)\le  w_1=w_1(\theta,\phi)\le1,\cr&&\cr
&&0\le \wo_2=w_2(\bT,\bP)\le \wo_1=w_1(\bT,\bP)\le1.
\tag{oh.13}
\end{eqnarray}
Note that the integrability condition (\ref{oh.11}) is fully compatible with
the\B transformation (\ref{oh.9}), so that other orderings of $w_1$ and $w_2$,
and of $\wo_1$ and $\wo_2$, correspond to equivalent domains.

\subsection{Critical region\label{sus.crr}}

We now consider the case with
\begin{equation}
\bK=K,\qquad\bD=p\Delta.
\tag{oh.14}
\end{equation}
Substituting (\ref{oh.6}) into (\ref{oh.4}) and (\ref{oh.5}), and using
similar equations for the barred variables, one can rewrite (\ref{oh.14})
to give two conditions between the $\theta$, $\phi$, $\bT$, and $\bP$.
As we also have (\ref{oh.11}), we find there is only one free parameter left.

At the Potts critical point, we find
\begin{equation}
\bK_{\rm c}=K_{\rm c}=\fracd23\log(1+\sqrt 3)=1.49245929\cdots,
\qquad\bD=\Delta=0.
\tag{oh.15}
\end{equation}
{}From (\ref{oh.5}) and (\ref{oh.6}), we see that for $\bD=p\Delta=0$ to hold,
we need to have
\begin{equation}
w_1=w_2,\quad\wo_1=\wo_2,\qquad\hbox{or}\qquad
\theta+\phi=\fracd13\pi,\quad\bT+\bP=\fracd13\pi.
\tag{oh.16}
\end{equation}
Now we use (\ref{oh.4}) and (\ref{oh.16}) to obtain
\begin{equation}
-\log w_1=-\log \wo_1=\fracd32 K_{\rm c},\quad\hbox{or}\quad
\phi=\bP=-\fracd1{12}\pi,\quad\theta=\bT=-\fracd14\pi.
\tag{oh.17}
\end{equation}
Near the critical point, we can make the changes of variables
\begin{eqnarray}
&&\phi=-\fracd1{12}\pi-\delta\phi,\qquad\theta=-\fracd14\pi-\delta\theta,
\cr&&\cr
&&\bP=-\fracd1{12}\pi-\delta\bP,\qquad\bT=-\fracd14\pi-\delta\bT,\cr&&\cr
&&\qquad\delta\bT=\delta\phi+\delta\bP-\delta\theta.
\tag{oh.18}
\end{eqnarray}
Substituting these expansions into (\ref{oh.6}), and then using (\ref{oh.4}),
(\ref{oh.5}), and (\ref{oh.14}), we find\mysingle{\pagebreak}
\begin{eqnarray}
\delta\theta&=&\delta\phi+(p^2-1)\,r_{\rm c}\,\delta\phi^2+
(p^2-1)^2\,r_{\rm c}^2\,\delta\phi^3\cr&&\cr
&&+\fracd1{12}(p^2-1)\left[\left(15\,p^4-p^2(142+64\,\sqrt3)+15\right)
r_{\rm c}^3\right.\cr&&\cr
&&\qquad+\left.\left(p^2(250+122\,\sqrt3)-
174-78\,\sqrt3\right)r_{\rm c}^2\right.\cr&&\cr
&&\qquad+\left.\left(p^2(110+78\,\sqrt3)+
470+222\,\sqrt3\right)r_{\rm c}\right.\cr&&\cr
&&\qquad-\left.(p^2+1)(261+171\,\sqrt3)\right]\delta\phi^4
+{\rm O}(\delta\phi^5)
\tag{oh.19}
\end{eqnarray}
and
\begin{eqnarray}
&&\delta\bP-p\,\delta\phi-\fracd12(p+1)(\delta\theta-\delta\phi)
=\delta\bT-p\,\delta\phi-\fracd12(p-1)(\delta\theta-\delta\phi)\cr&&\cr
&&\qquad=-\fracd13p\,(p^2-1)\left((14+8\,\sqrt3)\,r_{\rm c}^2
-(53+25\,\sqrt3)\,r_{\rm c}+45+18\,\sqrt3\right)\cr&&\cr
&&\qquad\qquad\times\quad\delta\phi^3\,
\left(1+\fracd32(p^2-1)\,r_{\rm c}\,\delta\phi\right)+{\rm O}(\delta\phi^5),
\tag{oh.20}
\end{eqnarray}
where we have defined
\begin{equation}
r_{\rm c}=\fracd12(1+\sqrt3)\left(1+
\frac{7\,\sqrt3-12}{\log(1+\sqrt3)}\right)
=1.535044409\cdots,
\tag{oh.21}
\end{equation}
in order to simplify the first three orders in (\ref{oh.19}). One can also
easily verify from (\ref{oh.19}) and (\ref{oh.20}) that $\delta\bT$ expressed
as a function of $\delta\bP$ is of the form (\ref{oh.19}) with $p$ replaced by
$1/p$.

Substituting (\ref{oh.19}) and (\ref{oh.20}) back into (\ref{oh.6}) and using
\begin{equation}
K-K_{\rm c}=-\frac{K_{\rm c}\,t}{1+t}\approx-K_{\rm c}\,t,\quad\mbox{where}
\quad t\equiv\frac{T-T_{\rm c}}{T_{\rm c}},
\tag{oh.23}
\end{equation}
we find from (\ref{oh.4}) and (\ref{oh.5}) that
\begin{equation}
K_{\rm c}\,t=-\fracd13(p^2+1)(3+\sqrt3)\,r_{\rm c}\,\delta\phi^2\,
\left(1+(p^2-1)\,r_{\rm c}\,\delta\phi\right)+{\rm O}(\delta\phi^4),
\tag{oh.22}
\end{equation}
and
\begin{equation}
\Delta=\bD/p=\frac{3-\sqrt3}{\pi K_{\rm c}}\,
\delta\phi\,\left(1+\fracd12\,(p^2-1)\,r_{\rm c}\,\delta\phi\right)
+{\rm O}(\delta\phi^3).
\tag{oh.24}
\end{equation}
Eliminating $\delta\phi$ from the above equations we find, near the Potts
critical point on the integrable curve, where
$K=2K_1\approx K_{\rm c}\,(1-t)=\fract23\log(1+\sqrt3)(1-t)$, the result
\begin{eqnarray}
&&t=-C_t\,(1+p^2)\,\Delta^2+{\rm O}(t^2,t\Delta^2,\Delta^4),\cr&&\cr
&&\qquad C_t\equiv\fract1{18}\,\pi^2\,\Bigl(2+(7\sqrt3+12)K_{\rm c}\Bigr)=
9.959616982\cdots,
\tag {oh.25}
\end{eqnarray}
Since the modulus $k$ is given by (\ref{e.26}), we can use (\ref{oh.18}) to
(\ref{oh.24}) to express it in terms $t=T/T_{\rm c}-1$ as
\begin{equation}
k^2=-C_k\,t+{\rm O}(t^3),\quad C_k\equiv
\frac{54\,(2+\sqrt 3)\,K_{\rm c}^2}{2+(7\sqrt 3+12)K_{\rm c}}=
4.981050242\cdots,
\tag{oh.26}
\end{equation}
which is independent of $p$ at this order.

The above expansions for the critical region include the symmetric case for
$p=1$ and the Ostlund-Huse model for $p=0$. We do not have such universal
formulae in the low-temperature regime, which we shall discuss next.

\subsection {Low-Temperature Limit\label{ltl}}

Considering the three-state chiral clock model with $\bK=K$ and $\bD=p\Delta$
in the low-temperature limit, we find that the limits $T\to0$ and $p\to0$
(or $p\to1$) do not commute. The limit $p\to0$ does not reproduce the
Ostlund-Huse model ($p\equiv0$) results, nor does the limit $p\to1$
reproduce the symmetric model ($p\equiv1$) results. We shall have to
consider several regimes separately.

First, for $p\ne0$ and $p\ne1$, we let
\begin{equation}
\phi=-\frac13\pi+\delta\phi,\quad\theta=-\frac13\pi+\delta\theta,\quad
\bP=-\delta\bP,\quad\bT=-\frac13\pi+\delta\bT,
\tag{oh.27}
\end{equation}
and we find the leading low-temperature expansion results
\begin{eqnarray}
\delta\theta\approx\delta\phi\,\delta\psi\equiv
\delta\phi\,{\left(\frac2{\sqrt3}\,\delta\phi\right)}^{\sqrt{b^2+b+1}},
\tag{oh.28}\\
\delta\bP\approx\delta\phi,\qquad
\delta\bT\approx\delta\phi\,{\left(\frac2{\sqrt3}\,\delta\phi\right)}^{b-1},
\tag{oh.29}
\end{eqnarray}
with
\begin{equation}
b
=\frac{\sin(\frac16(1+p)\pi)}{\sin(\frac16(1-p)\pi)},\qquad
\sqrt{b^2+b+1}=\frac{\frac12\sqrt3}{\sin(\frac16(1-p)\pi)}.
\tag{oh.30}
\end{equation}
Then we see from (\ref{oh.4}) and (\ref{oh.6}) that
\begin{equation}
K=\bK\approx-{2\over{3\sqrt3}}\log\delta\psi\to\infty,
\hbox{ as }\delta\phi\to0,
\tag{oh.31}
\end{equation}
whereas from (\ref{oh.5}) and (\ref{oh.6}) we find that $\bD=p\Delta$ and
\begin{equation}
\Delta\approx\frac14+\frac{3\,\delta\phi}{2\,\pi\,\log\delta\psi}
\approx\frac14-\frac1{2\,\pi\,K}
\exp\left[-3\,K\sin\left(\fracd16(1-p)\pi\right)\right].
\tag{oh.32}
\end{equation}
Further terms to the above asymptotic expansions follow by expanding
$\delta\theta$, $\delta\bP$, and $\delta\bT$ in powers of $\delta\phi$,
$\delta\psi\sim\delta\phi^{\sqrt{b^2+b+1}}$, $\delta\phi^{b-1}$, and
$1/\log\delta\psi$. This can be done for $1<b<\infty$, or $0<p<1$. For
$b\to1$ or $p\to0$, $\delta\phi^{b-1}\to1$ and is no longer small, so that
different asymptotic expansions arise. For $b\to\infty$ or $p\to1$, we
still have $\phi\to\fracd13\pi$ and $\bP\to0$, so that there is still
another regime at $p\approx1$ between the above regime and the symmetric
case $p\equiv1$.

For the three-state Ostlund-Huse case we have $p\equiv\bD\equiv0$, so that
\begin{equation}
\bT+\bP=-\fracd13\pi,\qquad\frac{\sin(3\bT)}{\sin(3\bP)}=1.
\tag{oh.33}
\end{equation}
Using also (\ref{oh.12}), we can then solve $\bT$ and $\bP$ in terms of
$\theta$ and $\phi$, namely
\begin{equation}
\bT=-\fracd13\pi-\half(\theta-\phi),\quad \bP=\half(\theta-\phi).
\tag{oh.34}
\end{equation}
In the low-temperature limit, with $\phi$ and $\theta$ near $-\fracd13\pi$,
\begin{equation}
\phi=-\fracd13\pi+\delta\phi,\qquad\theta=-\fracd13\pi+\delta\theta,
\tag{oh.35}
\end{equation}
we can find in a systematic way, from $K=\bK$, the asymptotic expansion for
$\delta\theta$ in terms of powers of $\delta\phi$, $\delta\phi^{\sqrt3}$,
and $1/\log\delta\phi$, in agreement with what one would expect from
(\ref{oh.29}) and (\ref{oh.30}) with $b-1=0$. The asymptotic expansion so
obtained can be equivalently (and more economically) expressed
as\pagebreak
\begin{eqnarray}
\log\delta\theta&=&\log(\delta\phi\,\delta\tilde\psi)
+\frac12\delta\phi-\sqrt3\,\delta\tilde\psi\cr&&\cr
&&+\left(\frac{4+\sqrt3}8-\frac1{2\,\lps}\right)\delta\phi^2
-\,\frac{1+\sqrt3}2\,\delta\phi\,\delta\tilde\psi
+\,\frac{6-\sqrt3}2\,\delta\tilde\psi^2\cr&&\cr
&&+\left(\frac1{18}+\frac1{4\,\lps^2}\right)\delta\phi^3-
\left(\frac{5+7\,\sqrt3}8-\frac{2+\sqrt3}{2\,\lps}
+\frac{\sqrt3}{2\,\lps^2}\right)\delta\phi^2\delta\tilde\psi\cr&&\cr
&&+\,\frac{6+\sqrt3}2\,\delta\phi\,\delta\tilde\psi^2
+\,\frac{27-29\,\sqrt3}6\,\delta\tilde\psi^3\,+\cdots\,,
\tag{oh.36}
\end{eqnarray}
where
\begin{equation}
\lps\equiv\log\,\delta\tilde\psi=\sqrt3\,\log\,(\delta\phi/\sqrt3),\qquad
\delta\tilde\psi\equiv\left({{\delta\phi}/{\sqrt3}\,}\right)^{\sqrt3}.
\tag{oh.37}
\end{equation}
Therefore the leading order is
\begin{equation}
\delta\theta\approx\delta\phi\,\delta\tilde\psi\sim\delta\phi^{1+\sqrt3},
\tag{oh.38}
\end{equation}
which is quoted below (\ref{bb.3}) in subsection \ref{sus.ltr}.
Now (\ref{oh.4}) to (\ref{oh.6}) can be used to give
\begin{equation}
K=-\fracd29\sqrt3\log\delta\tilde\psi+\cdots,\qquad
\Delta=\frac14-\frac1{K\pi }e^{-3K/2}+\cdots,
\tag{oh.39}
\end{equation}
which was plotted as line 3 in Fig.\ 7b.

For the symmetric case with $p\equiv1$, the low-temperature limit is given by
\begin{equation}
\theta\equiv\bT=-\fracd13\pi+\delta\theta,\qquad
\phi\equiv\bP=\theta+\fracd16\pi=-\fracd16\pi+\delta\theta,\qquad
\tag{oh.40}
\end{equation}
with $\delta\theta\to0$. We find from (\ref{oh.4}) to (\ref{oh.6}), the
asymptotic expansion
\begin{equation}
K=-\fracd29\sqrt3\log\delta\theta+\cdots,\qquad
\Delta=\frac14-\frac{\log2}{2\,\pi\,K}+\cdots,
\tag{oh.41}
\end{equation}
which was plotted as line 1 in Fig.\ 7b. Thus $\Delta$ decreases from
$\fracd14$ linearly as $T$ increases from zero, which is a behavior very
different from that shown in (\ref{oh.41}) for the Ostlund-Huse case, or from
(\ref{oh.32}).

The symmetric case can be extended by replacing the products in (\ref{b.12}),
in leading order, by their limiting values and assuming that the front
factors, which are powers of $1/\Lambda$ and $1/\bL$, are small. The weights
then take on the form (\ref{b.17}), which was also considered by Baxter.
We shall call this case the ``quasi-symmetric" case. More precisely, with
\begin{equation}
\theta=-\fracd13\pi+\delta\theta,\quad\bT=-\fracd13\pi+\delta\bT,\quad
\phi\to\phi_0\ne0\hbox{ or }-\fracd13\pi,
\tag{oh.42}
\end{equation}
we find from (\ref{oh.12}) that
\begin{equation}
\bP=-\fracd13\pi-\phi+\delta\theta+\delta\bT\to-\fracd13\pi-\phi_0.
\tag{oh.43}
\end{equation}
Now, if we let
\begin{equation}
g\equiv\frac{\sin^2(\fracd13\pi-\phi_0)}{\sin(\fracd13\pi)
\sin(\fracd13\pi+\phi_0)\sin(-\phi_0)},
\tag{oh.44}
\end{equation}
and
\begin{equation}
s\equiv-\log\frac{\sin(-\phi_0)}{\sin(\fracd13\pi-\phi_0)},\quad
\bS\equiv-\log\frac{\sin(\fracd13\pi+\phi_0)}{\sin(\fracd13\pi-\phi_0)},
\tag{oh.45}
\end{equation}
then we find from (\ref{oh.6}) for $\delta\theta$ and $\delta\bT$ small, the
simple formulae
\begin{equation}
w_n\approx e^{-s}\,{(g\,\delta\theta\vphantom{\bT})}^{n/3},\quad
\wo_n\approx e^{-\bS}\,{(g\,\delta\bT)}^{n/3},\quad\hbox{for }n=1\hbox{ or }2.
\tag{oh.46}
\end{equation}
Here we have ignored terms that are one order higher in $\delta\theta$ or
$\delta\bT$. If we substitute (\ref{oh.46}) in (\ref{oh.4}) and
(\ref{oh.5}), we find
\begin{eqnarray}
&{\displaystyle K\approx\fracd13\,\sqrt{(u+2\,s)^2+\fracd13u^2},}\quad
&{\displaystyle\bK\approx\fracd13\,\sqrt{(\bU+2\,\bS)^2+\fracd13\bU^2}},
\cr&&\tag{oh.47}\\
&{\displaystyle\Delta\approx\frac3{2\pi}\,
\arctan\frac{\tan(\fracd16\pi)\,u}{u+2\,s}},\quad
&{\displaystyle\bD\approx\frac3{2\pi}\,
\arctan\frac{\tan(\fracd16\pi)\,\bU}{\bU+2\,\bS}},
\tag{oh.48}
\end{eqnarray}
where
\begin{equation}
u\equiv-\log(g\,\delta\theta),\qquad\bU\equiv-\log(g\,\delta\bT).
\tag{oh.49}
\end{equation}
Formulae (\ref{oh.47}) and (\ref{oh.48}) are correct up to exponentially
small corrections in the temperature $T$, as we have ignored terms of order
$\delta\theta$ and $\delta\bT$. Therefore,
\begin{equation}
u=\fracd12\,\sqrt{27K^2-3s^2}-\fracd32s
+\hbox{exponentially small},
\tag{oh.50}
\end{equation}
and a similar equation with $u$, $K$, and $s$ replaced by $\bU$, $\bK$, and
$\bS$. Expanding these to a few orders in $1/K$ and $1/\bK$, we find
\begin{equation}
u=\fracd32\sqrt3\,K-\fracd32\,s-\fracd1{12}\sqrt3\,
\frac{s^2}K+{\rm O}\left(\frac{s^4}{{K\vphantom{\bK}}^3}\right),
\tag{oh.51}
\end{equation}
and a similar equation with bars for $\bU$. Substituting (\ref{oh.51}) into
(\ref{oh.48}), we find\mydouble{\pagebreak}
\begin{eqnarray}
&&\displaystyle{\Delta=\frac14-\frac s{2\pi K}-
\frac{s^3}{108\pi{K\vphantom{\bK}}^3}+
{\rm O}\left(\frac{s^5}{{K\vphantom{\bK}}^5}\right)},\cr&&\cr
&&\displaystyle{\bD=\frac14-\frac{\bS}{2\pi\bK}-
\frac{{\bS\,}^3}{108\pi{\bK\,}^3}+
{\rm O}\left(\frac{{\bS\,}^5}{{\bK\,}^5}\right)}.
\tag{oh.52}
\end{eqnarray}
These again exhibit linear behavior in $T$, as in the symmetric case. In fact,
for $\phi=\bP=\fracd16\pi$ and $\theta=\bT$, we have $s=\bS=\log2$, so
(\ref{oh.52}) reduces smoothly to (\ref{oh.41}).

On the other hand, when $\phi\to-\fracd13\pi$, $\bP\to0$, we have $s\to0$ and
$\bS\to\infty$. Then $\Delta$ tends to $\fracd14$ extremely fast, whereas
$\bD$ tends to $\fracd14$ very slowly. In the limit, there is a crossover from
the linear behavior of (\ref{oh.52}) to the exponential behavior of the first
low-temperature regime described by (\ref{oh.27}) to (\ref{oh.32}).

Finally, solving $\phi_0$ from $\bS=\gamma\,s$ and $\bT$ from $\bU=\gamma\,u$,
we find from (\ref{oh.47}) and (\ref{oh.48}) that $\bK=\gamma\,K$ and
$\bD=\Delta$ to all algebraic orders in $T$. So the quasi-symmetric case is
closely related to integrable 3-state chiral clock\B models with symmetric
chiral field, $\bD\equiv\Delta$, and we can use this as a\B definition,
suitable also outside the low-temperature regime.

\setcounter{equation}{0}
\section{SYMMETRIC {\mybmi N}-STATE CHIRAL POTTS MODEL\label{sec.sym}}

In this section we study the symmetric case for general $N$ and
give graphs for the chiral clock model parameters.

\subsection{Symmetric Lattice\label{sus.sym}}

For the symmetric case with $W=\wb$, the integrability condition (\ref{2.18})
or (\ref{e.23}) becomes
\begin{equation}
\phi=\theta+{\pi\over2N}.
\tag {4.9}
\end{equation}
Substituting this equation into (\ref{4.3}), we find from (\ref{4.7}) and
(\ref{4.8}) that $K_j$ and $\Delta_j$ depend on only one parameter $\theta$.
We only need to study the regime (fundamental domain) with
\begin{equation}
-{\pi\over N}\le\theta\le-{3\pi\over4N}.
\tag {e.21}
\end{equation}
First, we can see from (\ref{4.3}) that if we let $\theta\to\theta+\pi/N$,
then $B_n\to B_{n+1}$. Consequently, combining the first part of (\ref{4.6})
and the second equation of (\ref{4.5}), we find that $K_j\,\omega^{\Delta_j}
\to\omega^j K_j\,\omega^{\Delta_j}$. This means that by shifting the domain to
$0\le\theta\le\quarter\pi/N$, the amplitudes $K_j$ remain unchanged but
$\Delta_j\to\Delta_j+j$. Secondly, we note the reflection symmetry
$\theta\to-\theta-\half\theta/N$, changing $B_n\to-B_{-n}$. Thus we
can shift the domain to $-\threequarter\pi/N\ge\theta\ge-\half\pi/N$, with
the amplitudes $K_j$ remaining unchanged but now $\Delta_j\to-\Delta_j$.
Finally, for $-\half\pi/N\le\theta\le0$ (modulo $\pi/N$) the Boltzmann
weights are not all positive, as can be seen easily from (\ref{4.1}) or
(\ref{e.4}), with condition (\ref{4.9}).

For $\theta\!\downarrow\!-\pi/N$, we can see from (\ref{4.3}) that
$B_1\to-\infty$, hence $K_j\to\infty$ and temperature $T\to 0$. {}From
(\ref{4.5}) and (\ref{4.6}) we find
\begin{equation}
K_j\,\omega^{\Delta_j}\to {B_1\over N(\omega^j-1)}.
\tag {4.10}
\end{equation}
Since the ratios are finite, we obtain the zero-temperature results
\begin{equation}
\omega^{2\Delta_j}=-\omega^{-j},\qquad\Delta_j=\quarter(N-2j),
\qquad{K_j\over K_1}={\sin(\pi/N)\over\sin(\pi j/N)}.
\tag {4.11}
\end{equation}
Comparing with (\ref{e.17}) we see that we are at a superwetting point.

For the other end of the fundamental domain,
\begin{equation}
\theta=\theta_{\rm FZ}\equiv-\frac{3\pi}{4N},
\tag {4.12}
\end{equation}
we find using (\ref{4.3}) that $B_n=-B_{N-n+1}$. Hence, we conclude from
(\ref{4.7}) that $C_j=0$, or $\Delta_j=0$. In fact, by putting
$\theta=\theta_{\rm FZ}-\lambda$, it is easy to show that
$B_n(\lambda)\to-B_{N-n+1}(-\lambda)$ for $\lambda\to-\lambda$. Consequently,
from (\ref{4.7}) we find that the $S_j$ are unchanged but $C_j \to-C_j$, as
$\lambda\to-\lambda$. This means that the $K_j$ are unchanged, but the
$\Delta_j$ flip signs as $\lambda$ changes its sign.

To summarize, we find, as we increase $\theta$ from $-\pi/N$, that the
integrable line plotted as a function of $1/K_j$ versus $\Delta_j$ starts
from the zero-temperature values given in (\ref{4.11}) and the $\Delta_j$
decrease and the $1/K_j$ increase and for $\theta=\theta_{\rm FZ}$, we have
$\Delta_j=0$ and the $1/K_j$ reach their maximum values,
\begin{equation}
K_{j\rm c}={1\over N\sin({\pi j/ N})}\,\sum_{n=1}^{[\halfs N]}\,
\sin\biggl((2n\!-\!1){j\pi\over N}\biggr)
\log{\sin\Bigl((n-\frac14)\pi/N\Bigr)\over
\sin\Bigl((n-\frac34)\pi/N\Bigr)}.
\tag {4.13}
\end{equation}
This corresponds to the symmetric case of the Fateev-Zamolodchikov self-dual
solution, and is therefore critical. Needless to say that (\ref{4.13}) gives
the critical temperature for the symmetric solvable chiral Potts model.
As $\theta$ increases further the values of $\Delta_j$ are now negative. The
curves are symmetric with respect to the vertical axis.

\subsection{Graphs for Integrable Symmetric Case with
{\mybmi N}\mysp=\mysp4,\mysp.\mysp.\mysp.\mysp,\mysp7\label{sus.gra}}

\myfH

\myfI

For $N=4$, using (\ref{2.12}), we have
\begin{equation}
-\,\betaE{\cE(n)}=2K_1\cos\left({2\pi\over N}(jn+\Delta_1)\right)
+K_2\,(-1)^n.
\tag {4.14}
\end{equation}
In Fig.\ 8a we plot $1/K_1$ and $1/K_2$ versus $\Delta_1$ and
in Fig.\ 8b we plot $1/K_2$ versus $1/K_1$, which is almost a straight line
with slope approximately given by $\sqrt2\sim1.41$. At temperature $T=0$, we
have from (\ref{4.11}) that $\Delta_1=\half$ and $K_2/K_1=\half\sqrt2$.
Therefore, if we let $k_{\rm B}TK_1=1$ in the limit $T\to0$ by a suitable
choice of units, then the energies (\ref{4.14}) for the different states are
\begin{eqnarray}
&\cE(0)=-\fract32\sqrt2,\qquad&\cE(1)=\fract32\sqrt2,\cr
&\cE(2)=\half\sqrt2,\phantom{-}\qquad&\cE(3)=-\half\sqrt2.
\tag {4.15}
\end{eqnarray}
Hence, $\cE(-n)-\cE(0)=n\sqrt2$. This means that at zero temperature
the integrable line is at a superwetting transition. As $T$ or
$1/K_j$ increases, $\Delta_1$ decreases. When $\Delta_1=0$, we find
from (\ref{4.13}) that
\begin{equation}
K_{\rm1c}=0.3029227993,\quad K_{\rm2c}=0.2203433968,\quad
K_{\rm2c}/K_{\rm1c}=0.7273912604.
\tag {4.16}
\end{equation}
Comparing with the ratio $K_2/K_1=\half\sqrt2$ at zero temperature, we know
that the curve in Fig.\ 8b is not a straight line. We may magnify the effect
by plotting in Fig.\ 8c the ratio $K_2/K_1$ versus $\Delta_1$ and indeed we
find it not to be constant, with $\kappa_{jc}\lsim0.02$ for $\kappa_j$ defined
in (\ref{2.21}).

\myfJ

For different $N$, we find more or less the same situation as for $N=4$.
For $N$ odd, there are equal numbers of $K_j$ and $\Delta_j$ with
$1\le j\le\half(N-1)$. At $T=0$, we find then from (\ref{4.11}) that
$\Delta_{\halfs(N-1)}=\quarter$ is the smallest of the $\Delta_j$. For even
$N$, there are $\half N$ different amplitudes $K_j$ and $\half N-1$ angles
$\Delta_j$, not counting $\Delta_{\halfs N}\equiv0$. Now at $T=0$,
$\Delta_{\halfs N-1}=\half$ is the smallest.

\myfK

\myfP

\myfS

Again without loss of generality we can let $k_{\rm B}TK_1=1$. Then at $T=0$,
we have
\begin{equation}
\cE(n)=\sum_{j=1}^{N-1}\,{\sin(\pi/N)\over\sin(\pi j/N)}\,
\sin\left((2n-1){\pi j\over N}\right).
\tag {4.17}
\end{equation}
{}From this it is straightforward to show that
\begin{equation}
\cE(-n)-\cE(0)=2n\sin(\pi/N).
\tag {4.18}
\end{equation}
It also follows by comparing (\ref{4.18}) with (\ref{e.18}) and (\ref{4.17})
with (\ref{e.17}). Hence, it is at the endpoint of a superwetting line. In
Fig.\ 9a, (respectively 10a, or 11a), we plot $1/K_j$ with
$1\le j\le [\half N]$ versus the smallest angle $\Delta_{[\halfs(N-1)]}$ for
$N=5$, (6, or 7). In Fig.\ 9b, (10b, or 11b), we plot
$1/K_j$ with $2\le j\le [\half N]$ versus $1/K_1$ for $N=5$, (6, or 7). Again
we find that the curves look very much like straight lines.

\myfL

\myfQ

\myfT

For $\theta=\theta_{\rm FZ}$, we have $\Delta_j=0$, and the $K_j$ given by
(\ref{4.13}) have the numerical values
\begin{eqnarray}
N=5:&\hskip-1em\quad K_{\rm1c}=0.2878960239,&\hskip-1em\quad K_{\rm2c}=
0.1845136297,\cr
N=6:&\hskip-1em\quad K_{\rm1c}=0.2793167217,&\hskip-1em\quad
K_{\rm2c}=0.1675087565,\hskip-.3em\quad K_{\rm3c}=0.1468955978,\cr
N=7:&\hskip-1em\quad K_{\rm1c}=0.2738160609,&\hskip-1em\quad
K_{\rm2c}=0.1577390412,\hskip-.3em\quad K_{\rm3c}=0.1288523620,\cr
&&
\tag {4.19}
\end{eqnarray}
which are the critical-temperature values. The ratios are given
by\mysingle{\pagebreak}
\begin{eqnarray}
N=5:&\hskip-1em\quad K_{\rm2c}/K_{\rm1c}=0.6409037097,\cr
N=6:&\hskip-1em\quad K_{\rm2c}/K_{\rm1c}=0.5997090164,&\hskip-1em\quad
K_{\rm3c}/K_{\rm1c}=0.5259105038,\cr
N=7:&\hskip-1em\quad K_{\rm2c}/K_{\rm1c}=0.5760766578,&\hskip-1em\quad
K_{\rm3c}/K_{\rm1c}=0.4705800001.
\tag {4.20}
\end{eqnarray}
Comparing these numbers with the ratios at zero temperature,

\begin{eqnarray}
N=5:&\hskip-1em\quad K_2/K_1=0.6180339887,\cr
N=6:&\hskip-1em\quad K_2/K_1=0.5773502692,&\hskip-1em\quad K_3/K_1=0.5,\cr
N=7:&\hskip-1em\quad
K_2/K_1=0.5549581321,&\hskip-1em\quad K_3/K_1=0.4450418679,
\tag {4.21}
\end{eqnarray}
we find that the curves are not really straight lines.

\myfM

\myfR

\myfU

In Fig.\ 9c, (10c, or 11c), the $\Delta_j$ for $N=5$, (6, or 7) with
$2\le j\le [\half N]$ are plotted versus $\Delta_1$. These curves
also look like straight lines. In Fig.\ 9d we have plotted the ratio $K_2/K_1$
versus $\Delta_2$, while in Fig.\ 9e, we have plotted the ratio
$\Delta_1/\Delta_2$ versus $\Delta_2$, both for $N=5$. Again we find the
ratios not to be exactly constant, with $\kappa_{j},\delta_{j}\lsim0.02$.

\myfN

\myfO

\setcounter{equation}{0}
\section{LOW {\mybmi T}\ BETHE ANSATZ FOR DIAGONAL INTERFACES\label{sec.bal}}

In this section we shall present our Bethe Ansatz calculations with free
boundary conditions\footnote{Choosing fixed boundary conditions for the spins
in the first and last rows, corresponds to free boundary conditions for the
domain walls which live on the dual lattice, which is equivalent to demanding
that a domain wall can not move outside the system as is expressed in
(\ref{b.28}) and (\ref{b.29}) below.} for interfacial tensions of diagonal
interfaces in the low-temperature limit, using an extension of the
Bethe Ansatz\myref{\ref{Bethe}} similar to the ones introduced by
Gaudin\pmyref{\ref{Gaudin}}. These calculations also confirm our dilogarithmic
integral conjecture (\ref{b.5}) for the entropic corrections (linear in the
temperature $T$) to the ground-state interfacial tension results.

Using Baxter's notation\cmyref{\ref{Bax2},\ref{Bax3}} we start with the
two coupled eigenvalue equations
\begin{equation}
{\bf T}_q\,{\bf\cdot}\,{\bf y}=T_q\,{\bf x},\quad
{\bf{\hat T}}_q\,{\bf\cdot}\,{\bf x}={\hat T}_q\,{\bf y},
\tag{b.21}
\end{equation}
where ${\bf T}_q$ and ${\bf{\hat T}}_q$ are the two different diagonal
transfer matrices. We solve this pair of equations, by
assuming\mydouble{\pagebreak}
\begin{eqnarray}
&&x_{n_1,\ldots,n_r}=g(n_1,\ldots,n_r|\zeta_1,\ldots,\zeta_r),\cr&&\cr
&&\hspace{3cm}1\le n_1\le n_2\le\cdots\le n_r\le L,\cr&&\cr
&&y_{n_1,\ldots,n_r}=
g(n_1+\half,\ldots,n_r+\half|\zeta_1,\ldots,\zeta_r),\cr&&\cr
&&\hspace{3cm}1\le n_1\le n_2\le\cdots\le n_r\le L-1,
\tag{b.22}
\end{eqnarray}
with a common function $g$ which is a linear combination of exponentials.
Here $n_1,\ldots,n_r$ stand for the positions of the interfaces in a
given diagonal. We exclude overhangs, as in the M\"uller-Hartmann--Zittartz
approximation\cmyref{\ref{SP}} which is asymptotically exact at low
temperatures. If a sequence of, say, $m$ positions coincide, or
$n_j=\ldots=n_{j+m-1}$, then the interfaces merge to a single $m$-step
interface. The $\zeta_1,\ldots,\zeta_r$ are complex parameters, also often
written as $\zeta_j=\exp(ik_j)$. In terms of these, the eigenvalues are
\begin{equation}
T_q={\hat T}_q={(w_1\wo_1)}^{r/2}\,
\prod_{j=1}^r\,(\zeta_j^{-1/2}+\zeta_j^{1/2}).
\tag{b.23}
\end{equation}
{}From these, we can calculate the interfacial tension per two bonds via
\begin{equation}
\epsilon_r=-k_{\rm B}T\,\log(T_q{\hat T}_q),
\tag{b.24}
\end{equation}
in agreement with Baxter's normalization.

For the function $g$ we choose the Bethe Ansatz form
\begin{eqnarray}
&&g(m_1,\ldots,m_r|\zeta_1,\ldots,\zeta_r)\cr&&\cr
&&\qquad={\cal N}\,
\sum_{\varepsilon_1=\pm1}\cdots\sum_{\varepsilon_r=\pm1}\,
\prod_{j=1}^r\,{(-\zeta_j)}^{-\varepsilon_j/2}
\,\mathrel{\mathop{\prod\prod}_{1\le j<k\le r}}\,
{A(\zeta_j^{-\varepsilon_j},%
\zeta_k^{\varepsilon_{k\vphantom{j}}})}^{1/2}\cr&&\cr
&&\qquad\qquad\times\sum_{P\in S_r}\,
\mathrel{\mathop{\prod\prod}_{1\le j<k\le r}}\,
{A(\zeta_{Pj}^{\varepsilon_{Pj}},%
\zeta_{Pk}^{\varepsilon_{Pk\vphantom{j}}})}^{1/2}
\,\prod_{j=1}^r\,\zeta_{Pj}^{\varepsilon_{Pj}m_j},
\tag{b.25}
\end{eqnarray}
with the usual two-body scattering function
\begin{equation}
A(\zeta_1,\zeta_2)\equiv{1-2\Delta_{\rm6v}\,\zeta_2+\zeta_1\zeta_2\over
1-2\Delta_{\rm6v}\,\zeta_1+\zeta_1\zeta_2},
\quad2\Delta_{\rm6v}\equiv2\cos\lambda\equiv
{\Bigl({w_2\over w_1^2}-1\Bigr)}^{-1}.
\tag{b.26}
\end{equation}
This form guarantees that the ``two-interface scattering" is consistent, or
\begin{eqnarray}
&&(w_2-w_1^2)\,g(n_j\!\to\!n,n_{j+1}\!\to\! n)
\,-\,w_1\wo_1\,g(n_j\!\to\!n+1,n_{j+1}\!\to\!n) {\strut}_{\strut}\cr
&&\qquad\qquad+\,(\wo_2-\wo_1^2)\,g(n_j\!\to\!n+1,n_{j+1}\!\to\!n+1)=0,
{\strut}_{\strut}\cr
&&\hspace*{4.5cm}\hbox{for }j=1,\ldots r-1.
\tag{b.27}
\end{eqnarray}
One can next verify that three-interface and higher scattering processes
are consistent because of the special form (\ref{b.17}), which relates
to weights of fusion models generated from the six-vertex
model\pmyref{\ref{KRS}} This is also true for the case with $\Lambda\ne\bL$,
for which (\ref{b.25}) need be amended with trivial $\bL/\Lambda$ powers
which break the left-right symmetry.

In (\ref{b.25}), the last line corresponds to the usual
Bethe Ansatz\cmyref{\ref{Bethe}} whereas the summations over the
$\varepsilon$'s correspond to all waves reflected at the two boundaries.
The coefficients are chosen such that the first ``left" boundary condition,
\begin{equation}
g(m_1\!=\!\half,\ldots,m_r|\zeta_1,\ldots,\zeta_r)=0,
\tag{b.28}
\end{equation}
is automatically satisfied.

In order to solve the other ``right" boundary condition
\begin{equation}
g(m_1,\ldots,m_r\!=\!L\!+\!\half|\zeta_1,\ldots,\zeta_r)=0,
\tag{b.29}
\end{equation}
we have to require the ``Bethe Ansatz equations"
\begin{equation}
\zeta_k^{2L}\,\prod_{j\ne k}\,
\Big(A(\zeta_j,\zeta_k)A(\zeta_j^{-1},\zeta_k)\Bigr)=1,\quad k=1,\ldots,r.
\tag{b.30}
\end{equation}
We need to solve these equations (\ref{b.30}) in the thermodynamic limit
$L\to\infty$ at finite $r$ and substitute the solutions in (\ref{b.23})
and (\ref{b.24}).

The large-$L$ solution giving the largest eigenvalue is
\begin{equation}
\zeta_j={\cos\Bigl(({r\over2}-[{j-1\over2}])\lambda\Bigr)\over
\cos\Bigl(({r\over2}-[{j+1\over2}])\lambda\Bigr)},
\quad j=1,\ldots,r,
\tag{b.31}
\end{equation}
with $[x]$ the integer part of $x$. For even $r$ (\ref{b.31}) reduces to
\begin{equation}
r=2s,\quad\zeta_{2j-1}=\zeta_{2j}=
{\cos\Bigl((s-j+1)\lambda\Bigr)\over\cos\Bigl((s-j)\lambda\Bigr)},
\quad j=1,\ldots,s,
\tag{b.32}
\end{equation}
whereas for odd $r$ it reduces to\mydouble{\pagebreak}
\begin{eqnarray}
r=2s+1,\st&&\zeta_{2j-1}=\zeta_{2j}=
{\cos\Bigl((s-j+\fract32)\lambda\Bigr)\over\cos\Bigl((s-j+\half)\lambda\Bigr)},
\quad j=1,\ldots,s,\cr&&\cr
&&\zeta_{2s+1}=1.
\tag{b.33}
\end{eqnarray}
{}From these results we immediately obtain (\ref{b.4}) with $\mub_r$ given by
the last member of (\ref{b.5}). We see that we are precisely at the boundary
where the $\zeta$ pairs change from complex conjugate pairs to unequal real
pairs, which may be related to some pre-wetting phenomena on the integrable
line.

\setcounter{equation}{0}
\section{NEW MODEL\label{sec.new}}

There is already a great deal of exact results obtained for the chiral Potts
models. Yet not much of this has been utilized. One of the reasons is that the
temperature dependence in the integrable chiral Potts model, as given by
(\ref{2.7}) through (\ref{2.9}) is not very transparent. It is rather vague
what the critical temperature $T_{\rm c}$ is, thus making it harder to use the
exactly known exponents and other results for $N\ge4$.

Moreover, one would like to have a model which includes the chiral Potts model
as a special case, yet not so general as (\ref{2.2}) which has too many\B
variables. Such a model, with $N\ge3$, can be used to describe some\B
interesting physical phenomena, such as commensurate-incommensurate\B phase
transitions and wetting phenomena. It may not be exactly solvable in most
regions. Hence, numerical studies would have to be done on it.\B However, the
model may be in the same universality class as the integrable chiral Potts
model, and may be more interesting for $N\ge4$ than the chiral clock model
which has only a single cosine term in (\ref{2.12}).

\subsection{New 2-Parameter Chiral Potts Model for General
{\mybmi N}\label{sus.tpn}}

We therefore propose a new model with nearest-neighbor pair interactions
given by
\begin{equation}
-\cE(n)=E\,\sum_{j=1}^{N-1}\,{\sin(\pi/N)\over\sin(\pi j/N)}\,
\cos\left({2\pi\over N}\Bigl(nj+(N-2j)\Delta\Bigr)\right).
\tag {5.1}
\end{equation}
This has only two variables $E$ and $\Delta$. For $\Delta=\quarter$, it is
identical with the pair-interaction energy (\ref{4.17}) of the integrable
chiral Potts model at zero\B temperature and is thus on the superwetting line
there. For $\Delta\sim 0$, we find $\cE(0)$ is the ground state. In fact, it
is easy to show that for $\Delta<\Delta_{\rm CI}$, with
\begin{equation}
\Delta_{\rm CI}\equiv{1\over4}\,{(N+1)\over(N-1)},
\tag {e.22}
\end{equation}
and at low temperatures the system is in the ferromagnetic phase (or $W(0)$ is
the maximum). Within this regime, for $\Delta<\quarter$, the system is
unwetted, namely $\epsilon_p+\epsilon_{r-p}>\epsilon_r$; at
$\Delta=\quarter$, it goes through a superwetting transition, with
$\epsilon_p+\epsilon_{r-p}=\epsilon_r$; while for $\Delta>\quarter$, the
system is wet, with $\epsilon_p+\epsilon_{r-p}<\epsilon_r$. At
$\Delta=\Delta_{\rm CI}$, we find that $\cE(0)=\cE(-1)$ as in the chiral
clock model\pmyref{\ref{SY},\ref{ANNNI}} Therefore, the ground state is highly
degenerate and we believe that, at this point, the system goes through a
commensurate-incommensurate phase transition.

\subsection{Superwetting and Ground-State Degeneracy for
{\mybmi N}\mysp=\mysp5\label{sus.swf}}

To make all these statements clearer, let us take $N=5$ as an
illustration. We plot, in Fig.\ 12, the energy $\cE(n)$ given in (\ref{5.1})
as a function of $x=n/N$, for different values of $\Delta$.

\myfV

In Fig.\ 12a, we show the situation for $\Delta=\fract16<\fract14$. We find
that the increments in energy --- or the interfacial tension at zero
temperature --- $\epsilon_r=\cE(-r)-\cE(0)$ are given by
\begin{eqnarray}
&\epsilon_4=4.574329190,\quad&\epsilon_3=4.209056927,\cr
&\epsilon_2=3.315920618,\quad&\epsilon_1=2.036147842
\tag {5.2}
\end{eqnarray}
and are shown in Fig.\ 12a as the lengths of vertical straight lines
indicating the energy differences between the $\cE(-r)$ and the ground state
energy $\cE(0)$. By examining the lengths of these straight lines, we find
that $\epsilon_p+\epsilon_{r-p}>\epsilon_r$. Therefore it is energetically
unfavorable to have two interfaces with interfacial tensions $\epsilon_p$ and
$\epsilon_{r-p}$ instead of one with $\epsilon_r$. Hence, the system is not
wetted.

\myfW

In Fig.\ 12b, we plot the energy $\cE$ for $\Delta=\fract14$, and we find
$\epsilon_r=r\epsilon_1=2r\sin\pi/5$. This means that it is energetically
neutral to have one interface with $\epsilon_r$ or more than one, even $r$
interfaces with $\epsilon_1$. This is what we have called superwetting.

\myfX

In Fig.\ 12c, we plot $\cE$ for $\Delta=\fract3{10}>\fract14$. We now find
\begin{eqnarray}
&\epsilon_4=4.655997407,\quad&\epsilon_3=2.965539502,\cr
&\epsilon_2=1.703676245,\quad&\epsilon_1=0.6789144637
\tag {5.3}
\end{eqnarray}
and also that $r\epsilon_1<\epsilon_p+\epsilon_{r-p}<\epsilon_r$.
Consequently, it is energetically more favorable to have $r$ interfaces with
$\epsilon_1$ than just one with $\epsilon_r$ or another number less than $r$.
This means the system is wetted, allowing only one kind of interface with
interfacial tension $\epsilon_1$.

\myfY

We plot $\cE$ for $\Delta=\fract38$ in Fig.\ 13, and we find $\cE(N-1)=\cE(0)$.
This is the same situation as in the chiral-clock model, namely that the
ground state is highly degenerate for certain values of $\Delta$. Hence,
near $\Delta=\fract38$ and at\B low temperature, there exist floating
incommensurate phases. The model (\ref{5.1}) can then be used to describe
commensurate-incommensurate phase transitions. However, in these regimes,
exact solutions do not exist and numerical methods have to be be employed.

\subsection{Comparison with Integrable Model for
{\mybmi N}\mysp=\mysp5\label{sus.com}}

\myfZ

\myfz

Comparing with the values given by (\ref{4.11}) for the integrable model, we
find that the new model given by (\ref{5.1}) is integrable at zero temperature
at the special value $\Delta=\quarter$. On the other hand, unlike for the new
model, the ratios $K_j/K_1$ and $\Delta_j/\Delta_1$ for the integrable cases
are not constants, as shown in Fig.\ 8 to Fig.\ 11. We believe that the new
model may very well not be integrable at any other point. On the other hand, as
the ratios for the integrable cases are almost constants, it is interesting to
compare them. In Fig.\ 14a, we plot for $N=5$ the nearest-neighbor interaction
as a function of $x=n/N$ for the integrable case at the self-dual point with
\begin{equation}
\Delta_1=\Delta_2=0,\qquad K_1=1,\quad K_2/K_1=0.6409037097,
\tag {5.4}
\end{equation}
together with that of the new model at $\Delta=0$. Also, in Fig.\ 14b, we plot
the $\cE$ for the integrable model evaluated at $\theta=-9\pi/50$, with
\begin{eqnarray}
&\Delta_1=0.2943916857,\quad&\Delta_2=0.09778750304,\cr
&K_1=1,\phantom{.0000000000}\quad&K_2/K_1=0.6404446571,
\tag {5.5}
\end{eqnarray}
and the $\cE(n)$ of (\ref{5.1}) evaluated at $\Delta=\Delta_2=0.09778750304$.
We find that the curves are almost on top of each other.

As it is well known that by changing the strengths of the interactions, without
changing the symmetry (here $N$) or spatial dimension, the exponents are
universal, it is likely that the integrable model and the new model have the
same critical exponents for the same $N$. On the other hand, the models must be
different in some other sense and it would be interesting to find out more
about them.

It may also be interesting to study variations of model (\ref{5.1}), which
agree with the symmetric integrable model at a non-zero temperature and
to\B investigate if and how superwetting sets in. In the most general
chiral Potts model there are enough parameters to allow the existence
of solutions of the conditions for a superwetting line. It may be of
interest to see, by low-temperature expansion for example, if there are
solutions with real positive Boltzmann weights and what their physical
implications are.

\section*{\normalsize\bf ACKNOWLEDGMENTS}

We would like to thank Dr.\ A.\ J.\ Guttmann and Dr.\ W.\ Selke for their
interest in the chiral Potts model and for stimulating discussions during the
IUPAP conference in Berlin. We also like to thank Dr.\ B.\ M.\ McCoy for
encouragement. We especially would like to thank Dr.\ R.\ J.\ Baxter for
most helpful correspondence, which greatly enhanced our paper and helped
us correcting some of our errors.
\eject

\section*{\normalsize\bf REFERENCES}

\catcode`\@=11
\def\footnotesize{\@setsize\footnotesize{12pt}\xpt\@xpt
\abovedisplayskip 10pt plus2pt minus5pt\belowdisplayskip \abovedisplayskip
\abovedisplayshortskip \z@ plus3pt\belowdisplayshortskip 6pt plus3pt minus3pt
\def\@listi{\leftmargin\leftmargini \topsep 6pt plus 2pt minus 2pt\parsep 0pt
plus 0pt minus 0pt
\itemsep \parsep}}
\catcode`\@=12
\footnotesize\def\Bl{\B}

\begin{enumerate}
\item{}{L.\ Onsager, \jour{Phys.\ Rev.} \rpages{65}{1944}{117}{149}.}\label{O1}
\item{}{L.\ Onsager, \jour{Discussion, Nuovo Cimento}, {ser.\ 9},
\rspages{{6}{ \rm Suppl.}}{1949}{261}.}\label{O2}
\item{}{L.\ Onsager, in \book{Critical Phenomena in Alloys, Magnets and
Superconductors},\Bl R.\ E.\ Mills, E.\ Ascher, and R.\ I.\ Jaffee, eds.\
(McGraw-Hill, New York, 1971), \pppages{xix}{xxiv}, \nppages{3}{12}.}\label{O5}
\item{}{B.\ Kaufman,
\jour{Phys.\ Rev.} \rspages{76}{1949}{1232}.}
\label{O3}
\item{}{B.\ Kaufman and L.\ Onsager,
\jour{Phys.\ Rev.} \rspages{76}{1949}{1244}.}\label{O4}
\item{}{ A.\ E.\ Kennelly,
\jour{Electrical World and Engineer} \rpages{34}{1899}{413}{414}.}
\label{Kennelly}
\item{}{J.\ B.\ McGuire,
\jour{J.\ Math.\ Phys.} \rpages{5}{1964}{622}{636}.}\label{McGuire}
\item{}{C.\ N.\ Yang,
\jour{Phys.\ Rev.\ Lett.} \rpages{19}{1967}{1312}{1315}.}\label{Yang}
\item{}{R.\ J.\ Baxter, \book{Exactly Solved Models in Statistical Mechanics}
(Academic, London, 1982).}\label{BaxterBook}
\item{}{H.\ Au-Yang and J.\ H.\ H.\ Perk,
in \book{Advanced Studies in Pure Mathematics}, Vol.\ 19
(Kinokuniya-Academic, Tokyo, 1989), \ppages{57}{94}.}\label{APTa}\label{AP1}
\item{}{V.\ G.\ Kac, in \book{Infinite Dimensional Lie Algebras}
(Cambridge University Press, 1985).}\label{VKac}
\item{}{H.\ Bethe,
\jour{Z.\ Phys.} \rpages{71}{1931}{205}{226}.}\label{Bethe}
\item{}{P.\ P.\ Kulish and E.\ K.\ Sklyanin,
in \book{Integrable Quantum Field Theories},
J.\ Hietarinta\Bl and C.\ Montonen, eds.\
\book{Lecture Notes in Physics} {\bf 151}
(Springer, Berlin, 1981),\Bl \ppages{61}{119}.}\label{Kulish}
\item{}{M.\ Jimbo,
\jour{Lett.\ Math.\ Phys.} \rpages{10}{1985}{63}{69}.}\label{Jimbo}
\item{}{V.\ V.\ Bazhanov,
\jour{Phys.\ Lett.\ B} \rpages{159}{1985}{321}{324}.}\label{Bazhanov}
\item{}{V.\ G.\ Drinfel'd,
in \book{Proceedings of the International Congress of Mathematicians}
(Berkeley, Calif., 1986), \ppages{798}{820}.}\label{Drinfeld}
\item{}{L.\ D.\ Faddeev, N.\ Yu.\ Reshetikhin, and L.\ A.\ Takhtajan,
in \book{Algebraic Analysis}, Vol.\ 1, M.\ Kashiwara and T.\ Kawai, eds.\
(Academic, San Diego, 1988), \ppages{129}{139}.}\label{FaReTa}
\item{}{T.\ T.\ Wu,
\jour{Phys.\ Rev.} \rspages{149}{1966}{380}.}\label{Wu}
\item{}{L.\ P.\ Kadanoff,
\jour{Nuovo Cimento} \rspages{44B}{1966}{276}.}\label{Kadanoff}
\item{}{M.\ E.\ Fisher and R.\ J.\ Burford,
\jour{Phys.\ Rev.} \rspages{156}{1967}{583}.}\label{FisherBurford}
\item{}{T.\ T.\ Wu, B.\ M.\ McCoy, C.\ A.\ Tracy and E.\ Barouch,
\jour{Phys.\ Rev.\ B} \rspages{13}{1976}{316}.}\label{Wuetal}
\item{}{M.\ Sato, T.\ Miwa and M.\ Jimbo,
\jour{Proc.\ Japan Acad.} \rspages{53A}{1977}{6, 147, 153, 183}.}\label{Sato}
\item{}{J.\ H.\ H.\ Perk,
\jour{Phys.\ Lett.\ A} \rpages{79}{1980}{3}{5}.}\label{Perk2}
\item{}{E.\ H.\ Lieb,
\jour{Phys.\ Rev.\ Lett.} \rpages{18}{1967}{692}{694}.}\label{Lieb}
\item{}{E.\ H.\ Lieb and F.\ Y.\ Wu, in \book{Phase Transitions and
Critical Phenomena}, Vol.\ 1,\Bl C.\ Domb and M.\ S.\ Green, eds.\
(Academic, London, 1972), \ppages{331}{490}.}\label{LiebWu}
\item{}{G.\ von Gehlen and V.\ Rittenberg,
\jour{Nucl.\ Phys.\ B} \rpages{257 {\rm [FS14]}}{1985}{351}{370}.}\label{vGR}
\item{}{L.\ Dolan and M.\ Grady,
\jour{Phys.\ Rev.\ D} \rpages{25}{1982}{1587}{1604}.}\label{DoG}
\item{}{J.\ H.\ H.\ Perk,
in \book{Theta Functions Bowdoin 1987}, \jour{Proc.\ Symp.\ in Pure Math.}\Bl
Vol.\ 49 (Am.\ Math.\ Soc., Providence, RI, 1989).}\label{Perk}
\item{}{B.\ Davies,
\jour{J.\ Math.\ Phys.} \rpages{32}{1991}{2945}{2950}.}\label{Davies}
\item{}{H.\ Au-Yang, B.\ M.\ McCoy, J.\ H.\ H.\ Perk, S.\ Tang, and M.\ L.\
Yan, \jour{Phys.\ Lett.\ A} \rpages{123}{1987}{219}{223}.}\label{AMPTY}
\item{}{R.\ J.\ Baxter, J.\ H.\ H.\ Perk, and H.\ Au-Yang,
\jour{Phys.\ Lett.\ A} \rpages{128}{1988}{138}{142}.}\label{BPA}
\item{}{G.\ Albertini, B.\ M.\ McCoy, J.\ H.\ H.\ Perk, and S.\ Tang,
\jour{Nucl.\ Phys.\ B} \rpages{314}{1989}{741}{763}.}\label{AlMPT}
\item{}{A.\ O.\ Morris,
\jour{Quart.\ J.\ Math.\ Oxford}, ser.\ 2, \rpages{18}{1967}{7}{12};
\rpages{19}{1968}{289}{299}.}\label{Morris}
\item{}{A.\ J.\ Bracken and H.\ S.\ Green,
\jour{Nuovo Cimento} \rpages{9 {\rm A}}{1972}{349}{365},
and references quoted.}\label{Green}
\item{}{V.\ V.\ Bazhanov and Yu.\ G.\ Stroganov,
\jour{J.\ Stat.\ Phys.} \rpages{59}{1990}{799}{817}.}\label{BazhanovS}
\item{}{C.\ De Concini and V.\ G.\ Kac,
in \book{Operator Algebras, Unitary Representations,\Bl Enveloping
Algebras, and Invariant Theory}, A.\ Connes, M.\ Duflo, A.\ Joseph, and R.\
Rentschler, eds.\ (Birkh\"auser, Boston, 1990), \ppages{471}{506}.}\label{dCK}
\item{}{R.\ J.\ Baxter,
\jour{J.\ Stat.\ Phys.} \rpages{57}{1989}{1}{39}.}\label{Bax1}
\item{}{R.\ J.\ Baxter,
\jour{J.\ Stat.\ Phys.} \rpages{73}{1993}{461}{495}.}\label{Bax2}
\item{}{R.\ J.\ Baxter,
\jour{J.\ Phys.\ A} \rpages{27}{1994}{1837}{1849}.}\label{Bax3}
\item{}{R.\ J.\ Baxter,
\jour{J.\ Stat.\ Phys.} \rpages{52}{1988}{639}{667}.}\label{Bax4}
\item{}{G.\ Albertini, B.\ M.\ McCoy, and J.\ H.\ H.\ Perk,
in \book{Advanced Studies in Pure\Bl Mathematics}, Vol.\ 19
(Kinokuniya-Academic, Tokyo, 1989), \ppages{1}{55}.}\label{AlMP}
\item{}{R.\ B.\ Potts,
\jour{Proc.\ Camb.\ Phil.\ Soc.} \rpages{48}{1952}{106}{109}.}\label{Potts}
\item{}{F.\ Y.\ Wu,
\jour{Rev.\ Mod.\ Phys.} \rpages{54}{1982}{235}{268},
\rspages{55}{1983}{315}.}\label{WuRev}
\item{}{R.\ J.\ Baxter, \book{Exactly Solved Models in Statistical Mechanics}
(Academic, London, 1982), \pppages{322}{351}.}\label{Baxbk2}
\item{}{V.\ A.\ Fateev and A.\ B.\ Zamolodchikov,
\jour{Phys.\ Lett.\ A} \rpages{92}{1982}{37}{39}.}\label{FZ}
\item{}{B.\ M.\ McCoy, J.\ H.\ H.\ Perk, S.\ Tang, and C.\ H.\ Sah,
\jour{Phys.\ Lett.\ A} \rpages{125}{1987}{9}{14}.}\label{BPTS}
\item{}{H.\ Au-Yang, B.\ M.\ McCoy, J.\ H.\ H.\ Perk, and S.\ Tang,
in \book{Algebraic Analysis},\Bl Vol.\ 1, M.\ Kashiwara and
T.\ Kawai, eds.\ (Academic, San Diego, 1988), \ppages{29}{40}.}\label{AMPT}
\item{}{R.\ J.\ Baxter,
\jour{Phil.\ Trans.\ Roy.\ Soc.\ London Ser.\ A}
\rpages{289}{1978}{315}{346}.}\label{Bax-ZI}
\item{}{D.\ A.\ Huse, A.\ M.\ Szpilka, and M.\ E.\ Fisher,
\jour{Physica A} \rpages{121}{1983}{363}{398}.}\label{HSF}
\item{}{W.\ Selke and J.\ M.\ Yeomans,
\jour{Z.\ Phys.\ B} \rpages{46}{1982}{311}{318}.}\label{SY}
\item{}{W.\ Selke and W.\ Pesch,
\jour{Z.\ Phys.\ B} \rpages{47}{1982}{335}{340}.}\label{SP}
\item{}{M.\ E.\ Fisher,
\jour{J.\ Stat.\ Phys.} \rpages{34}{1984}{667}{729}.}\label{MEF}
\item{}{J.\ Yeomans and B.\ Derrida,
\jour{J.\ Phys.\ A} \rpages{18}{1985}{2343}{2355}.}\label{YD}
\item{}{W.\ Selke,
\jour{Phys.\ Repts.} \rpages{170}{1988}{213}{264}.}\label{ANNNI}
\item{}{A.\ L.\ Stella, X.-C.\ Xie, T.\ L.\ Einstein, and N.\ C.\ Bartelt,
\jour{Z.\ Phys.\ B} \rpages{67}{1987}{357}{361}.}\label{SXEB}
\item{}{R.\ B.\ Griffiths,
in \book{Phase Transitions in Surface Films}, J.\ G.\ Dash and J.\
Ruvalds, eds.\ (Plenum, New York and London, 1980), \ppages{1}{27},
see particularly eq.\ (5.13) and surrounding text.}\label{Griff}
\item{}{D.\ B.\ Abraham, L.\ F.\ Ko, and N.\ M.\ \v Svraki\'c,
\jour{Phys.\ Rev.\ B} \rpages{38}{1988}{12011}{12014}, and references
cited.}\label{AbrahamKoS}
\item{}{M.\ Gaudin,
\book{La Fonction d'Onde de Bethe} (Masson, Paris, 1983).}\label{Gaudin}
\item{}{A.\ N.\ Kirillov,
\jour{Zap.\ Nauch.\ Semin.\ LOMI} \rpages{164}{1987}{121}{133}
[\jour{J.\ Sov.\ Math.} \rpages{47}{1989}{2450}{2459}].}\label{Kir}
\item{}{W.\ Nahm, A.\ Recknagel, and M.\ Terhoeven,
\jour{Mod.\ Phys.\ Lett.\ A} \rpages{8}{1993}{1835}{1847}.}\label{NRT}
\item{}{M.\ den Nijs,
\jour{J.\ Phys.\ A} \rpages{17}{1984}{L295}{L300}.}\label{dN}
\item{}{P.\ P.\ Kulish, N.\ Yu.\ Reshetikhin, and E.\ K.\ Sklyanin,
\jour{Lett.\ Math.\ Phys.{}} \rpages{5}{1981}{393}{403}.}\label{KRS}
\end{enumerate}
\delete{\vfill\eject ~}
\mydouble {\vfill\eject
\section*{\normalsize\bf FIGURES CAPTIONS}
\begin{itemize}
\figcA \figcB \figcC \figcD \figcE \figcF \figcG \figcH \figcI \figcJ \figcK
\figcL \figcM \figcN \figcO \figcP \figcQ \figcR \figcS \figcT \figcU \figcV
\figcW \figcX \figcY \figcZ \figcz \end{itemize} }

\end{document}